\documentclass[journal=jacsat,manuscript=article]{achemso}

\usepackage[version=3]{mhchem} 
\usepackage[T1]{fontenc}       

\usepackage[labelfont=bf, font=small]{caption}

\usepackage{changepage}

\usepackage{amsmath,amssymb}

\usepackage{nameref,hyperref}

\usepackage{xparse}

\SectionNumbersOn 

\DeclareDocumentCommand \eref{oooo} {\IfNoValueTF{#2}{Eq~(\ref{#1})}{\IfNoValueTF{#3}{Eqs~(\ref{#1}) and (\ref{#2})}{\IfNoValueTF{#4}{Eqs~(\ref{#1})-(\ref{#3})}{Eqs~(\ref{#1})-(\ref{#4})}}}}
\newcommand{\bareEq}[1]{(\ref{#1})}

\DeclareDocumentCommand \fref{ooo} {\IfNoValueTF{#2}{Figure~\ref{#1}}{\IfNoValueTF{#3}{Figures~\ref{#1} and \ref{#2}}{Figures~\ref{#1}-\ref{#3}}}}
\newcommand{\letter}[1]{#1} 
\newcommand{\letterParen}[1]{(#1)} 

\newcommand{\sref}[1]{section~\ref{#1}}

\author{Tal Einav}
\affiliation{Department of Physics, California Institute of Technology, Pasadena, California 91125, United States}
\author{Linas Mazutis}
\affiliation{Institute of Biotechnology, Vilnius University, Vilnius, Lithuania}
\author{Rob Phillips}
\affiliation{Department of Applied Physics and Division of Biology, California Institute of Technology, Pasadena, California 91125, United States}
\email{phillips@pboc.caltech.edu}

\title[Statistical Mechanics of Allosteric Enzymes]{Statistical Mechanics of Allosteric Enzymes}

\begin{document}
	\begin{singlespace}
%
%
%
		
		\begin{abstract}
			The concept of allostery in which macromolecules switch between two
			different conformations is a central theme in biological processes ranging from gene
			regulation to cell signaling to enzymology. Allosteric enzymes pervade
			metabolic processes, yet a simple and unified treatment of the effects of
			allostery in enzymes has been lacking. In this work, we take the first step
			towards this goal by modeling allosteric enzymes and their interaction with
			two key molecular players - allosteric regulators and competitive inhibitors.
			We then apply this model to characterize existing data on enzyme activity,
			comment on how enzyme parameters (such as substrate binding affinity) can be
			experimentally tuned, and make novel predictions on how to control phenomena
			such as substrate inhibition.
		\end{abstract}
	\end{singlespace} 		
\renewcommand{\thepage}{S\arabic{page}}  

\title[Statistical Mechanics of Allosteric Enzymes Supplementary Information]{Statistical Mechanics of Allosteric Enzymes Supplementary Information}

\begin{document} 				

\begin{singlespace}


\section{Introduction}

All but the simplest of cellular reactions are catalyzed by enzymes,
macromolecules that can increase the rates of reactions by many orders of
magnitude. In some cases, such as phosphoryl transfer reactions, rate
enhancements can be as large as $10^{20}$-fold or more \cite{Lim2014}. A deeper
understanding of how enzymes work can provide insights into biological phenomena
as diverse as metabolic regulation or the treatment of disease
\cite{Hidestrand2001, Brattstrom1990, Zelezniak2010}. The basic principles of
enzyme mechanics were first proposed by Michaelis and Menten
\cite{Michaelis1913} and later extended by others \cite{Briggs1925, MONOD1965,
	Koshland1966}. While the earliest models considered enzymes as single-state
catalysts, experiments soon revealed that some enzymes exhibit richer dynamics
\cite{Cockrell2013, Wales1999}. The concept of allosteric enzymes was introduced
by Monod-Wyman-Changeux (MWC) and independently by Pardee and Gerhart
\cite{MONOD1963, MONOD1965, Gerhart1962, Gerhart1965}, providing a much broader
framework for explaining the full diversity of enzyme behavior. Since then, the
MWC concept in which macromolecules are thought of as having both an inactive
and active state has spread into many fields, proving to be a powerful
conceptual tool capable of explaining many biological phenomena \cite{Daber2009,
	Changeux2012, Marzen2013}.

Enzymology is a well studied field, and much has been learned both theoretically
and experimentally about how enzymes operate \cite{Segel1993,
	Cornish-Bowden1979, Fersht1999, Price1999}. With the vast number of distinct
molecular players involved in enzymatic reactions (for example: mixed,
competitive, uncompetitive, and non-competitive inhibitors as well as cofactors,
allosteric effectors, and substrate molecules), it is not surprising that new
discoveries continue to emerge about the subtleties of enzyme action
\cite{Reuveni2014, Pinto2015, Cockrell2013}. In this paper, we use the MWC model
to form a unifying framework capable of describing the broad array of behaviors
available to allosteric enzymes.

Statistical mechanics is a field of physics that describes the collective
behavior of large numbers of molecules. Historically developed to understand the
motion of gases, statistical physics has now seen applications in many areas of
biology and has provided unexpected connections between distinct problems such
as how transcription factors are induced by signals from the environment, the
function of the molecular machinery responsible for detecting small gradients in
chemoattractants, the gating properties of ligand-gated ion channels, and even
the accessibility of genomic DNA in eukaryotes which is packed into nucleosomes
\cite{Keymer2006, Endres2006, Mello2005, Hansen2008, Phillips2015a, Mirny2010,
	Narula2010}. One of us (RP) owes his introduction to the many beautiful uses of
statistical mechanics in biology to Bill Gelbart to whom this special issue is
dedicated.  During his inspiring career, Gelbart has been a passionate and
creative developer of insights into a wide number of problems using the tools of
statistical mechanics and we hope that our examples on the statistical mechanics
of allosteric enzymes will please him.

The remainder of the paper is organized as follows. In section
\ref{1SubstrateSite}, we show how the theoretical treatment of the traditional
Michaelis-Menten enzyme, an inherently non-equilibrium system, can be stated in
a language remarkably similar to equilibrium statistical mechanics. This sets
the stage for the remainder of the paper by introducing key notation and the
states and weights formalism that serves as the basis for analyzing more
sophisticated molecular scenarios. In \sref{MWCEnzymeSection}, we discuss how
the states and weights formalism can be used to work out the rates for the
simplest MWC enzyme, an allosteric enzyme with a single substrate binding site.
This is followed by a discussion of how allosteric enzymes are modified by the
binding of ligands, first an allosteric regulator in
\sref{AllostericEffectorSectionNew} and then a competitive inhibitor in
\sref{competitiveInihbitorSectionNew}. We next generalize to the much richer
case of enzymes with multiple substrate binding sites in
\sref{sectionMultipleSubstrateBindingSites}. Lastly, we discuss how to combine
the individual building blocks of allostery, allosteric effectors, competitive
inhibitors, and multiple binding sites to analyze general enzymes in
\sref{MoreComplexEnzymes}. Having built up this framework, we then apply our
model to understand observed enzyme behavior. In \sref{sec:ExperimentalData}, we
show how disparate enzyme activity curves can be unified within our model and
collapsed onto a single curve. We close by examining the exotic phenomenon of
substrate inhibition in \sref{secSubstrateInhibition} and show how the
allosteric nature of some enzymes may be the key to understanding and
controlling this phenomenon.

\section{Models} 

\subsection{Michaelis-Menten Enzyme} \label{1SubstrateSite}

We begin by briefly introducing the textbook Michaelis-Menten treatment of
enzymes \cite{Cornish-Bowden1979}. This will serve both to introduce basic
notation and to explain the states and weights methodology which we will use
throughout the paper.

Many enzyme-catalyzed biochemical reactions are characterized by
Michaelis-Menten kinetics. Such enzymes comprise a simple but important class
where we can study the relationship between the traditional chemical kinetics
based on reaction rates with a physical view dictated by statistical mechanics.
According to the Michaelis-Menten model, enzymes are single-state catalysts that
bind a substrate and promote its conversion into a product. Although this scheme
precludes allosteric interactions, a significant fraction of non-regulatory
enzymes (e.g. triosephosphate isomerase, bisphosphoglycerate mutase, adenylate
cyclase) are well described by Michaelis-Menten kinetics
\cite{Cornish-Bowden1979}.

The key player in this reaction framework is a monomeric enzyme $E$ which binds
a substrate $S$ at the substrate binding site (also called the active site or
catalytic site), forming an enzyme-substrate complex $ES$. The enzyme then
converts the substrate into product $P$ which is subsequently removed from the
system and cannot return to its original state as substrate. In terms of
concentrations, this reaction can be written as
\begin{equation} \label{eq:simple_scheme}
\begin{aligned}
\includegraphics[scale=1]{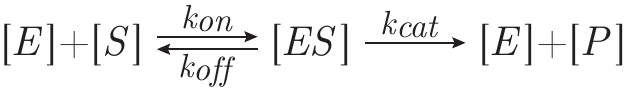}
\end{aligned}
\end{equation}
where the rate of product formation equals
\begin{equation}
\frac{d[P]}{dt}=[ES]k_{cat}. \label{eq:E_04}
\end{equation}

Briggs and Haldane assumed a time scale separation where the substrate and
product concentrations ($[S]$ and $[P]$) slowly change over time while the free
and bound enzyme states ($[E]$ and $[ES]$) changed much more rapidly
\cite{Briggs1925}. This allows us to approximate this system over  short time
scales by assuming that the slow components (in this case $[S]$) remain constant
and can therefore be absorbed into the $k_{on}$ rate \cite{Gunawardena2012},
\begin{equation} \label{eq:Equilibriumv2}
\begin{aligned}
\includegraphics[scale=1]{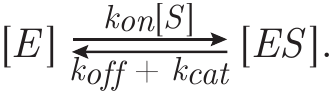}
\end{aligned}
\end{equation}
Assuming that the system \bareEq{eq:Equilibriumv2} reaches steady-state
(over the short time scale of this approximation) quickly enough that the
substrate concentration does not appreciably diminish, this implies
\begin{align} 
[E][S]k_{on}=[ES]\left(k_{off}+k_{cat}\right) \label{eq:E_07},
\end{align}
which we can rewrite as
\begin{equation} 
\frac{[ES]}{[E]}=\frac{[S]k_{on}}{k_{off}+k_{cat}}\equiv\frac{[S]}{K_M} \label{eq:E_07b}
\end{equation}
where $K_M=\frac{k_{off}+k_{cat}}{k_{on}}$ is called the \textit{Michaelis
	constant}. $K_M$ incorporates the binding and unbinding of ligand as well as the
conversion of substrate into product; in the limit $k_{cat}=0$, $K_M$ reduces to
the familiar dissociation constant $K_D=\frac{k_{off}}{k_{on}}$. Using
\eref[eq:E_07b] and the fact that the total enzyme concentration is conserved,
$[E]+[ES] = [E_{tot}]$, we can solve for $[E]$ and $[ES]$ separately as
\begin{align} 
[E] &= [E_{tot}]\frac{1}{1+\frac{[S]}{K_M}} \equiv [E_{tot}]p_E \label{eq:E_08a}\\
[ES] &= [E_{tot}]\frac{\frac{[S]}{K_M}}{1+\frac{[S]}{K_M}} \equiv [E_{tot}]p_{ES}, \label{eq:E_08b}
\end{align} 
where $p_E=\frac{[E]}{[E_{tot}]}$ and $p_{ES}=\frac{[ES]}{[E_{tot}]}$ are the
probabilities of finding an enzyme in the unbound and bound form, respectively.
Substituting the concentration of bound enzymes $[ES]$ from \eref[eq:E_08b] into
the rate of product formation \eref[eq:E_04],
\begin{equation} 
\frac{d[P]}{dt}=k_{cat}[E_{tot}]\frac{\frac{[S]}{K_M}}{1+\frac{[S]}{K_M}} \label{eq:E_09}.
\end{equation}
\fref[fig:1SubNonMWCcombined] shows the probability of free and bound enzyme as
well as the rate of product formation. The two parameters $k_{cat}$ and
$[E_{tot}]$ scale $\frac{d[P]}{dt}$ vertically (if $k_{cat}$ is increased by a
factor of 10, the $y$-axis values in \fref[fig:1SubNonMWCcombined]\letter{B}
will be multiplied by that same factor of 10), while $K_M$ effectively rescales
the substrate concentration $[S]$. Increasing $K_M$ by a factor of 10 implies
that 10 times as much substrate is needed to obtain the same rate of product
formation; on the semi-log plots in \fref[fig:1SubNonMWCcombined] this
corresponds to shifting all curves to the right by one power of 10.

\begin{figure}[h!]
	\centering \includegraphics[scale=1]{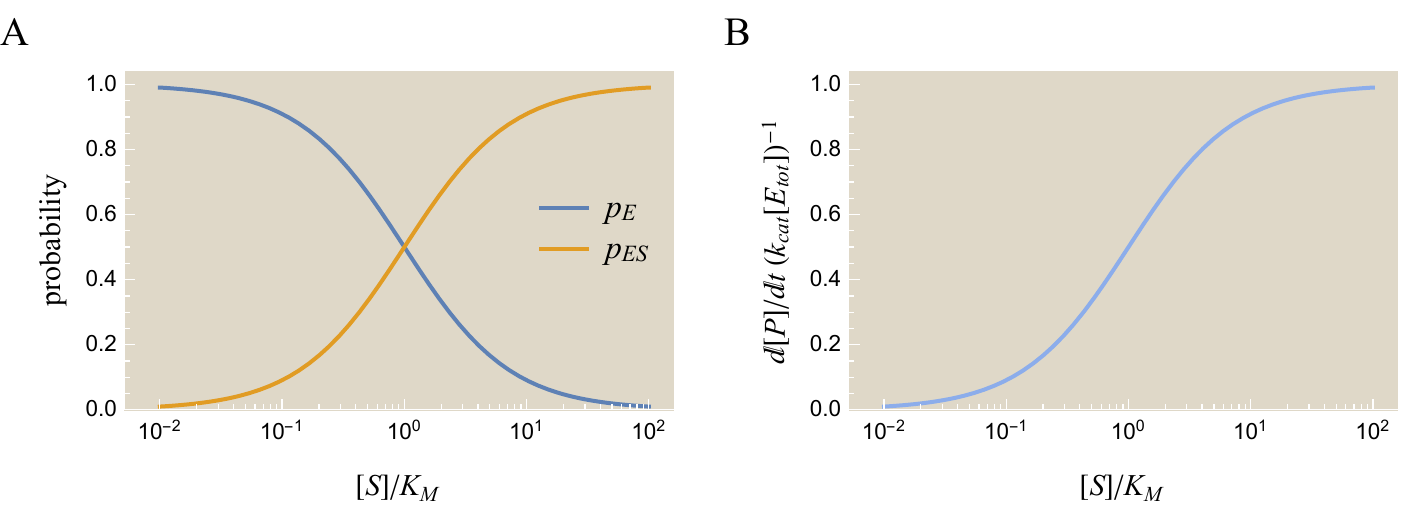}
	\caption{\textbf{Dynamics of the Michaelis-Menten enzyme.} \letterParen{A}
		Probabilities of the free enzyme $p_E$ and bound enzyme $p_{ES}$ states as a
		function of substrate concentration. As the amount of substrate $[S]$
		increases, more enzyme is found in the bound state rather than the free state.
		\letterParen{B} The rate of product formation for a non-allosteric enzyme. The
		rate of product formation has the same functional form as the probability
		$p_{ES}$ of the enzyme-substrate complex, as illustrated by
		\eref[eq:E_04][eq:E_08b].
	} \label{fig:1SubNonMWCcombined}
\end{figure}

\begin{figure}[h!]
	\centering \includegraphics[scale=1]{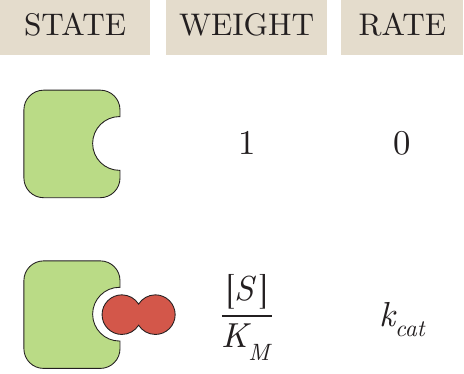} \caption{\textbf{States and
		weights for the Michaelis-Menten enzyme.} Each enzyme conformation is shown
		together with its weight and its catalytic rate. The probability of finding an
		enzyme (green) in either the free or bound state equals the weight of that
		state divided by the sum of all weights ($1+\frac{[S]}{K_M}$) where $[S]$ is
		the concentration of substrate (dark red) and
		$K_M=\frac{k_{off}+k_{cat}}{k_{on}}$ is the Michaelis constant. At $[S]=K_M$,
		half of the enzyme population exists in the free form and half exists in the
		bound form. For $[S]>K_M$, more than half of all enzymes will be bound to
		substrate.} \label{fig:CompetitorMWC3}
\end{figure}

We can visualize the microscopic states of the enzyme using a modified states
and weights diagram shown in \fref[fig:CompetitorMWC3] \cite{Bintu2005}. The
\textit{weight} of each enzyme state is proportional to the probability of its
corresponding state ($w_{E} \propto p_{E}$, $w_{ES} \propto p_{ES}$) - the
constant of proportionality is arbitrary but must be the same for all weights.
For example, from \eref[eq:E_08a][eq:E_08b] we can multiply the probability that
the enzyme will be unbound ($p_E$) or bound to substrate ($p_{ES}$) by
$1+\frac{[S]}{K_M}$ which yields the weights
\begin{align} 
w_{E}&=1\label{eq:E_10a}\\
w_{ES}&=\frac{[S]}{K_M}.\label{eq:E_10b}
\end{align} 
Given the weights of an enzyme state, we can proceed in the reverse direction
and obtain the probability for each enzyme state using
\begin{align} 
p_{E}&=\frac{w_{E}}{Z_{tot}} = \frac{1}{1+\frac{[S]}{K_M}}\label{eq:E_11a}\\
p_{ES}&=\frac{w_{ES}}{Z_{tot}} = \frac{\frac{[S]}{K_M}}{1+\frac{[S]}{K_M}}\label{eq:E_11b}
\end{align} 
where 
\begin{equation} \label{eq:partitionFunctionMichaelis}
Z_{tot}=w_{E}+w_{ES}
\end{equation}
is the sum of all weights. Dividing by $Z_{tot}$ ensures the total probability
of all enzyme states equals unity, $p_{E} + p_{ES} = 1$. The rate of product
formation \eref[eq:E_09] is given by the product of the enzyme concentration
$[E_{tot}]$ times the average catalytic rate over all states, weighed by
each state's (normalized) weights. In the following sections, we will find this
trick of writing states and weights very useful for modeling other molecular
players.

The weights in \fref[fig:CompetitorMWC3] allow us to easily understand
\fref[fig:1SubNonMWCcombined]\letter{A}: when $[S]<K_M$, $w_E>w_{ES}$ so that an
enzyme is more likely to be in the substrate-free state; when $[S]>K_M$,
$w_E<w_{ES}$ and an enzyme is more likely to be found as an enzyme-substrate
complex. Increasing $K_M$ shifts the tipping point of how much substrate is
needed before the bound $ES$ enzyme state begins to dominate over the free $E$
state.

It should be noted that the formal notion of states and weights employed in
physics applies only to equilibrium systems. For example, a ligand binding to a
receptor in equilibrium will yield states and weights similar to
\fref[fig:CompetitorMWC3] but with the Michaelis constant $K_M$ replaced by the
dissociation constant $K_D$ \cite{Phillips2010}. Yet the ligand-receptor states
and weights can also be derived from the Boltzmann distribution (where the
weight of any state $j$ with energy $E_j$ is proportional to $e^{-\beta E_j}$)
while the enzyme states and weights cannot be derived from the Boltzmann
distribution (because the enzyme system is not in equilibrium). Instead, the
non-equilibrium kinetics of the system are described by the modified states and
weights in \fref[fig:CompetitorMWC3], where the $K_D$ for substrate must be
replaced with $K_M$. These modified states and weights serve as a mathematical
trick that compactly and correctly represents the behavior of the enzyme,
enabling us to apply the well established tools and intuition of equilibrium
statistical mechanics when analyzing the inherently non-equilibrium problem of
enzyme kinetics. In the next several sections, we will show how to generalize
this method of states and weights to MWC enzymes with competitive inhibitors,
allosteric regulators, and multiple substrate binding sites.

\subsection{MWC Enzyme} \label{MWCEnzymeSection}

Many enzymes are not static entities, but dynamic macromolecules that constantly
fluctuate between different conformational states. This notion was initially
conceived by Monod-Wyman-Changeux (MWC) to characterize complex multi-subunit
proteins such as hemoglobin and aspartate transcarbamoylase (ATCase)
\cite{MONOD1963, MONOD1965, Gerhart1962}. The authors suggested that the ATC
enzyme exists in two supramolecular states: a relaxed ``R'' state, which has
high-affinity for substrate and a tight ``T'' state, which has low-affinity for
substrate. Although in the case of ATCase, the transition between the T and R
states is induced by an external ligand, recent experimental advances have shown
that many proteins intrinsically fluctuate between these different states even
in the absence of ligand \cite{Gardino2003, Milligan2003, Kern2003}. These
observations imply that the MWC model can be applied to a wide range of enzymes
beyond those with multi-subunit complexes.

We will designate an enzyme with two possible states (an \underline{A}ctive
state $E_A$ and an \underline{I}nactive state $E_I$) as an MWC enzyme. The
kinetics of a general MWC enzyme are given by
\begin{equation} \label{eq:ratesEnzymeMWC2}
\begin{aligned}
\includegraphics[scale=1]{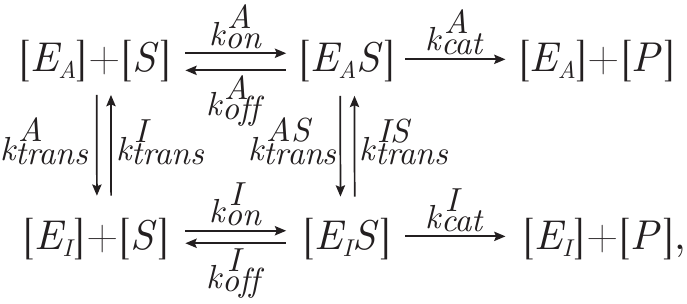}
\end{aligned}
\end{equation}
which relates the active and inactive enzyme concentrations ($[E_A]$, $[E_I]$)
to the active and inactive enzyme-substrate complexes ($[E_AS]$, $[E_IS]$). In
this two-state MWC model, similar to that explored by Howlett et
al.\cite{Howlett1977}, the rate of product formation is given by
\begin{equation} \label{basicMWCdPdt}
\frac{d[P]}{dt}=k_{cat}^{A}[E_AS]+k_{cat}^{I}[E_IS].
\end{equation}
The active state will have a faster catalytic rate (often much faster) than the inactive state, $k_{cat}^A > k_{cat}^I$.

As in the case of a Michaelis-Menten enzyme, we will assume that all four forms
of the enzyme ($E_A$, $E_I$, $E_AS$, and $E_IS$) quickly reach steady state on
time scales so short that the substrate concentration $[S]$ remains nearly
constant. Therefore, we can incorporate the slowly-changing quantities $[S]$ and
$[P]$ into the rates, a step dubbed the \textit{quasi-steady-state
	approximation} \cite{Gunawardena2012}. This allows us to rewrite the scheme
\bareEq{eq:ratesEnzymeMWC2} in the following form,
\begin{equation} \label{eq:ratesEnzymeMWCBetter}
\begin{aligned}
\includegraphics[scale=1]{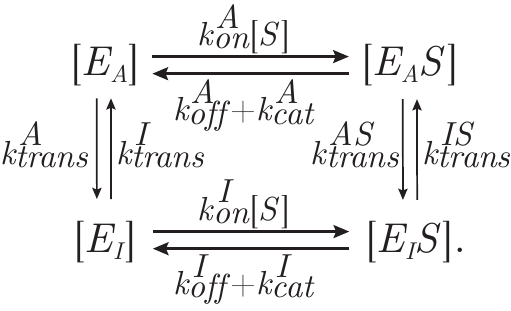}
\end{aligned}
\end{equation}

Assuming the quasi-steady-state approximation holds, the four enzyme states will
rapidly attain steady-state values
\begin{equation} \label{eq:quasiSteadyStateMWCEnzyme}
\frac{d[E_AS]}{dt}=\frac{d[E_A]}{dt}=\frac{d[E_IS]}{dt}=\frac{d[E_I]}{dt}=0.
\end{equation} 
In addition, a separate constraint on the system that is necessary and
sufficient to apply the method of states and weights is given by the
\textit{cycle condition}: the product of rates going clockwise around any cycle
must equal the product of rates going counterclockwise \cite{Gunawardena2012}.
It should be noted that to violate the cycle condition, a system must
continuously pay energy since at least one step in any cycle must be
energetically unfavorable. We shall proceed with the assumption that there are
no such cycles in our system. For the MWC enzyme
\bareEq{eq:ratesEnzymeMWCBetter}, this implies
\begin{equation} \label{eq:cycleLawMWCRawForm}
\left(k_{on}^{A}[S]\right) k_{trans}^{AS} \left( k_{off}^{I}+k_{cat}^{I} \right) k_{trans}^{I}= \left({k_{off}^{A}}+k_{cat}^{A}\right) k_{trans}^{A} \left(k_{on}^{I}[S]\right) k_{trans}^{IS}
\end{equation}
or equivalently
\begin{equation} \label{eq:cycleLawMWC}
\underbrace{\frac{k_{on}^{A}[S]}{{k_{off}^{A}}+k_{cat}^{A}}}_{\frac{[E_{A}S]}{[E_{A}]}} \underbrace{\vphantom{\frac{k_{on}^{A}[S]}{{k_{off}^{A}}+k_{cat}^{A}}}\frac{k_{trans}^{I}}{k_{trans}^{A}}}_{\frac{[E_{A}]}{[E_{I}]}}= \underbrace{\frac{k_{on}^{I}[S]}{k_{off}^{I}+k_{cat}^{I}}}_{\frac{[E_{I}S]}{[E_{I}]}} 
\underbrace{\vphantom{\frac{k_{on}^{A}[S]}{{k_{off}^{A}}+k_{cat}^{A}}}\frac{k_{trans}^{IS}}{k_{trans}^{AS}}}_{\frac{[E_{A}S]}{[E_{I}S]}}.
\end{equation}

The validity of both the quasi-steady-state approximation
\bareEq{eq:quasiSteadyStateMWCEnzyme} and the cycle condition
\bareEq{eq:cycleLawMWC} will be analyzed in Appendix
A. Assuming both statements hold, we
can invoke \textit{detailed balance} - the ratio of concentrations between two
enzyme states equals the inverse of the ratio of rates connecting these two
states. For example, between the active states $[E_AS]$ and $[E_A]$ in
\bareEq{eq:ratesEnzymeMWCBetter},
\begin{equation}
\frac{[E_{A}S]}{[E_{A}]}=\frac{k_{on}^{A}[S]}{k_{off}^{A}+k_{cat}^{A}} \equiv \frac{[S]}{K_M^{A}}\label{MWCKmDefa}
\end{equation}
where we have defined the Michaelis constant for the active state, $K_M^{A}$.
Similarly, we can write the equation for detailed balance between the inactive
states $[E_IS]$ and $[E_I]$ as
\begin{equation}
\frac{[E_{I}S]}{[E_{I}]}=\frac{k_{on}^{I}[S]}{k_{off}^{I}+k_{cat}^{I}} \equiv \frac{[S]}{K_M^{I}}\label{MWCKmDefb}.
\end{equation}
An enzyme may have a different affinity for substrate or a different catalytic
rate in the active and inactive forms. Typical measured values of $K_M$ fall
into the range $10^{-7}-10^{-1}\,\text{M}$ \cite{Wolfenden2006}. Whether $K_M^A$
or $K_M^I$ is larger depends on the specific enzyme.

As a final link between the language of chemical rates and physical energies, we
can recast detailed balance between $[E_A]$ and $[E_I]$ as
\begin{equation} \label{MWCEpsDef}
\frac{[E_A]}{[E_I]}=\frac{k_{trans}^{I}}{k_{trans}^{A}} \equiv e^{-\beta \left(\epsilon_A-\epsilon_I\right)},
\end{equation}
where $\epsilon_A$ and $\epsilon_I$ are the free energies of the enzyme in the
active and inactive state, respectively, and $\beta = \frac{1}{k_BT}$ where
$k_B$ is Boltzmann's constant and $T$ is the temperature of the system. Whether
the active state energy is greater than or less than the inactive state energy
depends on the enzyme. For example, $\epsilon_I < \epsilon_A$ in ATCase whereas
the opposite holds true, $\epsilon_A < \epsilon_I$, in chemoreceptors
\cite{Cockrell2013, Phillips2010}.

\begin{figure}[t]
	\centering \includegraphics[scale=1]{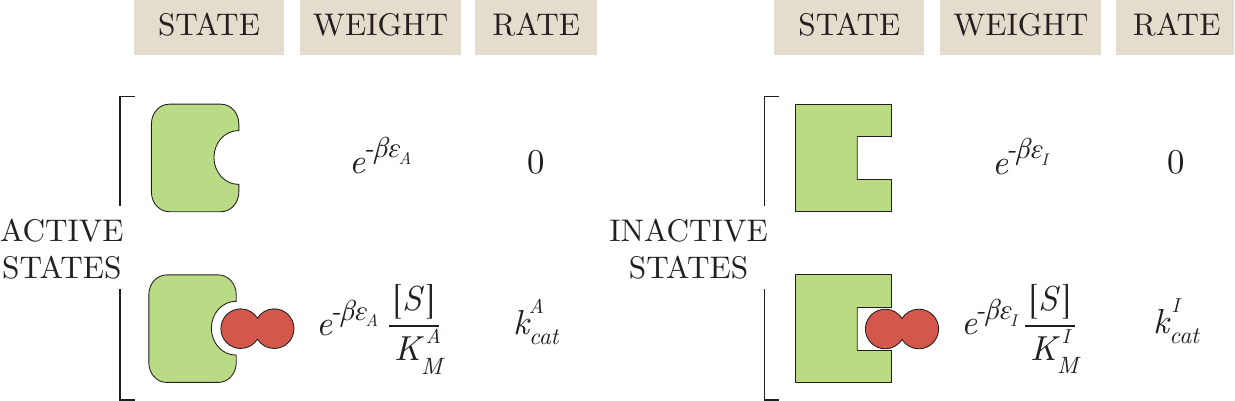} 
	\caption{\textbf{States and weights for an MWC enzyme.} The energies $\epsilon_A$ and
		$\epsilon_I$ provide the free energy scale for the substrate-free
		conformations, dictating their relative probabilities. Decreasing the energy
		$\epsilon_A$ of the active state would increase the probability of all the
		active enzyme conformations relative to the inactive conformations. $K_M^A$
		denotes the substrate concentration at which half of the active enzymes are
		bound and half the active enzymes are unbound, as indicated by the crossing of
		the ($p_{E_A}$, blue) and ($p_{E_AS}$, gold) curves at $[S]=K_M^A$ in
		\fref[fig:1SubNonMWCdPdt]. $K_M^I$ serves an analogous role for the inactive
		states.} \label{fig:statesWeightsMWCEnzyme2}
\end{figure}

Using \eref[MWCKmDefa][MWCKmDefb][MWCEpsDef], we can recast the cycle condition
\bareEq{eq:cycleLawMWC} (as shown in the under-braces) into a simple
relationship between the steady-state enzyme concentrations. Additionally, we
can use these equations to define the weights of each enzyme state in
\fref[fig:statesWeightsMWCEnzyme2]. Following \sref{1SubstrateSite}, the
probability of each state equals its weight divided by the sum of all weights,
\begin{align}
p_{E_A}&=e^{-\beta \epsilon_A}\frac{1}{Z_{tot}}\label{MWCBasicResults1a}\\
p_{E_AS}&=e^{-\beta \epsilon_A}\frac{\frac{[S]}{K_M^{A}}}{Z_{tot}}\label{MWCBasicResults1b}\\
p_{E_I}&=e^{-\beta \epsilon_I}\frac{1}{Z_{tot}}\label{MWCBasicResults1c}\\
p_{E_IS}&=e^{-\beta \epsilon_I}\frac{\frac{[S]}{K_M^{I}}}{Z_{tot}}\label{MWCBasicResults1d},
\end{align}
where 
\begin{equation} \label{eq:partitionFunctionMWC}
Z_{tot}=e^{-\beta \epsilon_A}\left(1+\frac{[S]}{K_M^{A}}\right)+e^{-\beta
	\epsilon_I}\left(1+\frac{[S]}{K_M^{I}}\right).
\end{equation}
Note that multiplying all of the
weights by a constant $c$ will also multiply $Z_{tot}$ by $c$, so that the
probability of any state will remain unchanged. That is why in
\fref[fig:CompetitorMWC3] we could neglect the $e^{-\beta \epsilon}$ factor
that was implicitly present in each weight.

The total amount of enzyme is conserved among the four enzyme states,
$[E_{tot}]=[E_A]+[E_AS]+[E_I]+[E_IS]$. Using this fact together with
\eref[MWCKmDefa][MWCKmDefb][MWCEpsDef] enables us to solve for the concentrations
of both types of bound enzymes, namely, 
\begin{align}
[E_AS] &= [E_{tot}]\frac{e^{-\beta \epsilon_A}\frac{[S]}{K_M^{A}}}{e^{-\beta \epsilon_A}\left(1+\frac{[S]}{K_M^{A}}\right)+e^{-\beta \epsilon_I}\left(1+\frac{[S]}{K_M^{I}}\right)} = [E_{tot}]p_{E_AS}\label{MWCBasicResults2a}\\
[E_IS] &= [E_{tot}]\frac{e^{-\beta \epsilon_I}\frac{[S]}{K_M^{I}}}{e^{-\beta \epsilon_A}\left(1+\frac{[S]}{K_M^{A}}\right)+e^{-\beta \epsilon_I}\left(1+\frac{[S]}{K_M^{I}}\right)} = [E_{tot}]p_{E_IS}.\label{MWCBasicResults2b}
\end{align}
Substituting these relations into \bareEq{basicMWCdPdt} yields the rate of
product formation,
\begin{align} \label{eq:basicMWCFinaldPdtEqquation}
\frac{d[P]}{dt} = [E_{tot}]\frac{k_{cat}^{A}e^{-\beta \epsilon_A}\frac{[S]}{K_M^{A}} + k_{cat}^{I} e^{-\beta \epsilon_I}\frac{[S]}{K_M^{I}}}{e^{-\beta \epsilon_A}\left(1+\frac{[S]}{K_M^{A}}\right)+e^{-\beta \epsilon_I}\left(1+\frac{[S]}{K_M^{I}}\right)}.
\end{align} 
The probabilities \bareEq{MWCBasicResults1a}-\bareEq{MWCBasicResults1d} of the
different states and the rate of product formation
\bareEq{eq:basicMWCFinaldPdtEqquation} are shown in \fref[fig:1SubNonMWCdPdt].
Although we use the same parameters from \fref[fig:1SubNonMWCcombined] for the
active state, the $p_{E_A}$ and $p_{E_AS}$ curves in
\fref[fig:1SubNonMWCdPdt]\letter{A} look markedly different from the $p_{E}$ and
$p_{ES}$ Michaelis-Menten curves in \fref[fig:1SubNonMWCcombined]\letter{A}.
This indicates that the activity of an MWC enzyme \textit{does not} equal the
activity of two independent Michaelis-Menten enzymes, one with the MWC enzyme's
active state parameters and the other with the MWC enzyme's inactive state
parameters. The interplay of the active and inactive states makes an MWC enzyme
inherently more complex than a Michaelis-Menten enzyme.

When $[S]=0$ the enzyme only exists in the unbound states $E_A$ and $E_I$ whose
relative probabilities are given by $\frac{p_{E_A}}{p_{E_I}}=e^{-\beta
	\left(\epsilon_A-\epsilon_I\right)}$. When $[S] \to \infty$, the enzyme spends
all of its time in the bound states $E_AS$ and $E_IS$ which have relative
probabilities $\frac{p_{E_AS}}{p_{E_IS}}=e^{-\beta
	\left(\epsilon_A-\epsilon_I\right)}\frac{K_M^I}{K_M^A}$. The curves for the
active states (for free enzyme $p_{E_A}$ and bound enzyme $p_{E_AS}$) intersect
at $[S]=K_M^A$ while the curves of the two inactive states intersect at
$[S]=K_M^I$. For the particular parameters shown, even though the unbound
\textit{inactive} state (green) dominates at low substrate concentrations, the
\textit{active} state (gold) has the largest statistical weights as the
concentration of substrate increases. Thus, adding substrate causes the enzyme
to increasingly favor the active state.

\begin{figure}[h!]
	\centering \includegraphics[scale=1]{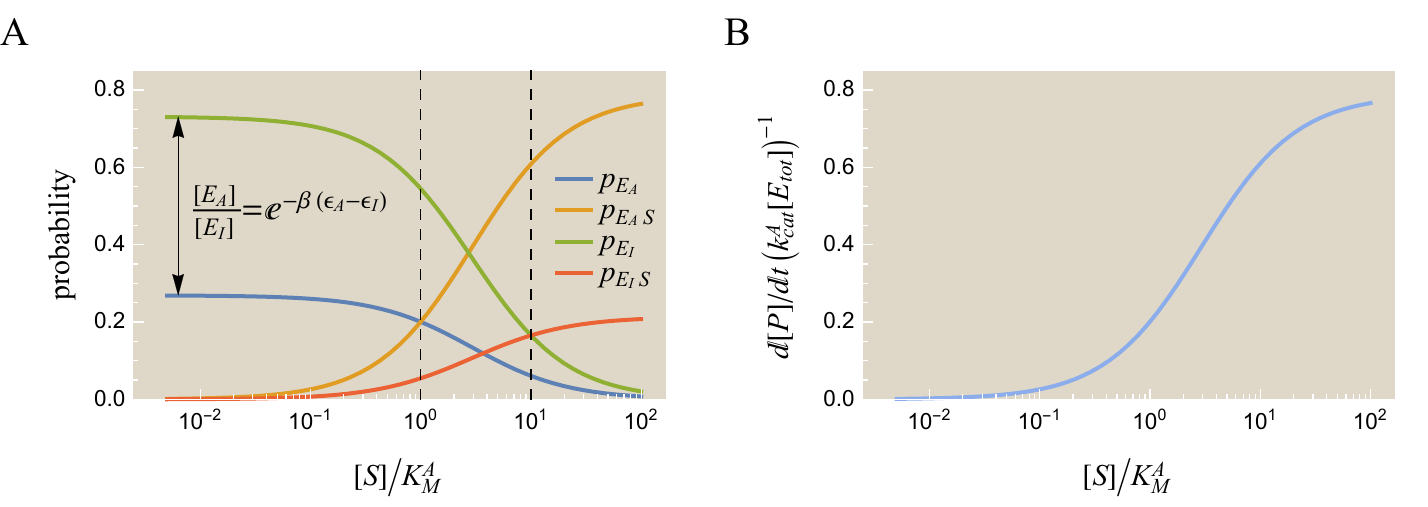} \caption{\textbf{Quantitative
		description of an MWC enzyme.} \letterParen{A} Probabilities of each enzyme
		state. While the active state has the same catalytic rate $k_{cat}^A$ and
		Michaelis constant $K_M^A$ as the Michaelis-Menten enzyme in
		\fref[fig:1SubNonMWCcombined]\letter{A}, the inactive state significantly
		alters the forms of $p_{E_A}$ and $p_{E_AS}$. The dashed vertical lines
		indicate where the substrate concentration equals $K_M^A$ and $K_M^I$,
		respectively. \letterParen{B} The rate of product formation, $\frac{d[P]}{dt}$.
		Assuming $\frac{k_{cat}^A}{k_{cat}^I} \gg 1$, $\frac{d[P]}{dt}$ (blue curve in
		\letterParen{B}) is dominated by the active enzyme-substrate complex,
		$p_{E_AS}$ (gold curve in \letterParen{A}). Parameters were chosen to reflect
		``typical'' enzyme kinetics values: $\frac{k_{cat}^A}{k_{cat}^I}=10^2$,
		$\frac{K_M^A}{K_M^I}=10^{-1}$, and $e^{-\beta
			\left(\epsilon_A-\epsilon_I\right)}=e^{-1}$ \cite{Phillips2015}. Substrate
		concentrations are shown normalized relative to the active state parameter
		$\frac{[S]}{K_M^A}$, although the inactive state parameter $\frac{[S]}{K_M^I}$
		could also have been used.} \label{fig:1SubNonMWCdPdt}
\end{figure}

Using this framework, we can compute properties of the enzyme kinetics curve
shown in \fref[fig:1SubNonMWCdPdt]\letterParen{B}. One important property is the
dynamic range of an enzyme, the difference between the maximum and minimum rate
of product formation. In the absence of substrate ($[S] \to 0$) and a saturating
concentration of substrate ($[S] \to \infty$), the rate of product formation
\eref[eq:basicMWCFinaldPdtEqquation] becomes
\begin{align}
\lim_{[S] \to 0} \frac{d[P]}{dt} &= 0 \\
\lim_{[S] \to \infty} \frac{d[P]}{dt} &= [E_{tot}]\frac{k_{cat}^A \frac{e^{-\beta \epsilon_A}}{K_M^{A}} + k_{cat}^I \frac{e^{-\beta \epsilon_I}}{K_M^{I}}}{\frac{e^{-\beta \epsilon_A}}{K_M^{A}}+\frac{e^{-\beta \epsilon_I}}{K_M^{I}}}.
\end{align}
From these two expressions, we can write the dynamic range as
\begin{align}
\text{dynamic range} &= \left( \lim_{[S] \to \infty} \frac{d[P]}{dt} \right) - \left( \lim_{[S] \to 0} \frac{d[P]}{dt} \right) \nonumber \\
&= [E_{tot}] k_{cat}^A \left(1 - \frac{1 - \frac{k_{cat}^I}{k_{cat}^A}}{1 + e^{-\beta (\epsilon_A - \epsilon_I)}\frac{K_M^{I}}{K_M^{A}}}\right) \label{eqDynamicRangeSimplified}
\end{align}
where every term in the fraction has been written as a ratio of the active and
inactive state parameters. We find that the dynamic range increases as
$\frac{k_{cat}^I}{k_{cat}^A}$, $e^{-\beta (\epsilon_A - \epsilon_I)}$, and
$\frac{K_M^{I}}{K_M^{A}}$ increase (assuming $k_{cat}^A > k_{cat}^I$).

Another important property is the concentration of substrate at which the rate
of product formation lies halfway between its minimum and maximum value, which
we will denote as $[S_{50}]$. It is straightforward to show that the definition
\begin{equation}
\lim_{[S] \to [S_{50}]} \frac{d[P]}{dt} = \frac{1}{2} \left( \lim_{[S] \to \infty} \frac{d[P]}{dt} + \lim_{[S] \to 0} \frac{d[P]}{dt} \right)
\end{equation}
is satisfied when
\begin{equation}
[S_{50}] = K_M^A \frac{e^{-\beta (\epsilon_A - \epsilon_I)} + 1}{e^{-\beta (\epsilon_A - \epsilon_I)} + \frac{K_M^{A}}{K_M^{I}}}. \label{eqS50}
\end{equation}
With increasing $e^{-\beta (\epsilon_A - \epsilon_I)}$, the value of $[S_{50}]$
increases if $K_M^A > K_M^I$ and decreases otherwise. $[S_{50}]$ always
decreases as $\frac{K_M^{A}}{K_M^{I}}$ increases. Lastly, we note that in the
limit of a Michaelis-Menten enzyme, $\epsilon_I \to \infty$, we recoup the
familiar results
\begin{alignat}{2}
\text{dynamic range} &= [E_{tot}] k_{cat}^A && \;\;\;\;\;\;\;\; (\epsilon_I \to \infty)\\
[S_{50}] &= K_M^A && \;\;\;\;\;\;\;\; (\epsilon_I \to \infty).
\end{alignat}

\subsection{Allosteric Regulator} \label{AllostericEffectorSectionNew}

The catalytic activity of many enzymes is controlled by molecules that bind to
regulatory sites which are often different from the active sites themselves. As
a result of ligand-induced conformational changes, these molecules alter the
substrate binding site which modifies the rate of product formation,
$\frac{d[P]}{dt}$. Allosterically controlled enzymes represent important
regulatory nodes in metabolic pathways and are often responsible for keeping
cells in homeostasis. Some well-studied examples of allosteric control include
glycogen phosphorylase, phosphofructokinase, glutamine synthetase, and aspartate
transcarbamoylase (ATCase). In many cases the data from these systems are
characterized phenomenologically using Hill functions, but the Hill coefficients
thus obtained can be difficult to interpret \cite{Frank2013}. In addition, Hill
coefficients do not provide much information about the organization or
regulation of an enzyme, nor do they reflect the relative probabilities of the
possible enzyme conformations, although recent results have begun to address
these issues \cite{Dyachenko2013}. In this section, we add one more layer of
complexity to our statistical mechanics framework by introducing an allosteric
regulator.

\begin{figure}[h!]
	\centering \includegraphics[scale=1]{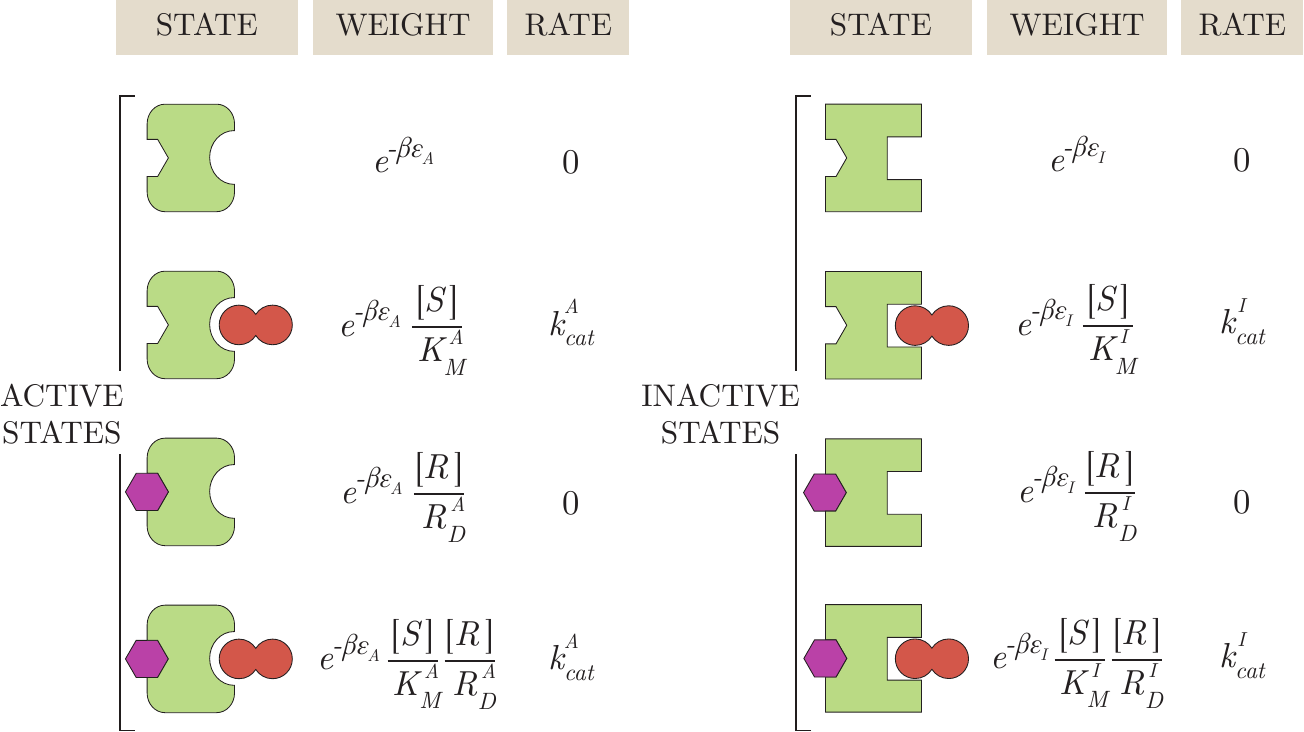} \caption{\textbf{States and
		weights for an MWC enzyme with an allosteric regulator.} The allosteric
		regulator (purple) does not directly interact with the substrate (dark red) but
		instead introduces a factor $\frac{[R]}{R_D}$ into the weights where $R_D$ is a
		\textit{dissociation} constant. Note that the regulator can only associate to
		and dissociate from the enzyme, whereas substrate can be turned into product as
		shown by the \textit{Michaelis} constant $K_M$. An allosteric activator binds
		more tightly to the active state enzyme, $R_D^A<R_D^I$, which leads to an
		increased rate of product formation because the active state catalyzes
		substrate at a faster rate than the inactive state, $k_{cat}^{A} >
		k_{cat}^{I}$. An allosteric inhibitor would satisfy $R_D^A>R_D^I$.}
	\label{fig:One_site_Enzyme_MWC_Reg}
\end{figure}

Consider an MWC enzyme with one site for an allosteric regulator $R$ and a
different site for a substrate molecule $S$ that will be converted into product.
We can define the effects of the allosteric regulator directly through the
states and weights. As shown in \fref[fig:One_site_Enzyme_MWC_Reg], the
regulator $R$ contributes a factor $\frac{[R]}{R_D^A}$ when it binds to an
active state and a factor $\frac{[R]}{R_D^I}$ when it binds to an inactive state
where $R_D^A$ and $R_D^I$ are the dissociation constants between the regulator
and the active and inactive states of the enzyme, respectively. Unlike the
\textit{Michaelis} constants $K_M^A$ and $K_M^I$ for the substrate, the
\textit{dissociation} constants $R_D^A$ and $R_D^I$ enter the states and weights
because the regulator can only bind and unbind to the enzyme (and cannot be
transformed into product). In other words, if we were to draw a rates diagram
for this enzyme system, detailed balance between the two states where the
regulator is bound and unbound would yield a dissociation constant
($\frac{k_{off}}{k_{on}}$) rather than a Michaelis constant
($\frac{k_{off}+k_{cat}}{k_{on}}$).

Using the states and weights in \fref[fig:One_site_Enzyme_MWC_Reg], we can
compute the probability of each enzyme state. For example, the probabilities of
the four states that form product are given by
\begin{align}
p_{E_AS}&=e^{-\beta \epsilon_A}\frac{\frac{[S]}{K_M^{A}}}{Z_{tot}}\label{MWCActprobabilitiesa}\\
p_{E_ASR}&=e^{-\beta \epsilon_A}\frac{\frac{[S]}{K_M^{A}}\frac{[R]}{R_D^{A}}}{Z_{tot}}\label{MWCActprobabilitiesb}\\
p_{E_IS}&=e^{-\beta \epsilon_I}\frac{\frac{[S]}{K_M^{I}}}{Z_{tot}}\label{MWCActprobabilitiesc}\\
p_{E_ISR}&=e^{-\beta \epsilon_I}\frac{\frac{[S]}{K_M^{I}}\frac{[R]}{R_D^{I}}}{Z_{tot}}\label{MWCActprobabilitiesd}
\end{align}
where 
\begin{equation} \label{eq:partitionFunctionRegulator}
Z_{tot}=e^{-\beta
	\epsilon_A}\left(1+\frac{[S]}{K_M^{A}}\right)\left(1+\frac{[R]}{R_D^{A}}\right)+e^{-\beta \epsilon_I}\left(1+\frac{[S]}{K_M^{I}}\right)\left(1+\frac{[R]}{R_D^{I}}\right)
\end{equation}
is the sum of all weights in \fref[fig:One_site_Enzyme_MWC_Reg]. An allosteric
activator has a smaller dissociation constant $R_D^A<R_D^I$ for binding to the
active state enzyme, so that for larger $[R]$ the probability that the enzyme
will be in the active state increases. Because the active state catalyzes
substrate at a faster rate than the inactive state, $k_{cat}^{A} > k_{cat}^{I}$,
adding an activator increases the rate of product formation $\frac{d[P]}{dt}$.
An allosteric inhibitor has the flipped relation $R_D^A>R_D^I$ and hence causes
the opposite effects.

Proceeding analogously to \sref{MWCEnzymeSection}, the total enzyme
concentration $[E_{tot}]$ is a conserved quantity which equals the sum of all
enzyme states ($[E_A]$, $[E_AS]$, $[E_AR]$, $[E_ASR]$, and their inactive state
counterparts). Using the probabilities in
\eref[MWCActprobabilitiesa][MWCActprobabilitiesb][MWCActprobabilitiesc][MWCActprobabilitiesd], we can write these concentrations as $[E_AS]=[E_{tot}]p_{E_AS}$, $[E_ASR]=[E_{tot}]p_{E_ASR},...$ so that the rate of product formation is given by
\begin{align}
\frac{d[P]}{dt}&=k_{cat}^A \left( [E_AS]+[E_ASR] \right)+k_{cat}^I \left( [E_IS]+[E_ISR] \right) \nonumber \\
&=[E_{tot}]\frac{k_{cat}^A e^{-\beta \epsilon_A}\frac{[S]}{K_M^{A}}\left(1+\frac{[R]}{R_D^{A}} \right)+k_{cat}^I e^{-\beta \epsilon_I}\frac{[S]}{K_M^{I}}\left(1+\frac{[R]}{R_D^{I}} \right)}{e^{-\beta \epsilon_A}\left(1+\frac{[S]}{K_M^{A}}\right)\left(1+\frac{[R]}{R_D^{A}}\right)+e^{-\beta \epsilon_I}\left(1+\frac{[S]}{K_M^{I}}\right)\left(1+\frac{[R]}{R_D^{I}}\right)}. \label{MWCActdPdt}
\end{align}

The rate of product formation \bareEq{MWCActdPdt} for different $[R]$ values is
shown in \fref[fig:1Act1SubdPdt]. It is important to realize that by choosing
the weights in \fref[fig:One_site_Enzyme_MWC_Reg], we have selected a particular
model for the allosteric regulator, namely one in which the regulator binds
equally well to an enzyme with or without substrate. There are many other
possible models. For example, we could add an interaction energy between an
allosteric regulator and a bound substrate. However, the simple model in
\fref[fig:One_site_Enzyme_MWC_Reg] already possesses the important feature that
adding more allosteric activator yields a larger rate of product formation
$\frac{d[P]}{dt}$, as shown in \fref[fig:1Act1SubdPdt].

\begin{figure}[h!]
	\centering \includegraphics[scale=1]{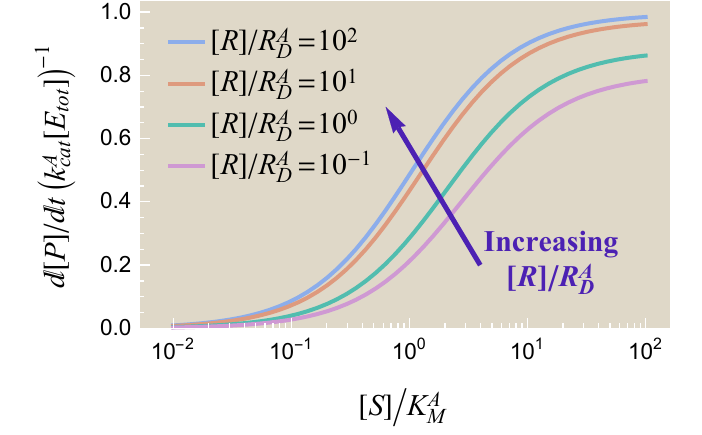} \caption{\textbf{Effects of an
		allosteric regulator $\boldsymbol{R}$ on the rate of product formation $\boldsymbol{\frac{d[P]}{dt}}$.}
		The regulator's greater affinity for the active enzyme state increases the
		fraction of the active conformations and hence $\frac{d[P]}{dt}$. Parameters
		used were $\frac{R_D^A}{R_D^I}=10^{-2}$ and the parameters from
		\fref[fig:1SubNonMWCdPdt].} \label{fig:1Act1SubdPdt}
\end{figure}

An allosteric regulator effectively tunes the energies of the active and
inactive states. To better understand this, consider the probability of an
active state enzyme-substrate complex (with or without a bound regulator).
Adding \eref[MWCActprobabilitiesa][MWCActprobabilitiesb],
\begin{align}
p_{E_AS}+p_{E_ASR}&=\frac{e^{-\beta \epsilon_A}\frac{[S]}{K_M^{A}}\left( 1+\frac{[R]}{R_D^{A}} \right)}{e^{-\beta \epsilon_A}\left(1+\frac{[S]}{K_M^{A}}\right)\left(1+\frac{[R]}{R_D^{A}}\right)+e^{-\beta \epsilon_I}\left(1+\frac{[S]}{K_M^{I}}\right)\left(1+\frac{[R]}{R_D^{I}}\right)}\nonumber \\
&\equiv\frac{e^{-\beta \tilde{\epsilon}_A}\frac{[S]}{K_M^{A}}}{e^{-\beta \tilde{\epsilon}_A}\left(1+\frac{[S]}{K_M^{A}}\right)+e^{-\beta \tilde{\epsilon}_I}\left(1+\frac{[S]}{K_M^{I}}\right)}\label{explainingRegulatorFunction}
\end{align}
where 
\begin{align}
\tilde{\epsilon}_A&=\epsilon_A-\frac{1}{\beta}\log{\left(1+\frac{[R]}{R_D^{A}}\right)}\label{eq:allostericEnzymeEquation112a}\\
\tilde{\epsilon}_I&=\epsilon_I-\frac{1}{\beta}\log{\left(1+\frac{[R]}{R_D^{I}}\right)}.\label{eq:allostericEnzymeEquation112b}
\end{align}
We now compare the total probability that an active state enzyme will be bound
to substrate in the presence of an allosteric regulator
(\eref[explainingRegulatorFunction]) to this probability in the absence of an
allosteric regulator (\eref[MWCBasicResults1b]). These two equations show that
an MWC enzyme in the presence of regulator concentration $[R]$ is equivalent to
an MWC enzyme with no regulator provided that we use the new energies
$\tilde{\epsilon}_A$ and $\tilde{\epsilon}_I$ for the active and inactive
states. An analogous statement holds for all the conformations of the enzyme, so
that the effects of a regulator can be completely absorbed into the energies of
the active and inactive states! In other words, adding an allosteric regulator
allows us to tune the parameters $\epsilon_A$ and $\epsilon_I$ of an allosteric
enzyme, and thus change its rate of product formation, in a quantifiable manner.
This simple result emerges from our assumptions that the allosteric regulator and
substrate bind independently to the enzyme and that the allosteric regulator
does not effect the rate of product formation.

One application of this result is that we can easily compute the dynamic
range of an enzyme as well as the concentration of substrate for half-maximal
rate of product formation discussed in \sref{MWCEnzymeSection}. Both of these
quantities follow from the analogous expressions for an MWC enzyme
(\eref[eqDynamicRangeSimplified][eqS50]) using the effective energies
$\tilde{\epsilon}_A$ and $\tilde{\epsilon}_I$, resulting in a dynamic range of the form
\begin{equation}
\text{dynamic range} = [E_{tot}] k_{cat}^A \left(1 - \frac{1 - \frac{k_{cat}^I}{k_{cat}^A}}{1 + e^{-\beta (\epsilon_A - \epsilon_I)} \frac{1+[R]/R_D^{A}}{1+[R]/R_D^{I}} \frac{K_M^{I}}{K_M^{A}}} \right)
\end{equation}
and an $[S_{50}]$ value of
\begin{equation}
[S_{50}] = K_M^A \frac{e^{-\beta (\epsilon_A - \epsilon_I)} \frac{1+[R]/R_D^{A}}{1+[R]/R_D^{I}} + 1}{e^{-\beta (\epsilon_A - \epsilon_I)} \frac{1+[R]/R_D^{A}}{1+[R]/R_D^{I}} + \frac{K_M^{A}}{K_M^{I}}}.
\end{equation}
As expected, the dynamic range of an enzyme increases with regulator
concentration $[R]$ for an allosteric activator ($R_D^A<R_D^I$). Adding more
activator will shift $[S_{50}]$ to the left if $K_M^{A} < K_M^{I}$ (as shown in
\fref[fig:1Act1SubdPdt]) or to the right if $K_M^{A} > K_M^{I}$. The opposite
effects hold for an allosteric inhibitor ($R_D^I<R_D^A$).

\subsection{Competitive Inhibitor} \label{competitiveInihbitorSectionNew}

Another level of control found in many enzymes is inhibition. A competitive
inhibitor $C$ binds to the same active site as substrate $S$, yet unlike the
substrate, the competitive inhibitor cannot be turned into product by the
enzyme. An enzyme with a single active site can either exist in the unbound
state $E$, as an enzyme-substrate complex $ES$, or as an enzyme-competitor
complex $EC$. As more inhibitor is added to the system, it crowds out the
substrate from the enzyme's active site which decreases product formation. Many
cancer drugs (e.g. lapatinib, sorafenib, erlotinib) are competitive inhibitors
for kinases involved in signaling pathways \cite{MellinghoffIngoK.Sawyers2012}.

\begin{figure}[h]
	\centering \includegraphics[scale=1]{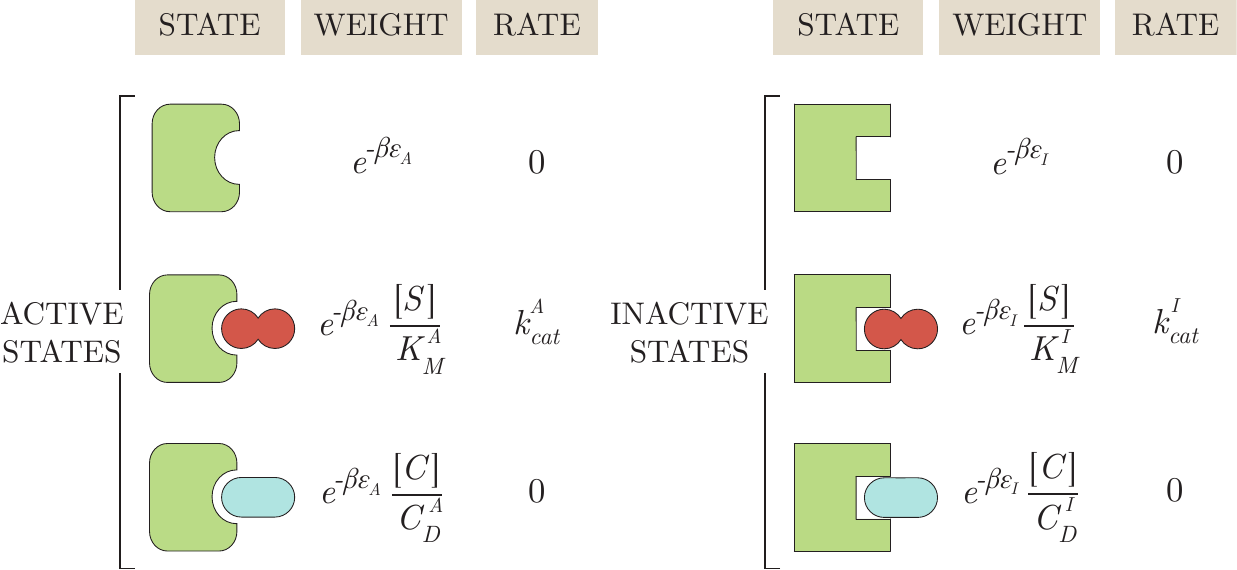} \caption{\textbf{States and
		weights for an MWC enzyme with a competitive inhibitor.} While the substrate $S$
		(dark red) can be transformed into product, the inhibitor $C$ (light blue) can
		occupy the substrate binding site but cannot be catalyzed. As seen with the
		allosteric regulator in \sref{AllostericEffectorSectionNew}, the competitive
		inhibitor contributes a factor $\frac{[C]}{C_D}$ to the statistical weight of a
		state where $C_D$ is the inhibitor's dissociation constant.}
	\label{fig:CompetitorStatesWeights}
\end{figure}

Starting from our model of an MWC enzyme in \fref[fig:statesWeightsMWCEnzyme2],
we can introduce a competitive inhibitor by drawing two new states (an
enzyme-competitor complex in the active and inactive forms) as shown in
\fref[fig:CompetitorStatesWeights]. Only the enzyme-substrate complex in the
active ($E_AS$) and inactive ($E_IS$) states form product. The probabilities of
each of these states is given by \eref[MWCBasicResults1b][MWCBasicResults1d] but
using the new partition function (which includes the competitive inhibitor
states),
\begin{equation} \label{eq:partitionFunctionCompetitor}
Z_{tot}=e^{-\beta
	\epsilon_A}\left(1+\frac{[S]}{K_M^{A}}+\frac{[C]}{C_D^{A}}\right)+e^{-\beta
	\epsilon_I}\left(1+\frac{[S]}{K_M^{I}}+\frac{[C]}{C_D^{I}}\right).
\end{equation}
Repeating the same analysis from \sref{MWCEnzymeSection}, we write the
concentrations of bound enzymes as $[E_AS]=[E_{tot}]p_{E_AS}$ and
$[E_IS]=[E_{tot}]p_{E_IS}$, where $[E_{tot}]$ is the total concentration of
enzymes in the system and $p_{E_{A,I}S}$ is the weight of the bound (in)active
state enzyme divided by the partition function,
\eref[eq:partitionFunctionCompetitor]. Thus the rate of product formation equals
\begin{align}
\frac{d[P]}{dt}&=k_{cat}^A [E_AS]+k_{cat}^I [E_IS] \nonumber \\
&=[E_{tot}]\frac{k_{cat}^Ae^{-\beta \epsilon_A}\frac{[S]}{K_M^{A}}+k_{cat}^Ie^{-\beta \epsilon_I}\frac{[S]}{K_M^{I}}}{e^{-\beta \epsilon_A}\left(1+\frac{[S]}{K_M^{A}}+\frac{[C]}{C_D^{A}}\right)+e^{-\beta \epsilon_I}\left(1+\frac{[S]}{K_M^{I}}+\frac{[C]}{C_D^{I}}\right)}. \label{eq:MWCActCompdPdt}
\end{align}

\begin{figure}[h!]
	\centering \includegraphics[scale=1]{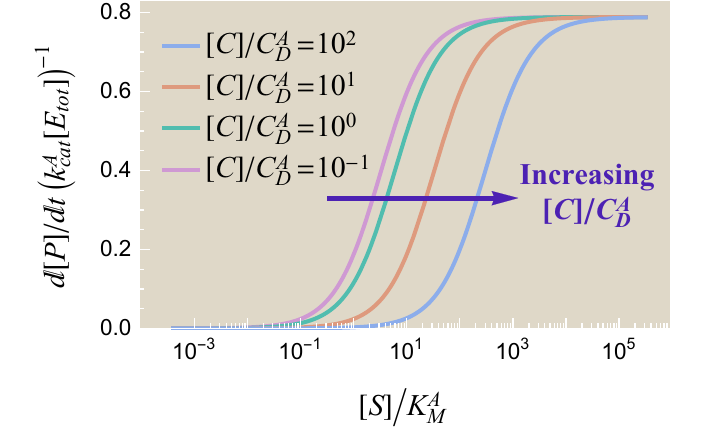} \caption{\textbf{Effects of a
		competitive inhibitor $\boldsymbol{C}$ on the rate of product formation $\boldsymbol{\frac{d[P]}{dt}}$.}
		When $[C] \lesssim C_D^A, C_D^I$, the inhibitor cannot out-compete the
		substrate at high substrate concentrations while the free form of enzyme
		dominates at low substrate concentrations. Therefore increasing $[C]$ up to
		values of $\approx C_D^A$ or $C_D^I$ has little effect on $\frac{d[P]}{dt}$.
		Once $[C] \gtrsim C_D^A, C_D^I$, the inhibitor can out-compete substrate at
		large concentrations, pushing the region where the enzyme-substrate complex
		dominates further to the right. Parameters used were $\frac{C_D^A}{C_D^I}=1$
		and the parameters from \fref[fig:1SubNonMWCdPdt].}
	\label{fig:1Sub1ActCompdPdt}
\end{figure}

\fref[fig:1Sub1ActCompdPdt] shows the rate of product formation for various
inhibitor concentrations $[C]$. Adding more competitive inhibitor increases the
probability of the inhibitor-bound states and thereby drains probability out of
those states competent to form product, as expected. Similarly to our analysis
of allosteric regulators, we can absorb the effects of the competitive inhibitor
($C_D^{A,I}$) in \eref[eq:MWCActCompdPdt] into the enzyme parameters
($\epsilon_{A,I}$, $K_M^{A,I}$),
\begin{align}
	\frac{d[P]}{dt} &= [E_{tot}] \frac{k_{cat}^A e^{-\beta \epsilon_A}\big(1+\frac{[C]}{C_D^{A}}\big)\frac{[S]}{K_M^{A}\big(1+\frac{[C]}{C_D^{A}}\big)}+k_{cat}^I e^{-\beta \epsilon_I}\big(1+\frac{[C]}{C_D^{I}}\big)\frac{[S]}{K_M^{I} \big(1+\frac{[C]}{C_D^{I}}\big)}}{e^{-\beta \epsilon_A}\big(1+\frac{[C]}{C_D^{A}}\big)\Big(1+\frac{[S]}{K_M^{A} \big(1+\frac{[C]}{C_D^{A}}\big)}\Big)+e^{-\beta \epsilon_I}\big(1+\frac{[C]}{C_D^{I}}\big)\Big(1+\frac{[S]}{K_M^{I} \big(1+\frac{[C]}{C_D^{I}}\big)}\Big)} \nonumber \\
	&\equiv [E_{tot}]\frac{k_{cat}^Ae^{-\beta \tilde{\epsilon}_A}\frac{[S]}{\tilde{K}_M^{A}}+k_{cat}^Ie^{-\beta \tilde{\epsilon}_I}\frac{[S]}{\tilde{K}_M^{I}}}{e^{-\beta \tilde{\epsilon}_A}\left(1+\frac{[S]}{\tilde{K}_M^{A}}\right)+e^{-\beta \tilde{\epsilon}_I}\left(1+\frac{[S]}{\tilde{K}_M^{I}}\right)}, \label{eq:ActJustLikeMWCNew}
\end{align}
where we have defined the new energies and Michaelis constants,
\begin{align}
\tilde{\epsilon}_A&=\epsilon_A-\frac{1}{\beta}\log{\left(1+\frac{[C]}{C_D^{A}}\right)}\label{eq:competitorEnzymeEquation112a}\\
\tilde{\epsilon}_I&=\epsilon_I-\frac{1}{\beta}\log{\left(1+\frac{[C]}{C_D^{I}}\right)}\label{eq:competitorEnzymeEquation112b}\\
\tilde{K}_M^{A}&=K_M^{A}\left(1+ \frac{[C]}{C_D^{A}}\right)\label{eq:competitorEnzymeEquation112c}\\
\tilde{K}_M^{I}&=K_M^{I}\left(1+ \frac{[C]}{C_D^{I}}\right)\label{eq:competitorEnzymeEquation112d}.
\end{align}
Note that \eref[eq:ActJustLikeMWCNew] has exactly the same form as the rate of
product formation of an MWC enzyme without a competitive inhibitor,
\eref[eq:basicMWCFinaldPdtEqquation]. In other words, a competitive inhibitor
modulates both the effective energies and the Michaelis constants of the active
and inactive states. Thus, an observed value of $K_M$ may not represent a true
Michaelis constant if an inhibitor is present. In the special case of a
Michaelis-Menten enzyme ($e^{-\beta \epsilon_I} \to 0$), we recover the
known result that a competitive inhibitor only changes the apparent Michaelis
constant \cite{Segel1993}.

As shown for the allosteric regulator, the dynamic range and the concentration
of substrate for half-maximal rate of product formation $[S_{50}]$ follow from
the analogous expressions for an MWC enzyme (\sref{MWCEnzymeSection},
\eref[eqDynamicRangeSimplified][eqS50]) using the parameters
$\tilde{\epsilon}_{A,I}$ and $\tilde{K}_M^{A,I}$. Hence an allosteric enzyme
with one active site in the presence of a competitive inhibitor has a dynamic
range given by
\begin{equation} \label{eqDynRangeInhibitor}
\text{dynamic range} = [E_{tot}] k_{cat}^A \left(1 - \frac{1 - \frac{k_{cat}^I}{k_{cat}^A}}{1 + e^{-\beta (\epsilon_A - \epsilon_I)} \frac{K_M^{I}}{K_M^{A}}} \right)
\end{equation}
and an $[S_{50}]$ value of
\begin{equation}
[S_{50}] = K_M^A \frac{e^{-\beta (\epsilon_A - \epsilon_I)} \left(1+ \frac{[C]}{C_D^{A}}\right) + \left(1+ \frac{[C]}{C_D^{I}}\right)}{e^{-\beta (\epsilon_A - \epsilon_I)} + \frac{K_M^{A}}{K_M^{I}}}.
\end{equation}

Notice that \eref[eqDynRangeInhibitor], the dynamic range of an MWC enzyme in
the presence of a competitive inhibitor, is exactly the same as
\eref[eqDynamicRangeSimplified], the dynamic range in the absence of an
inhibitor. This makes sense because in the absence of substrate ($[S] \to 0$)
the rate of product formation must be zero and at saturating substrate
concentrations ($[S] \to \infty$) the substrate completely crowds out any
inhibitor concentration. Instead of altering the rate of product formation at
these two limits, the competitive inhibitor shifts the $\frac{d[P]}{dt}$ curve,
and therefore $[S_{50}]$, to the right as more inhibitor is added.

Said another way, adding a competitive inhibitor effectively rescales the
concentration of substrate in a system. Consider an MWC enzyme in the absence of
a competitive inhibitor at a measured substrate concentration $[S_{\text{no\,}[C]}]$. Now
consider a separate system where an enzyme is in the presence of a competitive
inhibitor at concentration $[C]$ and at a measured substrate concentration
$[S_{\text{with}[C]}]$. It is straightforward to show that the rate of product
formation $\frac{d[P]}{dt}$ is the same for both enzymes,
\begin{align}
\frac{d[P]}{dt} &=[E_{tot}]\frac{k_{cat}^Ae^{-\beta \epsilon_A}\frac{[S_{\text{no\,}[C]}]}{K_M^{A}}+k_{cat}^Ie^{-\beta \epsilon_I}\frac{[S_{\text{no\,}[C]}]}{K_M^{I}}}{e^{-\beta \epsilon_A}\left(1+\frac{[S_{\text{no\,}[C]}]}{K_M^{A}}\right)+e^{-\beta \epsilon_I}\left(1+\frac{[S_{\text{no\,}[C]}]}{K_M^{I}}\right)} \nonumber \\
&=[E_{tot}]\frac{k_{cat}^Ae^{-\beta \epsilon_A}\frac{[S_{\text{with}[C]}]}{K_M^{A}}+k_{cat}^Ie^{-\beta \epsilon_I}\frac{[S_{\text{with}[C]}]}{K_M^{I}}}{e^{-\beta \epsilon_A}\left(1+\frac{[S_{\text{with}[C]}]}{K_M^{A}}+\frac{[C]}{C_D^{A}}\right)+e^{-\beta \epsilon_I}\left(1+\frac{[S_{\text{with}[C]}]}{K_M^{I}}+\frac{[C]}{C_D^{I}}\right)},
\end{align}
provided that
\begin{equation}
[S_{\text{with}[C]}] = \frac{e^{-\beta (\epsilon_A - \epsilon_I)} \left(1+ \frac{[C]}{C_D^{A}}\right) + \left(1+ \frac{[C]}{C_D^{I}}\right)}{e^{-\beta (\epsilon_A - \epsilon_I)} + 1} [S_{\text{no\,}[C]}]. \label{eqInhibitorSubstrateRelation}
\end{equation}
For any fixed competitive inhibitor concentration $[C]$, this rescaling amounts
to a constant multiplicative factor which results in a horizontal shift on a log
scale of substrate concentration $[S]$, as is indeed shown in
\fref[fig:1Sub1ActCompdPdt].

As we have seen, the effects of both an allosteric regulator and a competitive
inhibitor can be absorbed into the parameters of an MWC enzyme. This suggests
that experimental data from enzymes that titrate these ligands can be collapsed
into a 1-parameter family of curves where the single parameter is either the
concentration of an allosteric effector or a competitive inhibitor. Indeed, in
\sref{sec:ExperimentalData} we shall find that this theory matches well with
experimentally measured activity curves.

\subsection{Multiple Substrate Binding Sites} \label{sectionMultipleSubstrateBindingSites}

In 1965, Gerhart and Schachman used ultracentrifugation to determine that ATCase
can be separated into a large (100 kDa) catalytic subunit where substrate binds
and a smaller (30 kDa) regulatory subunit which has binding sites for the
allosteric regulators ATP and CTP \cite{Gerhart2014}. Their measurements
correctly predicted that ATCase had multiple active sites and multiple
regulatory sites, although their actual numbers were off (they predicted 2
active sites and 4 regulatory sites, whereas ATCase has 6 active sites and 6
regulatory sites) \cite{Gerhart1965}. Three years later, more refined sequencing
by Weber and crystallographic measurements by Wiley and Lipscomb revealed the
correct quaternary structure of ATCase \cite{Lipscomb2012, WEBER1968,
	WILEY1968}.

Many enzymes are composed of multiple subunits that contain substrate binding
sites (also called active sites or catalytic sites). Having multiple binding
sites grants the substrate more locations to bind to an enzyme which increases
the effective affinity between both molecules. A typical enzyme will have
between 1 and 6 substrate binding sites, and bindings sites for allosteric
regulators can appear with similar multiplicity. However, extreme cases exist
such as hemocyanin which can have as many as 48 active sites. \cite{Yokota1984}
Interestingly, across different species the same enzyme may possess different
numbers of active or regulatory sites, as well as be affected by other
allosteric regulators and competitive inhibitors \cite{Taylor2008, Wales1999}.
Furthermore, multiple binding sites may interact with each other in a complex
and often uncharacterized  manner \cite{Giroux1994}.

We now extend the single-site model of an MWC enzyme introduced in
\fref[fig:statesWeightsMWCEnzyme2] to an MWC enzyme with two substrate binding
sites. Assuming that both binding sites are identical and independent, the
states and weights of the system are shown in \fref[fig:Two-site_MWC]. When the
enzyme is doubly occupied $E_AS^2$, we assume that it forms product twice as
fast as a singly occupied enzyme $E_AS$.

\begin{figure}[h!]
	\centering
	\includegraphics[scale=1]{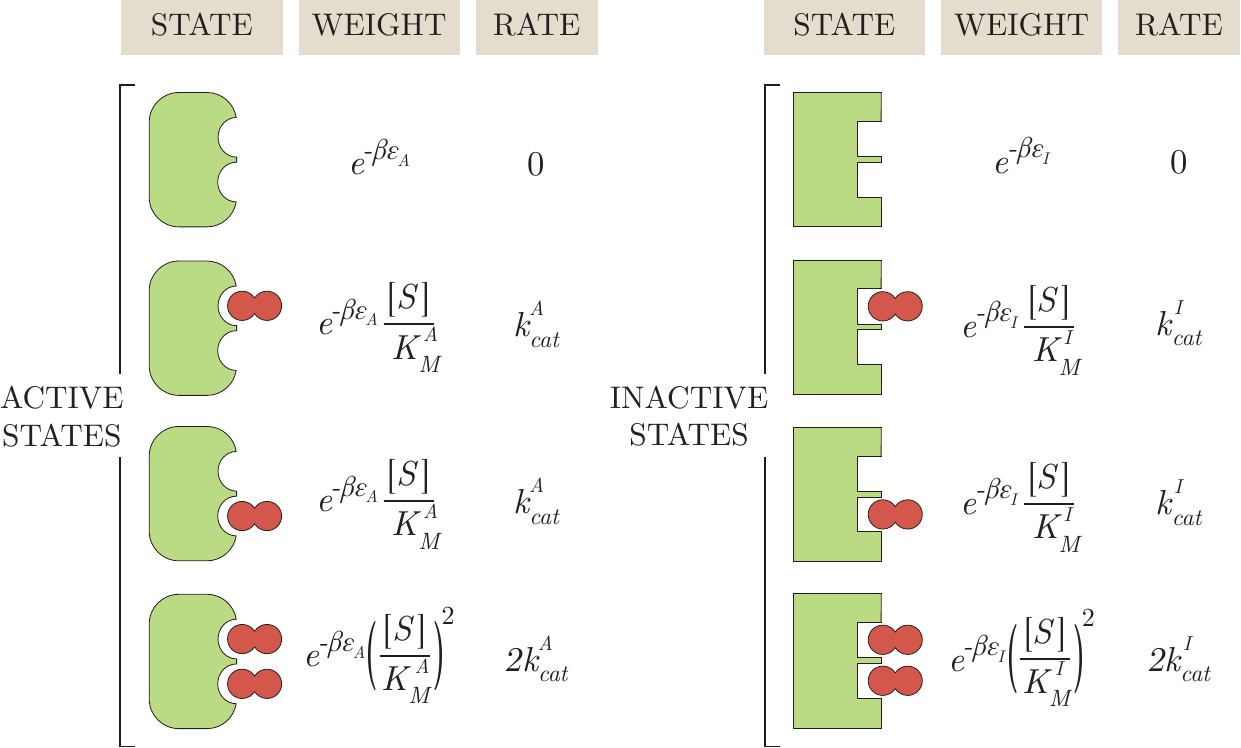}
	\caption{\textbf{States and weights for an MWC enzyme with two substrate binding sites.}
		Each binding site acts independently and the rate of product formation of a
		doubly bound state is twice the rate of the corresponding singly bound
		state.} \label{fig:Two-site_MWC}
\end{figure}

It has been shown that in MWC models, explicit cooperative interaction energies
are not required to accurately model biological systems; cooperativity is
inherently built into the fact that all binding sites switch concurrently from
an active state to an inactive state \cite{Marzen2013}. For example, suppose an
inactive state enzyme with two empty catalytic sites binds with its inactive
state affinity $K_M^I$ to a single substrate, and that this binding switches the
enzyme from the inactive to the active state. Then the second, still empty,
catalytic site now has the active state affinity $K_M^A$, an effect which can be
translated into cooperativity. Note that an explicit interaction energy, if
desired, can be added to the model very simply.

As in the proceeding sections, we compute the probability and concentration of
each enzyme conformation from the states and weights (see
\eref[MWCBasicResults1a][MWCBasicResults1b][MWCBasicResults2b]). Because the
active and inactive conformations each have two singly bound states and one
doubly bound state with twice the rate, the enzyme's rate of product formation
is given by
\begin{align} 
\frac{d[P]}{dt}&=k_{cat}^A\left(2p_{E_AS}\right)+2k_{cat}^A\left(p_{E_AS^2}\right)+k_{cat}^I\left(2p_{E_IS}\right)+2k_{cat}^I\left(p_{E_IS^2}\right) \nonumber \\
&=2[E_{tot}]\frac{k_{cat}^A e^{-\beta \epsilon_A}\frac{[S]}{K_M^{A}}\left(1+\frac{[S]}{K_M^{A}} \right)+k_{cat}^I e^{-\beta \epsilon_I}\frac{[S]}{K_M^{I}}\left(1+\frac{[S]}{K_M^{I}} \right)}{e^{-\beta \epsilon_A}\left(1+\frac{[S]}{K_M^{A}}\right)^2+e^{-\beta \epsilon_I}\left(1+\frac{[S]}{K_M^{I}}\right)^2} \label{rateEquationTwoSubstrateSites}
\end{align}
We will have much more to say about this model in
\sref{twoSiteMWCEnzymeSection}, where we will show that $\frac{d[P]}{dt}$ as a
function of substrate concentration $[S]$ may form a peak. For now, we mention
the well-known result that a Michaelis-Menten enzyme with two independent active
sites will act identically to two Michaelis-Menten enzymes each with a single
active site (as can be seen in the $\epsilon_I \to \infty$ limit of
\eref[rateEquationTwoSubstrateSites]) \cite{Segel1993}. It is intuitively clear
that this result does not extend to MWC enzymes: $\frac{d[P]}{dt}$ for a
two-site MWC enzyme, \eref[rateEquationTwoSubstrateSites], does not equal twice
the value of $\frac{d[P]}{dt}$ for a one-site MWC enzyme,
\eref[eq:basicMWCFinaldPdtEqquation].

\subsection{Modeling Overview} \label{MoreComplexEnzymes}

The above sections allow us to model a complex enzyme with any number of
substrate binding sites, competitive inhibitors, and allosteric regulators.
Assuming that the enzyme is in steady state and that the cycle condition holds,
we first enumerate its states and weights and then use those weights to
calculate the rate of product formation. Our essential conclusions about the
roles of the various participants in these reactions can be summarized as
follows:
\begin{enumerate}
	\item The (in)active state enzyme contributes a factor ($e^{-\beta
		\epsilon_I}$) $e^{-\beta \epsilon_A}$ to the weight. The mathematical
	simplicity of this model belies the complex interplay between the active and
	inactive states. Indeed, an MWC enzyme cannot be decoupled into two
	Michaelis-Menten enzymes (one for the active and the other for the inactive
	states).
	
	\item Each bound substrate contributes a factor ($\frac{[S]}{K_M^I}$)
	$\frac{[S]}{K_M^A}$ in the (in)active state where
	$K_M=\frac{k_{off}+k_{cat}}{k_{on}}$ is a Michaelis constant between the
	substrate and enzyme. It is this Michaelis constant, and not the dissociation
	constant, which enters the states and weights diagram.
	
	\item Each bound allosteric regulator or competitive inhibitor $X$ contributes
	a factor ($\frac{[X]}{X_d^I}$) $\frac{[X]}{X_d^A}$ in the (in)active state
	where $X_D=\frac{k_{off}^X}{k_{on}^X}$ is the dissociation constant between $X$
	and the enzyme. An allosteric regulator $R$ effectively tunes the energies of
	the active and inactive states as shown in
	\eref[eq:allostericEnzymeEquation112a][eq:allostericEnzymeEquation112b]. A
	competitive inhibitor $C$ effectively changes both the energies and Michaelis
	constants of the active and inactive states as described by
	\eref[eq:competitorEnzymeEquation112a][eq:competitorEnzymeEquation112b][eq:competitorEnzymeEquation112c][eq:competitorEnzymeEquation112d].
	
	\item The simplest model for multiple binding sites assumes that each site is
	independent of the others. The MWC model inherently accounts for the
	cooperativity between these sites, resulting in sigmoidal activity curves
	despite no direct interaction terms.
\end{enumerate}
In Appendix B, we simultaneously combine all of
these mechanisms by analyzing the rate of product formation of ATCase (which has
multiple binding sites) in the presence of substrate, a competitive inhibitor,
and allosteric regulators. In addition, the supplementary \textit{Mathematica}
notebook lets the reader specify their own enzyme and see its corresponding
properties.

Note that while introducing new components (such as a competitive inhibitor or
an allosteric regulator) introduces new parameters into the system, increasing
the number of sites does not. For example, an MWC enzyme with 1
(\fref[fig:statesWeightsMWCEnzyme2]), 2 (\fref[fig:Two-site_MWC]), or more
active sites would require the same five parameters: $e^{-\beta
	\left(\epsilon_A-\epsilon_I\right)}$, $K_M^A$, $K_M^I$, $k_{cat}^{A}$, and
$k_{cat}^{I}$.

\section{Applications} \label{sectionApplications}

Having built a framework to model allosteric enzymes, we now turn to some
applications of how this model can grant insights into observed enzyme
behavior. Experimentally, the rate of product formation of an enzyme is often
measured relative to the enzyme concentration, a quantity called
\textit{activity},
\begin{equation} \label{eq:activityEq}
A \equiv \frac{1}{[E_{tot}]}\frac{d[P]}{dt}.
\end{equation}
Enzymes are often characterized by their activity curves as substrate,
inhibitor, and regulator concentrations are titrated. Such data not only
determines important kinetic constants but can also characterize the nature of
molecular players such as whether an inhibitor is competitive, uncompetitive,
mixed, or non-competitive \cite{Cornish-Bowden1974, Berg2002, Li2005}. After
investigating several activity curves, we turn to a case study of the curious
phenomenon of substrate inhibition, where saturating concentrations of substrate
inhibit enzyme activity, and propose a new minimal mechanism for substrate
inhibition caused solely by allostery.

\subsection{Regulator and Inhibitor Activity Curves} \label{sec:ExperimentalData}

We begin with an analysis of $\alpha$-amylase, one of the simplest allosteric
enzymes, which only has a single catalytic site. $\alpha$-amylase catalyzes the
hydrolysis of large polysaccharides (e.g. starch and glycogen) into smaller
carbohydrates in human metabolism. It is competitively inhibited by isoacarbose
\cite{Li2005} at the active site and is allosterically activated by Cl$^-$ ions
at a distinct allosteric site \cite{Bussy1996}.

\begin{figure}[h!]
	\centering \includegraphics[scale=1]{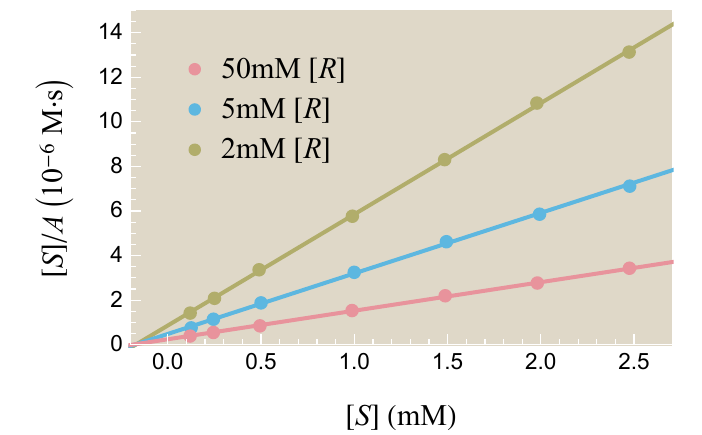} \caption{\textbf{Theoretically
		and experimentally probing the effects of an allosteric regulator on activity.}
		Data points show experimentally measured activity from Feller et al.~for the
		enzyme $\alpha$-amylase using substrate analog $[S]$ (EPS) and allosteric
		activator $[R]$ (NaCl) \cite{Bussy1996}. Best fit theoretical curves described
		by \eref[activatorEq2] are overlaid on the data. The best fit parameters are
		$e^{-\beta \left(\epsilon _A-\epsilon _I\right)} = 7.8 \times 10^{-4}$, $K_M^A
		= 0.6 \,\text{mM}$, $K_M^I = 0.2 \,\text{mM}$, $R_D^A = 0.03 \,\text{mM}$,
		$R_D^I = 7.9 \,\text{mM}$, $k_{cat}^{A} = 14 \,\text{s}^{-1}$, and $k_{cat}^{I}
		= 0.01 \,\text{s}^{-1}$.}\label{fig:amylaseRegulator}
\end{figure}

\fref[fig:amylaseRegulator] plots substrate concentration divided by
activity, $[S] / A$, as a function of substrate $[S]$. Recall from
\sref{AllostericEffectorSectionNew} that an enzyme with one active site and one
allosteric site has activity given by \eref[MWCActdPdt],
\begin{equation}
A = \frac{k_{cat}^A e^{-\beta \epsilon_A}\frac{[S]}{K_M^{A}}\left(1+\frac{[R]}{R_D^{A}} \right)+k_{cat}^I e^{-\beta \epsilon_I}\frac{[S]}{K_M^{I}}\left(1+\frac{[R]}{R_D^{I}} \right)}{e^{-\beta \epsilon_A}\left(1+\frac{[S]}{K_M^{A}}\right)\left(1+\frac{[R]}{R_D^{A}}\right)+e^{-\beta \epsilon_I}\left(1+\frac{[S]}{K_M^{I}}\right)\left(1+\frac{[R]}{R_D^{I}}\right)}. \label{activatorEq}
\end{equation}
Thus we expect the $[S] / A$ curves in \fref[fig:amylaseRegulator] to be linear in $[S]$,
\begin{equation}
\frac{[S]}{A} = \frac{e^{-\beta \epsilon_A}\left(1+\frac{[S]}{K_M^{A}}\right)\left(1+\frac{[R]}{R_D^{A}}\right)+e^{-\beta \epsilon_I}\left(1+\frac{[S]}{K_M^{I}}\right)\left(1+\frac{[R]}{R_D^{I}}\right)}{k_{cat}^A e^{-\beta \epsilon_A}\frac{1}{K_M^{A}}\left(1+\frac{[R]}{R_D^{A}} \right)+k_{cat}^I e^{-\beta \epsilon_I}\frac{1}{K_M^{I}}\left(1+\frac{[R]}{R_D^{I}} \right)}. \label{activatorEq2}
\end{equation}
\fref[fig:amylaseRegulator] shows that the experimental data is well
characterized by the theory so that the rate of product formation at any other
substrate and allosteric activator concentration can be predicted by this model.
The fitting procedure is discussed in detail in Appendix
B.

In the special case of a Michaelis-Menten enzyme ($\epsilon_I \to \infty$), the
above equation becomes
\begin{alignat}{2}
\frac{[S]}{A} &= \frac{K_M^{A}+[S]}{k_{cat}^A} && \;\;\;\;\;\;\;\; (\epsilon_I \to \infty).
\end{alignat}
The $x$-intercept of all lines in such a plot would intersect at the point
$(-K_M^{A},0)$ which allows an easy determination of $K_M^{A}$. This is why
plots of $[S]$ vs $[S] / A$, called Hanes plots, are often seen in
enzyme kinetics data. Care must be taken, however, when extending this analysis
to allosteric enzymes where the form of the $x$-intercept is more complicated.

We now turn to competitive inhibition. \fref[fig:dataCollapse]\letterParen{A} plots the inverse
rate of product formation $\left( \frac{d[P]}{dt} \right)^{-1}$ of $\alpha$-amylase as a
function of the competitive inhibitor concentration $[C]$. The competitive
inhibitor isoacarbose is titrated for three different concentrations of the
substrate $\alpha$-maltotriosyl fluoride ($\alpha$G3F).

\begin{figure}[h!]
	\centering \includegraphics[scale=1]{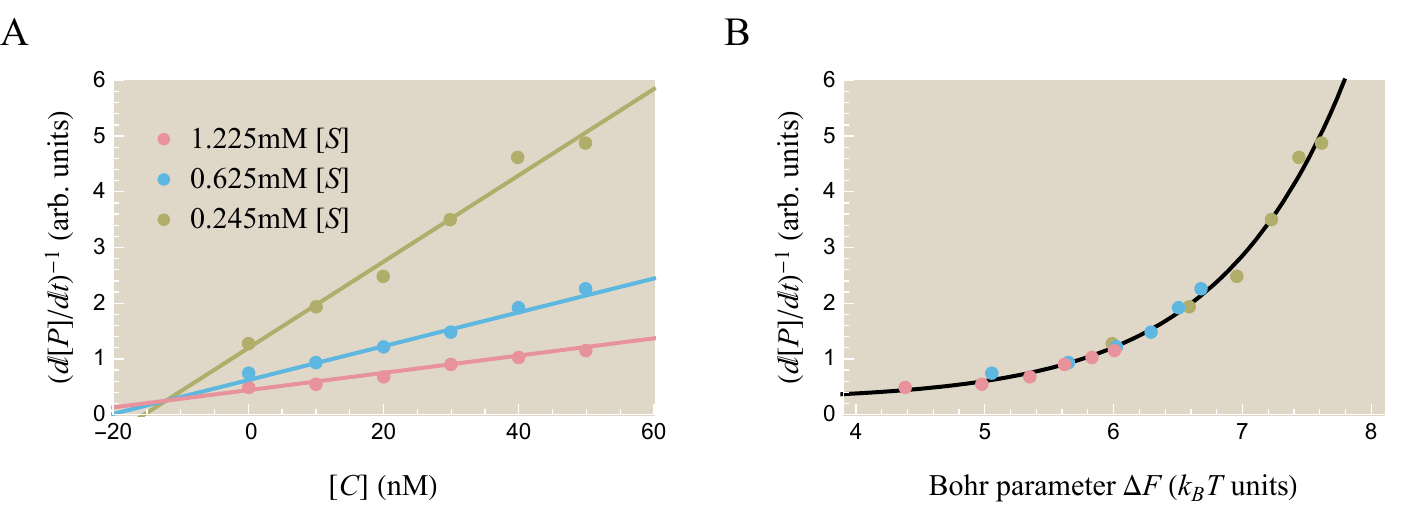} \caption{\textbf{Theoretically
		and experimentally probing the effects of a competitive inhibitor on activity.}
		\letterParen{A} Data points show experimentally measured activity in arbitrary
		units from Li et al.~for the enzyme $\alpha$-amylase using substrate analog
		$[S]$ ($\alpha$-maltotriosyl fluoride) and competitive inhibitor $[C]$
		(isoacarbose) \cite{Li2005}. Best fit theoretical curves described by the
		inverse of \eref[eqDataActivity] are overlaid on the data. The best fit
		parameters are $e^{-\beta \left(\epsilon _A-\epsilon _I\right)} = 36$, $K_M^A =
		0.9 \,\text{mM}$, $K_M^I = 2.6 \,\text{mM}$, $C_D^A = 12 \,\text{nM}$, $C_D^I =
		260 \,\text{nM}$, and $\frac{k_{cat}^{A}}{k_{cat}^{I}} = 1.4$. Note that the
		$x$-axis varies $[C]$ rather than $[S]$ as in most other plots. \letterParen{B}
		A data collapse of the three curves using the Bohr parameter $\Delta F$ from
		\eref[bohrParameter] which encompasses the effects of both the substrate and
		inhibitor upon the system.}\label{fig:dataCollapse}
\end{figure}

Recall from \sref{competitiveInihbitorSectionNew}, \eref[eq:MWCActCompdPdt] that
the rate of product formation for an allosteric enzyme with one active site in
the presence of a competitive inhibitor is given by
\begin{equation}
\left( \frac{d[P]}{dt} \right)^{-1}=\frac{1}{[E_{tot}]} \frac{e^{-\beta \epsilon_A}\left(1+\frac{[S]}{K_M^{A}}+\frac{[C]}{C_D^{A}}\right)+e^{-\beta \epsilon_I}\left(1+\frac{[S]}{K_M^{I}}+\frac{[C]}{C_D^{I}}\right)}{k_{cat}^Ae^{-\beta \epsilon_A}\frac{[S]}{K_M^{A}}+k_{cat}^Ie^{-\beta \epsilon_I}\frac{[S]}{K_M^{I}}}, \label{eqDataActivity}
\end{equation}
so that the best fit $\left( \frac{d[P]}{dt} \right)^{-1}$ curves in
\fref[fig:dataCollapse]\letterParen{A} are linear functions of $[C]$. Rather
than thinking of \eref[eqDataActivity] as a function of the competitive
inhibitor concentration $[C]$ and the substrate concentration $[S]$ separately,
we can combine these two quantities into a single natural parameter for the
system. This will enable us to collapse the different activity curves in
\fref[fig:dataCollapse]\letterParen{A} onto a single master curve as shown in
\fref[fig:dataCollapse]\letterParen{B}. Algebraically manipulating
\eref[eqDataActivity],
\begin{align} \label{eqActDataCollapse}
\frac{d[P]}{dt}&= [E_{tot}] \frac{\left(k_{cat}^A e^{-\beta \left(\epsilon_A-\epsilon_I\right)} \frac{K_M^{I}}{K_M^{A}}+k_{cat}^I\right) \frac{[S]}{K_M^{I}}}{\left( e^{-\beta \left(\epsilon_A-\epsilon_I\right)} \frac{K_M^{I}}{K_M^{A}} + 1 \right) \frac{[S]}{K_M^{I}} + e^{-\beta \left(\epsilon_A-\epsilon_I\right)} \left(1+\frac{[C]}{C_D^{A}}\right)+\left(1+\frac{[C]}{C_D^{I}}\right)} \nonumber \\
&\equiv [E_{tot}] \frac{\left(k_{cat}^AK +k_{cat}^I\right)e^{-\beta  \Delta F}}{(K+1)e^{-\beta  \Delta F}+1}
\end{align}
where
\begin{align}
K &= e^{-\beta \left(\epsilon_A-\epsilon_I\right)}\frac{K_M^I}{K_M^A} \\
\Delta F&=-\frac{1}{\beta }\text{Log}\left[\frac{\frac{[S]}{K_M^I}}{e^{-\beta \left(\epsilon_A-\epsilon_I\right)}\left(1+\frac{[C]}{C_D^A}\right)+\left(1+\frac{[C]}{C_D^I}\right)}\right] \label{bohrParameter}.
\end{align}
Therefore, $\left( \frac{d[P]}{dt} \right)^{-1}$ curves at any substrate and inhibitor
concentrations can be compactly shown as data points lying on a single curve in
terms of $\Delta F$, which is called the \textit{Bohr parameter}. Such a data
collapse is also possible in the case of allosteric regulators or enzymes with
multiples binding sites, although those data collapses may require more than one
variable $\Delta F$. In Appendix C, we show that the
Bohr parameter corresponds to a free energy difference between enzyme
states and examine other cases of data collapse.

\subsection{Substrate Inhibition} \label{secSubstrateInhibition}

We now turn to a striking phenomenon observed in the enzyme literature: not all
enzymes have a monotonically increasing rate of product formation. Instead peaks
such as those shown schematically in \fref[fig:peakIntroductoryCartoons] can
arise in various enzymes, displaying behavior that is impossible within
Michaelis-Menten kinetics. By exploring these two phenomena with the MWC model,
we gain insight into their underlying mechanisms and can make quantifiable
predictions as to how to create, amplify, or prevent such peaks.

\begin{figure}
	\centering \includegraphics[scale=1]{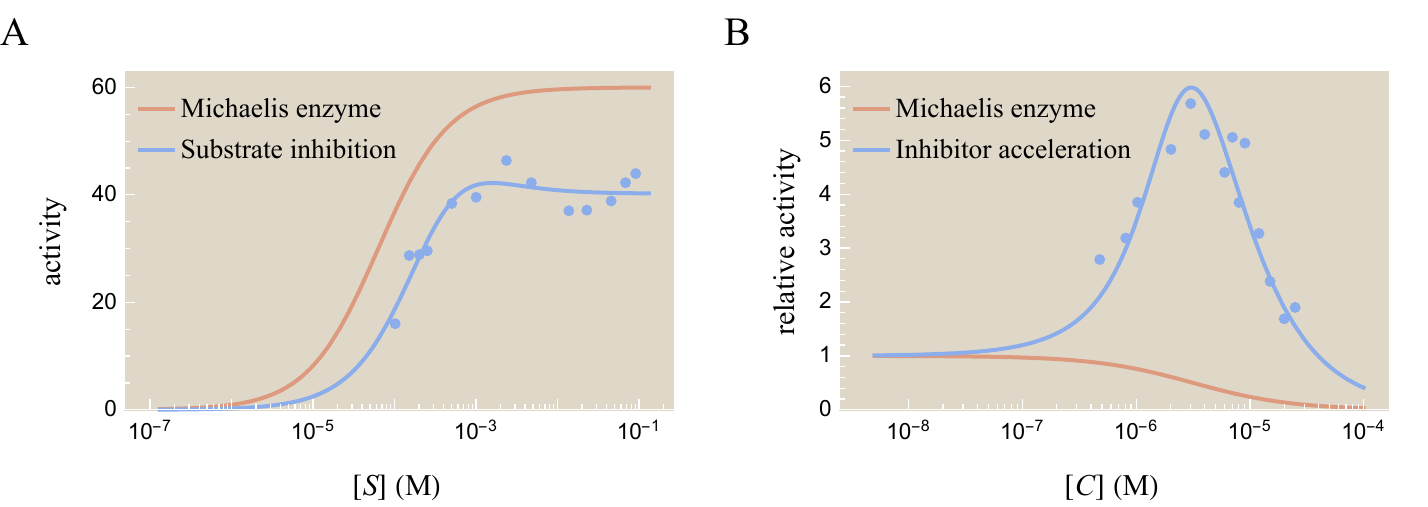} \caption{\textbf{Enzyme
		activity curves do not need to be monotonic as predicted by Michaelis-Menten
		enzyme kinetics.} \letterParen{A} As many as 20\% of enzymes exhibit substrate
		inhibition, where at high substrate concentrations activity decreases, in
		contrast to a Michaelis-Menten enzyme shown for reference \cite{Kaiser1980,
			Reed2010}. Activity for acetylcholinesterase is shown in units of
		$\text{(nanomoles product)} \cdot\text{min}^{-1}\cdot\text{(mL enzyme)}^{-1}$
		\cite{Changeux1966}. \letterParen{B} Some enzymes exhibit inhibitor
		acceleration, where adding a small amount of a competitive inhibitor increases
		the rate of product formation. This generates a peak in activity, in stark
		contrast to a Michaelis-Menten enzyme which only decreases its activity as more
		competitive inhibitor is added. Relative activity is shown for ATCase, where
		relative activity equals activity at $[C]$ divided by the activity with no
		competitive inhibitor \cite{Wales1999}. The data and best fit parameters for
		the substrate inhibition and inhibitor acceleration curves are discussed in
		Appendix C.} \label{fig:peakIntroductoryCartoons}
\end{figure}

In \fref[fig:peakIntroductoryCartoons]\letterParen{A}, the monotonically
increasing Michaelis-Menten curve makes intuitive sense - a larger substrate
concentration implies that at any moment the enzyme's active site is more likely
to be occupied by substrate. Therefore, we expect that the activity, $A =
\frac{1}{[E_{tot}]}\frac{d[P]}{dt}$, should increase with the substrate
concentration $[S]$. Yet many enzymes exhibit a peak activity, a behavior called
substrate inhibition \cite{Kaiser1980}.

Even more surprisingly, when a small amount of competitive inhibitor - a
molecule whose very name implies that it competes with substrate and decreases
activity - is mixed together with enzyme, it can \textit{increase} the rate of
product formation. This latter case, called inhibitor acceleration, is shown in
\fref[fig:peakIntroductoryCartoons]\letterParen{B} \cite{Wales1999, Miller2011}.
In contrast, a Michaelis-Menten enzyme shows the expected behavior that adding
more competitive inhibitor decreases activity. We will restrict our attention to
the phenomenon of substrate inhibition and relegate a discussion of inhibitor
acceleration to Appendix D.

Using the MWC enzyme model, we can make predictions about which enzymes can
exhibit substrate inhibition. We first formulate a relationship between the
fundamental physical parameters of an enzyme that are required to generate such
a peak and then consider what information about these underlying parameters can
be gained by analyzing experimental data.

\subsubsection{Single-Site Enzyme} 

As a preliminary exercise, we begin by showing that an enzyme with a
\textit{single} active site cannot exhibit substrate inhibition. Said another
way, the activity, \eref[eq:activityEq], of such an enzyme cannot have a peak as
a function of substrate concentration $[S]$. For the remainder of this paper, we
will use the fact that all Michaelis and dissociation constants ($K_M$'s,
$C_D$'s, and $R_D$'s) are positive and assume that both catalytic constants
($k_{cat}^A$ and $k_{cat}^I$) are strictly positive unless otherwise stated.

Consider the MWC enzyme with a single substrate binding site shown in
\fref[fig:statesWeightsMWCEnzyme2]. Using \eref[eq:basicMWCFinaldPdtEqquation],
it is straightforward to compute the derivative of activity with respect to
substrate concentration $[S]$, namely,
\begin{equation} \label{1SubMWCActivity}
\frac{dA}{d[S]}=\frac{(e^{-\beta \epsilon _A}+e^{-\beta \epsilon _I})\left(e^{-\beta \epsilon _A}\frac{k_{cat}^A}{K_M^A}+e^{-\beta \epsilon _I}\frac{k_{cat}^I}{K_M^I} \right)}{\left(e^{-\beta \epsilon _A} \left(1+\frac{[S]}{K_M^A}\right)+e^{-\beta \epsilon _I} \left(1+\frac{[S]}{K_M^I}\right)\right)^2}.
\end{equation}
Since the numerator cannot equal zero, this enzyme cannot have a peak in its
activity when $[S]$ is varied. Note that the numerator is positive, indicating
that enzyme activity will always increase with substrate concentration.

The above results are valid for an arbitrary MWC enzyme with a single-site. In
particular, in the limit $\epsilon_I \to \infty$, an MWC enzyme becomes a
Michaelis-Menten enzyme. Therefore, a Michaelis-Menten enzyme with a single
active site cannot exhibit a peak in activity. In Appendix
E, we discuss the generalization of this
result: a Michaelis-Menten enzyme with an arbitrary number of catalytic sites
cannot have a peak in activity. Yet as we shall now see, this generalization
cannot be made for an MWC enzyme, which can indeed exhibit a peak in its
activity when it has multiple binding sites.

\subsubsection{Substrate Inhibition} \label{twoSiteMWCEnzymeSection}

As many as 20\% of enzymes are believed to exhibit substrate inhibition, which
can offer unique advantages to enzymes such as stabilizing their activity amid
fluctuations, enhancing signal transduction, and increasing cellular efficiency
\cite{Reed2010}. This prevalent phenomenon has elicited various explanations,
many of which rely on non-equilibrium enzyme dynamics, although some equilibrium
mechanisms are known \cite{Kaiser1980}. An example of this latter case is seen
in the enzyme aspartate transcarbamoylase (ATCase) which catalyzes one of the
first steps in the pyrimidine biosynthetic pathway. Before ATCase can bind to
its substrate asparatate (Asp), an intermediate molecule carbamoyl phosphate
(CP) must first bind to ATCase, inducing a change in the enzyme's shape and
electrostatics which opens up the Asp binding slot \cite{Wang2005, Hsuanyu1987}.
Because Asp can weakly bind to the CP binding pocket, at high concentrations Asp
will outcompete CP and prevent the enzyme from working as efficiently, thereby
causing substrate inhibition \cite{Harris2011}.

To the list of such mechanisms, we add the  possibility that an enzyme may
exhibit substrate inhibition without any additional effector molecules. In
particular, an allosteric enzyme with two identical catalytic sites can exhibit
a peak in activity when the substrate concentration $[S]$ is varied. We will
first analyze the properties of this peak and then examine why it can occur. For
simplicity, we will assume $k_{cat}^I=0$ throughout this section and leave the
general case for Appendix E.

Using \eref[rateEquationTwoSubstrateSites][eq:activityEq], the activity of an MWC
enzyme with two active sites is given by
\begin{equation} \label{eq:twoSubstrateSiteActivityEquation}
A = \frac{1}{[E_{tot}]}\frac{d[P]}{dt} = \frac{2k_{cat}^A e^{-\beta \epsilon_A}\frac{[S]}{K_M^{A}}\left(1+\frac{[S]}{K_M^{A}} \right)}{e^{-\beta \epsilon_A}\left(1+\frac{[S]}{K_M^{A}}\right)^2+e^{-\beta \epsilon_I}\left(1+\frac{[S]}{K_M^{I}}\right)^2}.
\end{equation}
A peak will exist provided that $\frac{dA}{d[S]}=0$ has a positive $[S]$ root.
The details of differentiating and solving this equation are given in Appendix
E, the result of which is that a peak
in activity $A$ occurs as a function of $[S]$ provided that
\begin{equation}
1+e^{-\beta \left(\epsilon _A-\epsilon _I\right)}<\left(\frac{K_M^A}{K_M^I}-1\right)^2 \label{eq:TwoSiteMWCPeak1a} \;\;\;\;\;\;\;\; (k_{cat}^I=0).
\end{equation}

\begin{figure}[h!]
	\centering \includegraphics[scale=1]{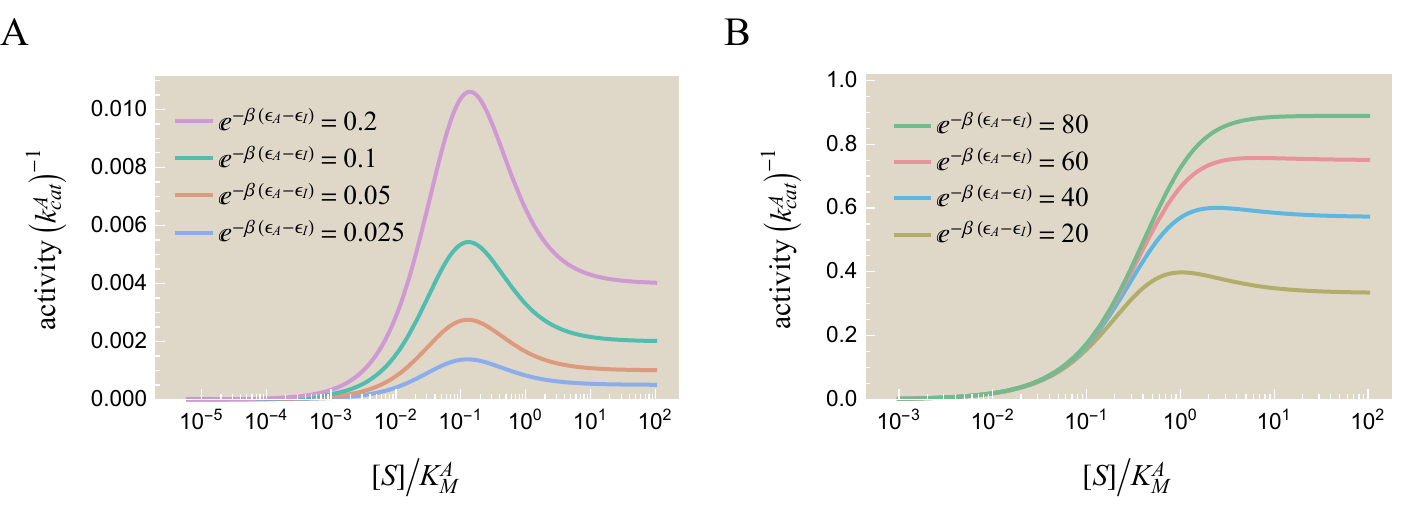} \caption{\textbf{Peaks in
		enzyme activity $\boldsymbol{A = \frac{1}{E_{tot}}\frac{d[P]}{dt}}$ as a function of
		substrate concentration $\boldsymbol{[S]}$.} Activity is shown in units of $k_{cat}^A$, which
		rescales the activity curves vertically. The peak for \letterParen{A} small and
		\letterParen{B} large ratios of the enzyme's energy in the active versus
		inactive state, $e^{-\beta \left(\epsilon _A-\epsilon _I\right)}$. The height
		of the peak increases with $e^{-\beta \left(\epsilon _A-\epsilon _I\right)}$.
		The activity is computed from \eref[eq:twoSubstrateSiteActivityEquation] using
		the parameters $k_{cat}^I=0$, $\frac{K_M^A}{K_M^I}=10$, and the different
		values of $e^{-\beta \left(\epsilon _A-\epsilon _I\right)}$ shown. As predicted
		by \eref[eq:TwoSiteMWCPeak1a], every value in the range $e^{-\beta
			\left(\epsilon _A-\epsilon _I\right)} < \left(\frac{K_M^A}{K_M^I}-1\right)^2$
		will yield a peak in activity. While the peak is more pronounced when the
		active state is energetically favorable ($e^{-\beta \left(\epsilon _A-\epsilon
			_I\right)} < 1$) in \letterParen{A}, the maximum peak height is much larger in
		\letterParen{B} as seen by the different scale of the \textit{y}-axis.}
	\label{fig:TwoSiteMWCPeakPlot}
\end{figure}

The height of such a peak is given by
\begin{equation} \label{eq:peakHeightSubstrateAcceleration}
A_{peak} = k_{cat}^A \frac{K_M^I}{K_M^A-K_M^I} \left( \sqrt{1+e^{-\beta \left(\epsilon _A-\epsilon _I\right)}} - 1 \right).
\end{equation}
Examples of peaks in activity are shown in \fref[fig:TwoSiteMWCPeakPlot] for
various values of $e^{-\beta \left(\epsilon _A-\epsilon _I\right)}$.
Substituting in the peak condition \eref[eq:TwoSiteMWCPeak1a], the maximum peak
height is at most
\begin{equation} \label{eq:peakHeightSubstrateAccelerationRelation}
A_{peak} < k_{cat}^A \frac{\frac{K_M^A}{K_M^I}-2}{\frac{K_M^A}{K_M^I}-1}.
\end{equation}

If we consider the maximum value of $e^{-\beta \left(\epsilon _A-\epsilon
	_I\right)}$ allowed by the peak condition \eref[eq:TwoSiteMWCPeak1a], the peak
height approaches $k_{cat}^A$ for large $\frac{K_M^A}{K_M^I}$ (as seen by the
green curve $e^{-\beta \left(\epsilon _A-\epsilon _I\right)}=80$ in
\fref[fig:TwoSiteMWCPeakPlot]\letterParen{B}). In this limit, the active
bound state dominates over all the other enzyme states so that the activity
reaches its largest possible value, $k_{cat}^A$. Although the ``peak height'' is
maximum in this case, the activity curve is nearly sigmoidal, making the peak
hard to distinguish. To that end, it is reasonable to compare the peak height to
the activity at large substrate concentrations,
\begin{equation} \label{eq:saturatingSubstrateConcentration}
A_{[S] \to \infty} = 2k_{cat}^A \frac{e^{-\beta \left(\epsilon _A-\epsilon _I\right)}}{\left(\frac{K_M^A}{K_M^I}\right)^2 + e^{-\beta \left(\epsilon _A-\epsilon _I\right)}}.
\end{equation}
As the energy difference between the active and inactive state $e^{-\beta
	\left(\epsilon _A-\epsilon _I\right)}$ increases, the peak height $A_{peak}$
monotonically increases but the relative peak height $\frac{A_{peak}}{A_{[S] \to
		\infty}}$ monotonically decreases. These relations might be used to design enzymes
with particular activity curves; conversely, experimental data of substrate
inhibition can be used to fix a relation between the parameters $e^{-\beta
	\left(\epsilon _A-\epsilon _I\right)}$ and $\frac{K_M^A}{K_M^I}$ of an enzyme.

We now turn to the explanation of how such a peak can occur. One remarkable fact
is that a peak \textit{cannot} happen without allostery. If we consider a
Michaelis-Menten enzyme (by taking the limit $k_{cat}^I \to 0$ and $\epsilon _I
\to \infty$), then the peak condition \eref[eq:TwoSiteMWCPeak1a] cannot be
satisfied. 

\begin{figure}[h!]
	\centering \includegraphics[scale=1]{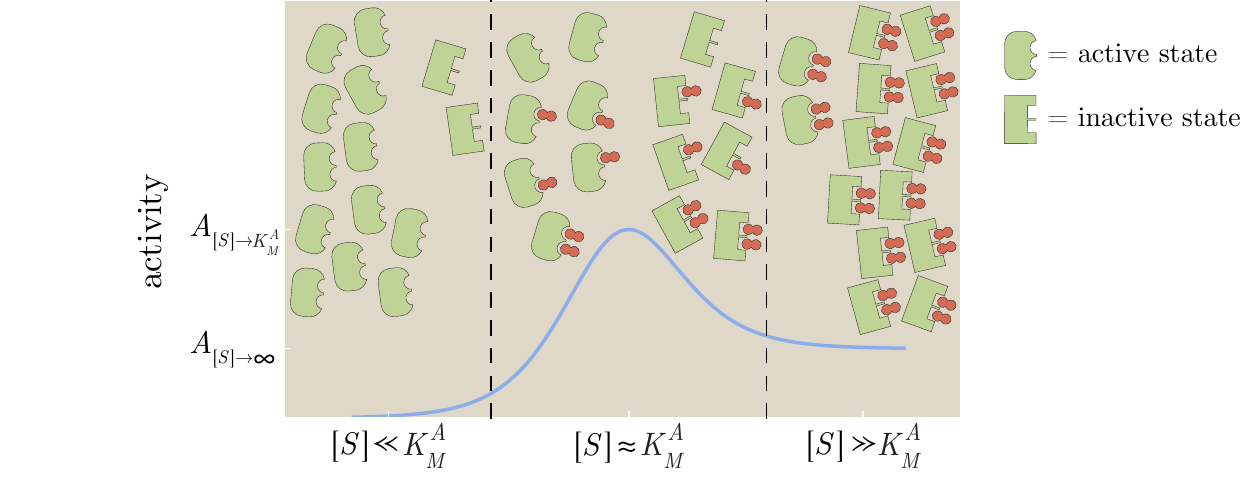} \caption{\textbf{Mechanism
		underlying peak in activation by substrate $\boldsymbol{S}$.} At low substrate concentrations
		(left region), all enzymes are unbound and are mostly in the active form
		(rounded, green). As the amount of substrate is increased (middle region), the
		probability that an enzyme is singly bound and then doubly bound increase.
		Because the substrate prefers to bind to an inactive state (sharp, green)
		enzyme-substrate complex, binding more substrate pushes the enzymes into the
		inactive state. At medium substrate concentrations, more active state
		enzyme-substrate complexes exist than at high substrate concentrations (right
		region) which yields a peak. Each enzyme fluctuates between its different
		configurations, and the cartoons show the distributions of the most prevalent
		states.} \label{fig:pictoralSubstrateAcceleration}
\end{figure}

To gain a qualitative understanding of how a peak can occur, consider an enzyme
that inherently prefers the active state ($e^{-\beta \left(\epsilon _A-\epsilon
	_I\right)} > 1$) but with substrate that preferentially binds to the inactive
state ($\frac{K_M^A}{K_M^I} > 1$). Such a system is realized in bacterial
chemotaxis, where the chemotaxis receptors are active when unbound but inactive
when bound to substrate \cite{Phillips2010}. This setup is shown schematically
in \fref[fig:pictoralSubstrateAcceleration]. At low substrate concentrations,
$[S] \ll K_M^A$, most enzymes will be unbound and therefore in the active state.
At intermediate substrate concentrations, $[S] \approx K_M^A$, many enzymes will
be singly bound. Because $\frac{K_M^A}{K_M^I} > 1$, the substrate will pull
these bound enzymes towards the inactive state. For large substrate
concentrations, $[S] \gg K_M^A$, most of the enzymes will be doubly bound and
hence will be predominantly in the inactive form. Because the inactive state
does not catalyze substrate ($k_{cat}^I=0$), only the number of substrate bound
to active state enzymes increase the rate of product formation, and because more
of these exist in the intermediate regime a peak forms.

To be more quantitative, the activity \eref[eq:twoSubstrateSiteActivityEquation] at the medium
substrate concentration ($[S] = K_M^A$) is given by
\begin{equation} \label{eq:mediumSubstrateConcentration}
A_{[S] \to K_M^A} = k_{cat}^A \frac{4 e^{-\beta \left(\epsilon _A-\epsilon
		_I\right)}}{\left(\frac{K_M^A}{K_M^I} + 1\right)^2 + 4 e^{-\beta \left(\epsilon
		_A-\epsilon _I\right)}}.
\end{equation}
Comparing this to $A_{[S] \to \infty}$ in \eref[eq:saturatingSubstrateConcentration], we find that $A_{[S] \to K_M^A} > A_{[S] \to \infty}$ provided that 
\begin{equation} \label{eq:storySubstrateAcceleration}
1+e^{-\beta \left(\epsilon _A-\epsilon _I\right)} < \frac{1}{4} \left(\frac{K_M^A}{K_M^I}-1\right)^2.
\end{equation}
This is in close agreement with the peak condition \eref[eq:TwoSiteMWCPeak1a]
and the factor of $\frac{1}{4}$ is due to the fact that the peak need not occur
precisely at $[S] = K_M^A$. 

Note that the peak condition \eref[eq:TwoSiteMWCPeak1a] does not necessarily
force the unbound enzyme to favor the active state ($e^{-\beta \left(\epsilon
	_A-\epsilon _I\right)} > 1$), since this condition can still be satisfied if
$e^{-\beta \left(\epsilon _A-\epsilon _I\right)} < 1$. However, the peak
condition does require that substrate preferentially binds to the inactive state
enzyme (in fact, we must have $\frac{K_M^A}{K_M^I} > 2$ to satisfy the peak
condition).

Recall that as many as 20\% of enzymes exhibit substrate inhibition, and this
particular mechanism will not apply in every instance. To be concrete, an
allosteric enzyme that obeys the mode of substrate inhibition proposed above
must: (1) have at least two catalytic sites and (2) must be driven towards the
inactive state upon substrate binding. Therefore, an enzyme such as ATCase which
exhibits substrate inhibition but where the substrate preferentially binds to
the active state enzyme must have a different underlying mechanism
\cite{Wang2007}. Various alternative causes including the effects of pH due to
substrate or product buildup \cite{Masson2002, Segel1993} or the sequestering
effects of ions \cite{Otero2011, Penner1969} may also be responsible for
substrate inhibition. Yet the mechanism of substrate inhibition described above
exactly matches the conditions of acetylcholinesterase whose activity, shown in
\fref[fig:peakIntroductoryCartoons]\letterParen{A}, is well categorized by the
MWC model \cite{Changeux1966}. It would be interesting to test this theory by
taking a well characterized enzyme, tuning the MWC parameters so as to satisfy
the peak condition \eref[eq:TwoSiteMWCPeak1a] (or an analogous relationship for
an enzyme with more than two catalytic sites), and checking whether the system
then exhibits substrate inhibition. Experimentally, tuning the parameters can be
undertaken by introducing allosteric regulators or competitive inhibitors as
described by
\eref[eq:allostericEnzymeEquation112a][][eq:allostericEnzymeEquation112b] and
\eref[eq:competitorEnzymeEquation112a][eq:competitorEnzymeEquation112b][eq:competitorEnzymeEquation112c][eq:competitorEnzymeEquation112d], respectively. For example, in Appendix E, we describe an enzyme system where introducing a competitive inhibitor induces a peak in activity.

%

\section{Discussion}

Allosteric molecules pervade all realms of biology from chemotaxis receptors to
chromatin to enzymes \cite{Kantrowitz1988, Sprang1988, Boettcher2011,
	Changeux2012}. There are various ways to capture the allosteric nature of
macromolecules, with the MWC model representing one among many
\cite{Koshland1966, Gill1986, Yifrach1995}. In any such model, the simple
insight that molecules exist in an active and inactive state opens a rich new
realm of dynamics.

The plethora of molecular players that interact with enzymes serve as the
building blocks to generate complex behavior. In this paper, we showed the
effects of competitive inhibitors, allosteric regulators, and multiple binding
sites, looking at each of these factors first individually and then combining
separate aspects. This framework matched well with experimental data and enabled
us to make quantifiable predictions on how the MWC enzyme parameters may be
tuned upon the introduction of an allosteric regulator
\eref[eq:allostericEnzymeEquation112a][][eq:allostericEnzymeEquation112b] or a
competitive inhibitor
\eref[eq:competitorEnzymeEquation112a][eq:competitorEnzymeEquation112b][eq:competitorEnzymeEquation112c][eq:competitorEnzymeEquation112d].

As an interesting application, we used the MWC model to explore the unusual
behavior of substrate inhibition, where past a certain point adding more
substrate to a system decreases its rate of product formation. This mechanism
implies that an enzyme activity curve may have a peak (see
\fref[fig:peakIntroductoryCartoons]), a feat that is impossible for a
Michaelis-Menten enzyme. We explored a novel minimal mechanism for substrate
inhibition which rested upon the allosteric interactions of the active and
inactive enzyme states, with suggestive evidence for such a mechanism in
acetylecholinesterase.

The power of the MWC model stems from its simple description, far-reaching
applicability, and its ability to unify the proliferation of data gained over
the past 50 years of enzymology research. A series of activity curves at
different concentrations of a competitive inhibitor all fall into a 1-parameter
family of curves, allowing us to predict the activity at any other inhibitor
concentration. Such insights not only shed light on the startling beauty of
biological systems but may also be harnessed to build synthetic circuits and
design new drugs.  We close by noting our gratitude and admiration to Prof. Bill
Gelbart to whom this special is dedicated and who has inspired us with his
clever use of ideas from statistical physics to understand biological systems.

\begin{acknowledgement}

The authors thank T. Biancalani, J.-P. Changeux,  A. Gilson, J. Kondev, M.
Manhart, R. Milo, M. Morrison, N. Olsman, and J. Theriot for helpful discussions
and insights on this paper. All plots were made entirely in \textit{Mathematica}
using the CustomTicks package \cite{Caprio2005}. This work was supported in the
RP group by the National Institutes of Health through DP1 OD000217 (Director's
Pioneer Award) and R01 GM085286, La Fondation Pierre-Gilles de Gennes, and the
National Science Foundation under Grant No. NSF PHY11-25915 at the Kavli Center
for Theoretical Physics. The work was supported in the LM group by Research and
Development Program (Grant No. CH-3-SMM-01/03).

\end{acknowledgement}

\begin{suppinfo}
Supporting Information includes derivations aforementioned. Also included is a
\textit{Mathematica} notebook which reproduces the figures in the paper and
includes an interactive enzyme modeler.
\end{suppinfo}

\end{singlespace}
\setcounter{page}{2}

\appendix

\begin{singlespace}

\section{Validity of Approximations} \label{AppendixSectionValidityApproximations}

In \sref{MWCEnzymeSection}, we showed the generalization of the Michaelis-Menten
model by granting the enzyme access to an active and inactive conformation. We
then analyzed this system using two assumptions: the quasi-steady-state
approximation \eref[eq:quasiSteadyStateMWCEnzyme] and the cycle condition
\eref[eq:cycleLawMWC]. In this section, we will formally determine when these
approximations are valid for an MWC enzyme and discuss what happens when we
relax these assumptions. It is straightforward to extend these results to the
more complicated MWC enzyme models where we introduce allosteric regulators, add
competitive inhibitors, and consider enzymes with multiple binding sites.

\subsection{Definitions} \label{appendixDefinitions}

In \sref{MWCEnzymeSection}, we characterized an MWC enzyme using the reaction
scheme
\begin{equation} \label{appendixRatesEnzymeMWCBetter}
\begin{aligned}
\includegraphics[scale=1]{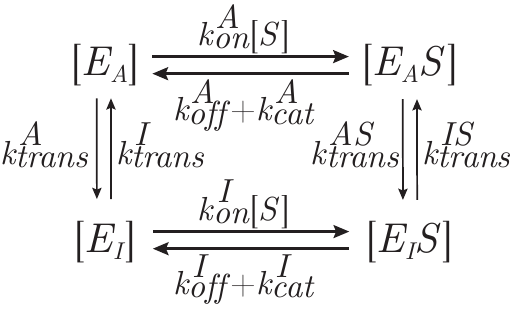}
\end{aligned}
\end{equation}
which we will now discuss in detail. We will use the following definitions
freely\cite{Gunawardena2012}:
\begin{itemize}
	\item An \textit{edge} of a reaction scheme denotes the value of an arrow from
	one enzyme state to another. The edges on the left of
	\bareEq{appendixRatesEnzymeMWCBetter} are $k_{trans}^A$ (linking $[E_A]$ to
	$[E_I]$) and $k_{trans}^I$ (linking $[E_I]$ to $[E_A]$).
	
	\item A \textit{path} along enzyme states is the product of edges along this
	path. For example, the path from $[E_I]$ to $[E_A]$ to $[E_AS]$ for the MWC
	scheme above is given by  $k_{trans}^{I} k_{on}^A [S]$.
	
	\item A system is in \textit{steady state} if the concentration of every enzyme
	conformation does not change over time. For the scheme
	above this implies
	$\frac{d[E_AS]}{dt}=\frac{d[E_A]}{dt}=\frac{d[E_IS]}{dt}=\frac{d[E_I]}{dt}=0$.
	
	\item The \textit{cycle condition} states that the product of edges going
	clockwise around any cycle must equal the product of edges going
	counterclockwise. For scheme \bareEq{appendixRatesEnzymeMWCBetter}, the product
	of edges clockwise equals $\left( k_{on}^A [S] \right) \left( k_{trans}^{AS}
	\right) \left( k_{off}^{I}+k_{cat}^{I} \right) \left( k_{trans}^{I} \right)$
	and the product of edges moving counter-clockwise equals $\left(
	k_{off}^{A}+k_{cat}^{A} \right) \left( k_{trans}^{A} \right) \left(
	k_{on}^{I}[S] \right) \left( k_{trans}^{IS} \right)$.
	
	\item \textit{Detailed balance} implies that the flow between two enzyme states
	is the same in the forward and backwards direction. For the scheme above, if
	the flow of enzymes from the $[E_A]$ state to the $[E_AS]$ state (given by
	$[E_A] [S] k_{on}^A$) equals the flow from $[E_AS]$ to $[E_A]$ (given by
	$[E_AS] \left( k_{off}^A + k_{cat}^A \right)$) then the pair of edges between
	$[E_A]$ and $[E_AS]$ obeys detailed balance. A reaction scheme is in
	\textit{equilibrium} if and only if every edge obeys detailed balance which
	occurs if and only if the system is in steady state and obeys the cycle
	condition.
\end{itemize}

\subsection{Cycle Condition} \label{understandingCycleCondition}

In this section we consider why the cycle condition is necessary to ensure that
a system in steady state is in equilibrium. Assume the MWC enzyme scheme
\bareEq{appendixRatesEnzymeMWCBetter} is in steady state,
\begin{equation} \label{appendixQuasiSteadyStateMWCEnzyme}
\frac{d[E_AS]}{dt}=\frac{d[E_A]}{dt}=\frac{d[E_IS]}{dt}=\frac{d[E_I]}{dt}=0.
\end{equation}
The cycle condition ensures that equilibrium holds around the cycle in
\bareEq{appendixRatesEnzymeMWCBetter} regardless of which path is traversed. For
example, suppose the system is in equilibrium and we want to use detailed
balance to determine the relation between $E_AS$ and $E_I$. Detailed balance
provides a relation between adjacent vertices (i.e. any two enzyme states
connected by arrows) such as $E_AS$ and $E_IS$ or $E_IS$ and $E_I$. Hence we can
find a relation between two non-adjacent edges such as $E_AS$ and $E_I$ by
following two different paths,
\begin{equation} \label{eq:ratesEnzymeMWCBetterWithArrows}
\begin{aligned}
\includegraphics[scale=1]{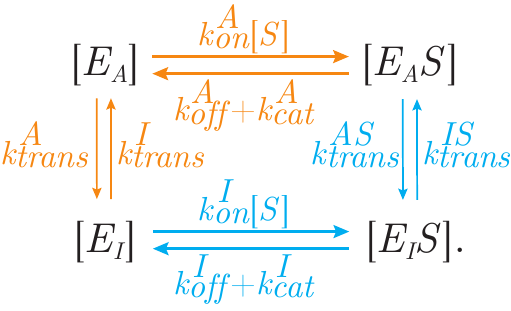}
\end{aligned}
\end{equation}
We could travel clockwise and follow the blue path around
\bareEq{eq:ratesEnzymeMWCBetterWithArrows}, first using detailed balance
between $E_AS$ and $E_IS$ and then between $E_IS$ and $E_I$,
\begin{equation} \label{explainingCycleCondition1}
\frac{[E_AS]}{[E_I]}=\frac{[E_AS]}{[E_IS]}\frac{[E_IS]}{[E_I]}=\frac{k_{trans}^{IS}}{k_{trans}^{AS}} \frac{k_{on}^{I}[S]}{k_{off}^{I}+k_{cat}^{I}}.
\end{equation}
On the other hand, we could have moved counter-clockwise around
\bareEq{eq:ratesEnzymeMWCBetterWithArrows} along the orange path, first using the
relationship between $E_AS$ and $E_A$ and then between $E_A$ and $E_I$,
\begin{equation} \label{explainingCycleCondition2}
\frac{[E_AS]}{[E_I]}=\frac{[E_AS]}{[E_A]}\frac{[E_A]}{[E_I]}=\frac{k_{on}^{A}[S]}{{k_{off}^{A}}+k_{cat}^{A}} \frac{k_{trans}^{I}}{k_{trans}^{A}}.
\end{equation}
Setting \eref[explainingCycleCondition1][explainingCycleCondition2] equal to
each other yields the cycle condition! 

\subsection{Quasi-Steady-State Approximation} \label{sec:QSSA}

We will now consider the dynamics of the MWC enzyme, 
\begin{equation} \label{eq:DiscussionRatesEnzymeMWC2}
\begin{aligned}
\includegraphics[scale=1]{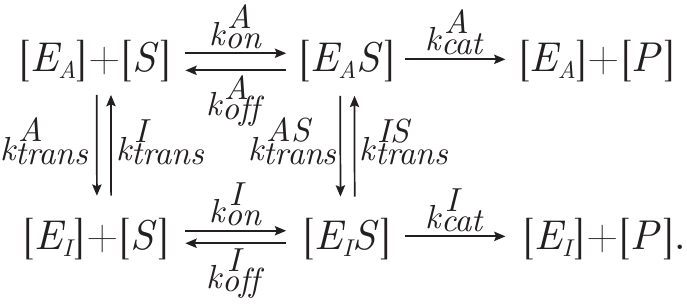}
\end{aligned}
\end{equation}
At time $t=0$, the enzyme and substrate are mixed together and the rate of
product formation is measured over time. The system starts off with all enzymes
in the unbound forms $E_A$ or $E_I$ and there are no enzyme-substrate complexes
$E_AS$ or $E_IS$.

To gain some intuition into this system, we first consider \fref[fig:transient]
which shows how this MWC enzyme can behave over time for reasonable parameter
values. On the long time scales in \fref[fig:transient]\letter{B}, the substrate
concentration will appreciably diminish to $1/e$ of its original value after a
long time $\tau _S$. On the other hand, \fref[fig:transient]\letter{A} shows
that within a time $\tau_E \ll \tau _S$ the enzymes reach $1/e$ of what appears
to be a ``steady state.'' Of course, this is not a true steady-state, since
after a time $\tau _S$ the substrate concentration will appreciably decrease and
the enzyme conformations will correspondingly change. Instead, we call the
situation after one second a quasi-steady-state, meaning that the enzyme
conformations have all reached a steady-state value \textit{assuming the current
	substrate concentration is fixed}.

When $\tau _E$ is significantly smaller than $\tau _S$ (typically \(\tau _E\)
only needs to be roughly 100 times smaller than \(\tau _S\)), the dynamics of
the enzymes and substrate can be separated. In other words, we can assume that
the fast step (where the enzymes equilibrate to the current concentration of
substrate) happens instantly when considering the slow dynamics of the substrate
concentration diminishing over time. 
This is the quasi-steady-state approximation that we formally made in
\eref[eq:quasiSteadyStateMWCEnzyme] of \sref{MWCEnzymeSection}. We will next
show what relationship between the rate constants must hold so that the
quasi-steady-state approximation is valid.

\begin{figure}[h!]
	\centering \includegraphics[scale=1]{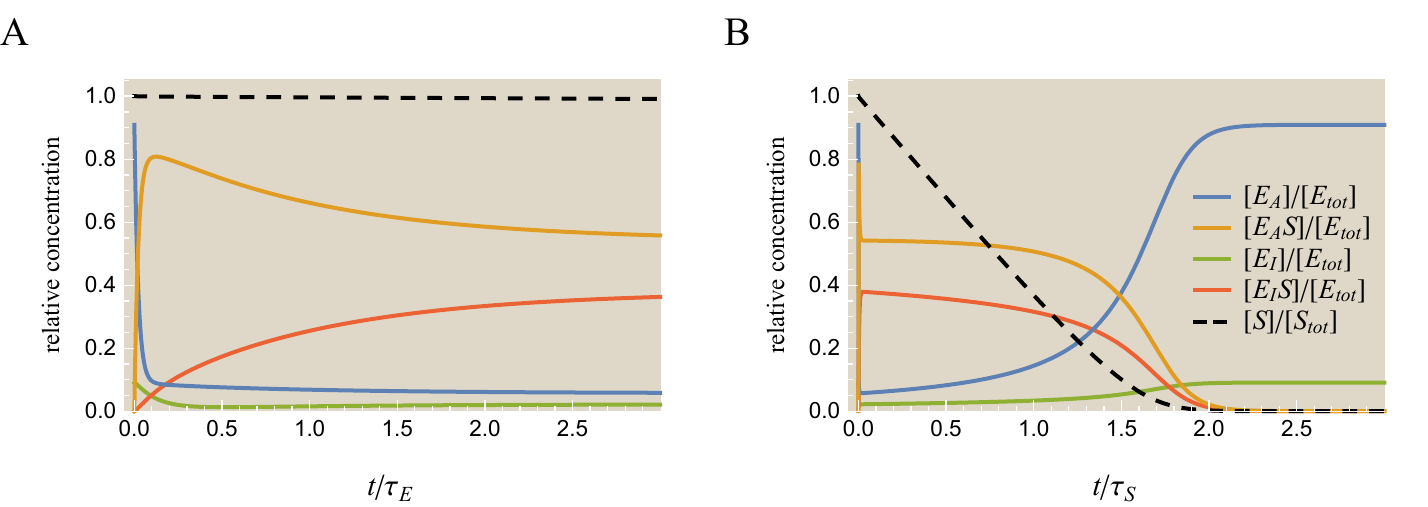} \caption{\textbf{The
		quasi-steady-state approximation.} \letterParen{A} The fast dynamics of the
		system in \eref[eq:DiscussionRatesEnzymeMWC2] begins by mixing unbound enzymes
		($E_A$ and $E_I$) and substrate. The enzyme conformations quickly reach steady
		state on a time scale of $\tau_E \approx 0.04\,\text{s}$. During this period,
		the substrate concentration remains very nearly constant. \letterParen{B} The
		substrate changes appreciably over the much longer time scale $\tau_S \approx
		11\,\text{s}$. Over this longer time scale, we can assume the
		quasi-steady-state approximation: the enzymes conformations are always in
		quasi-steady-state with the slowly diminishing substrate concentration.
		Concentrations used were $\left[E_{tot}\right]=1\,\mu\text{M}$,
		$\left[S_{tot}\right]=1\,\text{mM}$, $\left[E_AS\right]=\left[E_IS\right]=0$,
		and $\frac{\left[E_A\right]}{\left[E_I\right]}=\frac{k_{trans}^{I}}{k_{trans}^{A}} \equiv e^{-\beta \left(\epsilon_A-\epsilon_I\right)}$. The rate constants used were $k_{on}^A=1\,\text{s}^{-1}\text{M}^{-1}$, $k_{on}^I=10^{-1}\,\text{s}^{-1}\text{M}^{-1}$, $k_{off}^A=1\,\text{s}^{-1}$, $k_{off}^I=10^{-3}\,\text{s}^{-1}$, $k_{cat}^A=10^2\,\text{s}^{-1}$, $k_{cat}^I=10\,\text{s}^{-1}$, $k_{trans}^{AS}=k_{trans}^{IS}=k_{trans}^{A}=10\,\text{s}^{-1}$, and $k_{trans}^{I}=10^2\,\text{s}^{-1}$.} \label{fig:transient}
\end{figure}

We first calculate the time scale \(\tau _E\) for the enzyme conformations to
equilibrate. We will assume that the substrate concentration equals the constant
value \([S_{tot}]\) throughout this short timescale (which, as shown in
\fref[fig:transient]\letter{A}, is reasonable) and then invoke a self-consistency
condition to ensure that the actual change in substrate concentration during the
period \(\tau _E\) was negligible.

As a warm up, we first consider the Michaelis-Menten enzyme which we redraw here
\begin{equation} \label{eq:Discussionsimple_scheme}
\begin{aligned}
\includegraphics[scale=1]{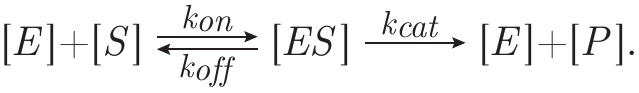}
\end{aligned}
\end{equation}
The Michaelis-Menten enzyme is governed by the multiple differential equations
\begin{equation} \label{eq:DiscussionMichaelisWarmUpEquation}
\frac{d[E]}{dt}=[ES]\left(k_{off}+k_{cat}\right)-[E]\left[S_{tot}\right]k_{on}=-\frac{d[ES]}{dt}
\end{equation}
and the constraint \([E]+[ES]=\left[E_{tot}\right]\). As stated above, we fix
the substrate concentration at \(\left[S_{tot}\right]\) and assume that the
system starts off with \([E]=\left[E_{tot}\right]\) and \([ES]=0\). Solving the
differential equation \eref[eq:DiscussionMichaelisWarmUpEquation] yields
\begin{align}
[E]&=\left[E_{tot}\right]\frac{K_M+\left[S_{tot}\right] e^{-t/\tau }}{K_M+\left[S_{tot}\right]} \\
[ES]&=\left[E_{tot}\right][S_{tot}]\frac{1-e^{-t/\tau }}{K_M+\left[S_{tot}\right]}
\end{align}
where \(\tau =\frac{1}{k_{on} [S_{tot}]+k_{off}+k_{cat}}\) is the time scale for
the system to equilibrate. Interestingly, $\frac{1}{\tau}$ equals the sum of all
rates between the states \([E]\) and \([ES]\) (i.e. the sum of all time scales
in this system). Furthermore, $\tau$ does not depend on the initial conditions
of the system.

We now turn to the harder case of the MWC enzyme whose kinetics
we describe using the scheme 
\begin{equation} \label{eq:ratesEnzymeMWCColored}
\begin{aligned}
\includegraphics[scale=1]{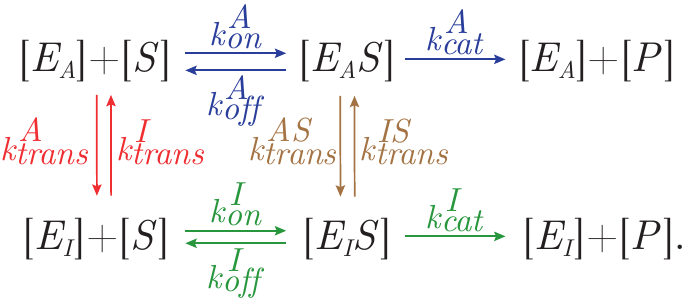}
\end{aligned}
\end{equation}
As we just saw for the Michaelis-Menten enzyme, if we just considered any edge
of the MWC enzyme separately, its corresponding time constant would be
$\frac{1}{\text{sum of rates along this edge}}$: \(\frac{1}{k_{on}^A
	[S_{tot}]+k_{off}^A+k_{cat}^A}\) between \(\left[E_A\right]\) and
\(\left[E_AS\right]\) (blue); \(\frac{1}{k_{trans}^A+k_{trans}^I}\) between
\(\left[E_A\right]\) and \(\left[E_I\right]\) (red); \(\frac{1}{k_{on}^I
	[S_{tot}]+k_{off}^I+k_{cat}^I}\) between \(\left[E_I\right]\) and
\(\left[E_IS\right]\) (green); and \(\frac{1}{k_{trans}^{AS} +k_{trans}^{IS}}\) between
\(\left[E_AS\right]\) and \(\left[E_IS\right]\) (brown). We can approximate the time
scale \(\tau _E\) of this system as the maximum of these four time scales between
adjacent edges, 
\begin{adjustwidth}{-0.15in}{0in}
	\begin{align} \label{eq:tauApproximation}
	\tau _E &\approx \max \left(\frac{1}{k_{on}^A [S_{tot}]+k_{off}^A+k_{cat}^A},\frac{1}{k_{trans}^A +k_{trans}^I},\frac{1}{k_{on}^I
		[S_{tot}]+k_{off}^I+k_{cat}^I},\frac{1}{k_{trans}^{AS} +k_{trans}^{IS}}\right) \nonumber \\
	&=\frac{1}{\min\left(k_{trans}^A+k_{trans}^I,k_{on}^A
		\left[S_{tot}\right]+k_{cat}^A+k_{off}^A,k_{trans}^{AS}+k_{trans}^{IS},k_{on}^I \left[S_{tot}\right]+k_{cat}^I+k_{off}^I\right)}.
	\end{align}
\end{adjustwidth}
This result is very similar (and in fact overestimates) the exact derivation of
\(\tau _E\) discussed in the next section, Appendix
\ref{quasiSteadyStateTimeConstants}.

With this form of \(\tau _E\) in hand, we could proceed in several ways to
determine when the quasi-steady-state approximation holds. For example, we could
compute the time scale \(\tau _S\) for the substrate to diminish and then
enforce \(\tau _E\ll \tau _S\) as the quasi-steady-state approximation. However,
Segel and Slemrod \cite{Segel2012} determined a tighter constraint by demanding that the amount of substrate converted into product during the transient
period $0<t<\tau_E$ only amounts to a tiny fraction of the
initial substrate concentration. The amount of substrate turned into product $\Delta [S]$ after time $\tau_E$ can be overestimated as 
\begin{equation}
\Delta [S]\approx \left| \frac{d[S]}{dt}\right| _{\max }\tau _E
\end{equation}
so that the quasi-steady-state approximation can be written as 
\begin{equation} \label{eq:QSSAconstraint}
\frac{\Delta [S]}{\left[S_{tot}\right]}\approx \frac{1}{\left[S_{tot}\right]}\left| \frac{d[S]}{dt}\right| _{\max }\tau
_E\ll 1.
\end{equation}
From \bareEq{eq:DiscussionRatesEnzymeMWC2}, the rate of change of substrate
concentration for the MWC enzyme is
\begin{equation}
\frac{d[S]}{dt}=-\left[E_A\right][S]k_{on}^A-\left[E_I\right][S]k_{on}^I+\left[E_AS\right]k_{off}^A+\left[E_IS\right]k_{off}^I.
\end{equation}
Recall that at \(t=0\), the system starts off with all
enzymes unbound: \(\left[E_AS\right]=\left[E_IS\right]=0\) and
\(\left[E_A\right]+\left[E_I\right]=[E_{tot}]\). Then \(\left|
\frac{d[S]}{dt}\right| _{\max }\) occurs at \(t=0\) (when $[S]=[S_{tot}]$) and
an upper bound is given by
\begin{equation} \label{eq:DiscussionOverestimatedSdt}
\left| \frac{d[S]}{dt}\right| _{\max } = [S_{tot}]\left(\left[E_A\right]k_{on}^A+\left[E_I\right]k_{on}^I \right) \leq \left[E_{tot}\right]\left[S_{tot}\right]\max \left(k_{on}^A,k_{on}^I\right).
\end{equation}
Substituting this result and the time scale \eref[eq:tauApproximation] into
\eref[eq:QSSAconstraint], we find a sufficient condition for the quasi-steady
state approximation to hold for an MWC enzyme:
\begin{adjustwidth}{-0.17in}{0in}
	\begin{equation} \label{eq:QSSAvalidityEquation}
	[E_{tot}]\frac{\max\left(k_{on}^A,k_{on}^I\right)}{\min\left(k_{trans}^A+k_{trans}^I,k_{on}^A
		\left[S_{tot}\right]+k_{cat}^A+k_{off}^A,k_{trans}^{AS}+k_{trans}^{IS},k_{on}^I \left[S_{tot}\right]+k_{cat}^I+k_{off}^I\right)} \ll 1.
	\end{equation}
\end{adjustwidth}

We could repeat this analysis for a Michaelis-Menten enzyme where only the
\(E_A\) and \(E_AS\) states exist. This is equivalent to disregarding all terms
except for \(k_{on}^A\), \(k_{off}^A\), and \(k_{cat}^A\) in the max and min of
\eref[eq:QSSAvalidityEquation], so that the quasi-steady-state conditions
reduces to \([E_{tot}]\frac{k_{on}^A}{k_{on}^A
	\left[S_{tot}\right]+k_{cat}^A+k_{off}^A}=\frac{[E_{tot}]}{\left[S_{tot}\right]+K_M^A}\ll 1\) which is identical to the condition found by Segel \cite{Segel2012}.

\subsection{Time Constants for the Quasi-Steady-State Approximation} \label{quasiSteadyStateTimeConstants}

In this section we derive an exact expression for the time constant for which
the MWC enzyme \bareEq{eq:ratesEnzymeMWCBetter} will attain its steady state
for each enzyme conformation assuming that the substrate concentration
$[S]=[S_{tot}]$ remains fixed. The rate of change of each enzyme conformation
can be written in matrix form (with bold denoting vectors and matrices) as
\begin{equation} \label{eq:transientRateOfChange}
\frac{d\boldsymbol{E}}{dt}= \boldsymbol{K} \boldsymbol{E}
\end{equation}
where 
\begingroup\makeatletter\def\f@size{10}\check@mathfonts
\def\maketag@@@#1{\hbox{\m@th\normalsize\normalfont#1}}
\begin{adjustwidth}{-0.5in}{0in}
	\begin{equation}
	\boldsymbol{K}=\left(
	\begin{array}{cccc}
	-k_{cat}^A-k_{off}^A-k_{trans}^{AS} & k_{on}^A\left[S_{tot}\right] & k_{trans}^{IS} & 0 \\
	k_{cat}^A+k_{off}^A & -k_{on}^A\left[S_{tot}\right]-k_{trans}^A & 0 & k_{trans}^I \\
	k_{trans}^{AS} & 0 & -k_{cat}^I-k_{off}^I-k_{trans}^{IS} & k_{on}^I\left[S_{tot}\right] \\
	0 & k_{trans}^A & k_{cat}^I+k_{off}^I & -k_{trans}^I-k_{on}^I\left[S_{tot}\right] \\
	\end{array}
	\right),\,\boldsymbol{E}=\left(
	\begin{array}{c}
	\left[E_AS\right] \\
	\left[E_A\right] \\
	\begin{array}{c}
	\left[E_IS\right] \\
	\left[E_I\right] \\
	\end{array}
	\\
	\end{array}
	\right).
	\end{equation}
\end{adjustwidth}
\endgroup
This matrix can be decomposed as
\begin{equation}
\boldsymbol{K}=\boldsymbol{V}^{-1} \boldsymbol{\Lambda} \boldsymbol{V}
\end{equation}
where \(\boldsymbol{V}\)'s columns are the eigenvectors of \(\boldsymbol{K}\)
and \(\boldsymbol{\Lambda}\) is a diagonal matrix whose entries are the
eigenvalues of \(\boldsymbol{K}\). In general, it is known that the eigenvalues
of such a matrix $\boldsymbol{K}$ representing the dynamics of any graph such as
\bareEq{eq:ratesEnzymeMWCBetter} from the text has one eigenvalue that is 0
while the remaining eigenvalues are non-zero and have negative real parts
\cite{Mirzaev2013}. (Indeed, because all of the columns of \(\boldsymbol{K}\)
add up to zero, \(\boldsymbol{K}\) is not full rank and hence one of its
eigenvalues must be zero.) Defining the vector
\begin{equation}
\boldsymbol{\tilde{E}} \equiv \boldsymbol{V} \boldsymbol{E}=\left(
\begin{array}{c}
\tilde{E}_1 \\
\tilde{E}_2 \\
\tilde{E}_3 \\
\tilde{E}_4 \\
\end{array}
\right),
\end{equation}
\eref[eq:transientRateOfChange] can be rewritten as
\begin{equation}
\frac{d\boldsymbol{\tilde{E}}}{dt}= \boldsymbol{\Lambda} \boldsymbol{\tilde{E}}.
\end{equation}
If the eigenvalues of $\Lambda$ are $\lambda _1$, $\lambda _2$, $\lambda _3$,
and 0, then \(\tilde{E}_j=c_je^{\lambda _jt}\) for \(j=1,2,3\) and
\(\tilde{E}_4=c_4\) where the \(c_j\)'s are constants determined by initial
conditions. Since the \(\tilde{E}_j\)'s are linear combinations of
\(\left[E_AS\right],\left[E_A\right],\left[E_IS\right],\) and
\(\left[E_I\right]\), this implies that the \(-\frac{1}{\lambda
	_1}$,$-\frac{1}{\lambda _2},\) and \(-\frac{1}{\lambda _3}\) (or
$-\frac{1}{\Re(\lambda_j)}$ if the eigenvalues are complex) are the time scales for
the system to come to equilibrium. Therefore, we can compute the overall time
scale for the system to come to equilibrium as
\begin{equation} \label{eq:tauExactAppendix}
\tau _E^{(exact)}=\max \left(-\frac{1}{\lambda _1},-\frac{1}{\lambda _2},-\frac{1}{\lambda _3}\right).
\end{equation}
Although the eigenvalues of this matrix can be calculated in closed form, they
are long and complicated expressions that contribute less intuition than the
approximation
\begin{equation} \label{eq:tauApproxAppendix}
\tau _E=\max \left(\frac{1}{k_{on}^A [S]+k_{off}^A+k_{cat}^A},\frac{1}{k_{trans}^A +k_{trans}^I},\frac{1}{k_{on}^I
	[S]+k_{off}^I+k_{cat}^I},\frac{1}{k_{trans}^{AS} +k_{trans}^{IS}}\right)
\end{equation}
used in \eref[eq:tauApproximation] in the text. However, given the exact form,
we can compare how well our approximation \eref[eq:tauApproxAppendix] matches the exact form \eref[eq:tauExactAppendix]. 

When the four time scales in \eref[eq:tauApproxAppendix] are comparable to each other, the approximation is very close to the exact form. However, when at least one pair of edges in the MWC enzyme rates diagram, 
\begin{equation} \label{eq:ratesEnzymeMWCAppendixVersion}
\begin{aligned}
\includegraphics[scale=1]{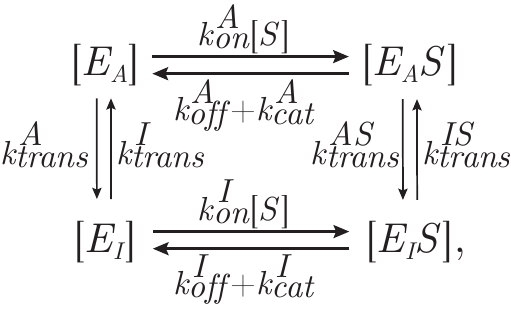}
\end{aligned}
\end{equation}
is very small the approximation tends to overshoot the exact value of $\tau _E$. For example, if
\(k_{trans}^A\approx k_{trans}^I\approx 0\)), \eref[eq:tauApproxAppendix]
implies $\tau _E \to \infty$ whereas \eref[eq:tauExactAppendix] can remain finite. 

\subsection{Generalizing the Cycle Condition}

We now consider what happens if an enzyme does not obey the cycle condition.
Provided that the quasi-steady-state approximation holds, then on the long time
scales the enzyme conformations quickly equilibrate to the current substrate
concentration. From \bareEq{eq:DiscussionRatesEnzymeMWC2}, the rate of change
of each enzyme species obeys
\begin{align}
\frac{d\left[E_A S\right]}{dt}&=0=\left[E_A\right][S]k_{on}^A-\left[E_AS\right]\left(k_{off}^A+k_{cat}^A+k_{trans}^{AS}\right)+\left[E_IS\right]k_{trans}^{IS} \\
\frac{d\left[E_A\right]}{dt}&=0=\left[E_AS\right]\left(k_{off}^A+k_{cat}^A\right)-\left[E_A\right][S]k_{on}^A-\left[E_A\right]k_{trans}^A+\left[E_I\right]k_{trans}^I \\
\frac{d\left[E_IS\right]}{dt}&=0=\left[E_I\right][S]k_{on}^I-\left[E_IS\right]\left(k_{off}^I+k_{cat}^I+k_{trans}^{IS}\right)+\left[E_AS\right]k_{trans}^{AS} \\
\frac{d\left[E_I\right]}{dt}&=0=\left[E_IS\right]\left(k_{off}^I+k_{cat}^I\right)-\left[E_I\right][S]k_{on}^I-\left[E_I\right]k_{trans}^I+\left[E_A\right]k_{trans}^A.
\end{align}
This system of equations, together with the conservation of total enzyme, \(\left[E_{tot}\right]=\left[E_AS\right]+\left[E_IS\right]+\left[E_A\right]+\left[E_I\right]\),
can be solved to obtain the quasi-steady-state values of each enzyme species. Using the Michaelis constants \(K_M^A=\frac{k_{off}^A+k_{cat}^A}{k_{on}^A}\)
and \(K_M^I=\frac{k_{off}^I+k_{cat}^I}{k_{on}^I}\), we can write the solutions as the three ratios
\begin{align}
\frac{\left[E_AS\right]}{\left[E_A\right]}&=\frac{[S]}{K_M^A}\frac{\left(K_M^Ik_{on}^I+k_{trans}^I\gamma +[S]k_{on}^I \gamma
	\right)+k_{trans}^A\alpha \gamma}{\left(K_M^Ik_{on}^I+k_{trans}^I\gamma +[S]k_{on}^I\gamma \right)+k_{trans}^A\alpha
	\beta \delta } \label{eq:fullyGeneralRatio1} \\
\frac{\left[E_IS\right]}{\left[E_I\right]}&=\frac{[S]}{K_M^I}\frac{\left(K_M^Ak_{on}^A+k_{trans}^A\delta +[S]k_{on}^A\delta
	\right)+k_{trans}^I\frac{\delta }{\alpha }}{\left(K_M^Ak_{on}^A+k_{trans}^A\delta +[S]k_{on}^A\delta \right)+k_{trans}^I\frac{\gamma
	}{\alpha \beta }} \label{eq:fullyGeneralRatio2} \\
\frac{\left[E_A\right]}{\left[E_I\right]}&=\frac{k_{trans}^I}{k_{trans}^A}\frac{\left(K_M^I k_{on}^I+k_{trans}^I\gamma
	+k_{trans}^A \alpha \beta \delta \right)+[S]k_{on}^I\gamma }{\left(K_M^I k_{on}^I+k_{trans}^I\gamma +k_{trans}^A
	\alpha \beta \delta \right)+[S]k_{on}^I\beta \delta } \label{eq:fullyGeneralRatio3}
\end{align}
where we have defined \(\alpha \equiv \frac{k_{on}^I}{k_{on}^A}\), \(\beta
\equiv \frac{K_M^I}{K_M^A}\), \(\gamma \equiv
\frac{k_{trans}^{IS}}{k_{trans}^I}\),
\(\delta \equiv \frac{k_{trans}^{AS}}{k_{trans}^A}\) to simplify the results. Notice that the terms in parenthesis in the numerator
and denominator of these three ratios are the same. Indeed, the large fractions in all three equations equal 1 if we set \(\gamma =\beta \delta\)
so that 
\begin{align}
\frac{\left[E_AS\right]}{\left[E_A\right]}&=\frac{[S]}{K_M^A} \\
\frac{\left[E_IS\right]}{\left[E_I\right]}&=\frac{[S]}{K_M^I} \\
\frac{\left[E_A\right]}{\left[E_I\right]}&=\frac{k_{trans}^I}{k_{trans}^A}.
\end{align}
This fortuitous choice of $\gamma $ is equivalent to the cycle condition
\eref[eq:cycleLawMWC], and so it is no surprise that these three ratios match
\eref[MWCKmDefa][MWCKmDefb][MWCEpsDef].

Invoking the cycle condition is a theoretical convenience which greatly
simplifies our equations. If the cycle condition does not hold, we can follow
our same procedure to turn
\eref[eq:fullyGeneralRatio1][eq:fullyGeneralRatio2][eq:fullyGeneralRatio3] into a
more general result for states and weights by only assuming the
quasi-steady-state approximation. While this more general procedure is
straightforward to implement numerically, it comes at the cost of introducing
more parameters into the model (for example, values for \(k_{on}^I\) and
\(k_{trans}^I\) must now be explicitly given whereas before we only needed to
determine the ratios $\frac{k_{on}^{I}}{k_{off}^{I}+k_{cat}^{I}}$ and
$\frac{k_{trans}^I}{k_{trans}^A}$) and the parameters will now depend upon the
substrate concentration.

Finally, we note that the cycle condition need not be invoked if a model does not contain any cycles. In other words, if we instead defined an MWC enzyme using the rates diagram
\begin{equation} \label{eq:modifiedMWCEnzymeRatesDiagram}
\begin{aligned}
\includegraphics[scale=1]{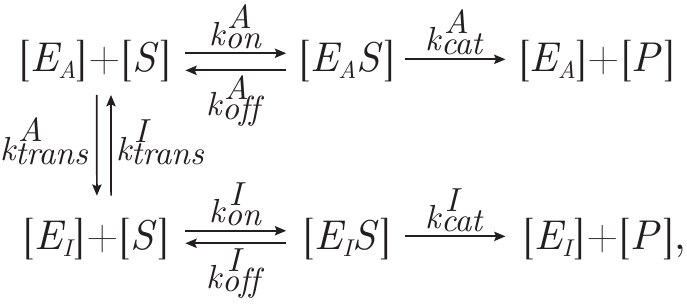}
\end{aligned}
\end{equation}
our analysis would proceed identically without needing to invoke the cycle
condition. Therefore, the cycle condition ensures that the system
\bareEq{eq:DiscussionRatesEnzymeMWC2} has the right value of
$\frac{k_{trans}^{AS}}{k_{trans}^{IS}}$ so that it can operate identically to
\bareEq{eq:modifiedMWCEnzymeRatesDiagram}.

\section{General Enzyme Models} \label{generalEnzymeModel}

In this section, we discuss the procedure used to fit the experimental enzyme
kinetics data to the theoretical framework we have developed for allosteric
enzymes. We then discuss the individual fits for each enzyme considered
throughout the paper. These fits may also be viewed directly in the
supplementary \textit{Mathematica} notebook which contains the code to generate
all of the plots in the paper as well as the experimental data for each enzyme.

All fitting was done using nonlinear regression (NonlinearModelFit in
\textit{Mathematica}) using the realistic constraints $K_M,C_D,R_D \in [10^{-2}
\mu\text{M}, 10^6 \mu\text{M}]$, $k_{cat} \in [10^{-2} \,\text{s}^{-1}, 10^5
\,\text{s}^{-1}]$, and $e^{-\beta \left(\epsilon _A-\epsilon _I\right)} \in
[-10,10]$ \cite{Phillips2015}. Initial conditions for the nonlinear regression
were chosen randomly from this parameter space until a sufficiently good fit
($R^2 > 0.99$) was found.

It must be noted that, as with nearly all models, there are serious ambiguities
in the best fit values since multiple sets of best fits values yield nearly
identical curves. In point of fact, if the nonlinear regression would be
performed without any constraints, it nearly always lands outside of the
physically relevant parameter space (although the qualitative form of the best
fit curves may be nearly indistinguishable from those that we show below). This
attribute of models, dubbed as ``sloppiness,'' is well known
\cite{Transtrum2015}. One of its implications may be that a biological system
can more easily evolve whichever activity profile it requires to maximize
fitness, since the system is more likely the stumble across the best possible
activity profile if it exists for numerous sets of parameters.

With this in mind, our results below demonstrate that our framework is
\textit{sufficient} to describe the complex interactions of allosteric enzymes,
but that the individual parameter values (i.e. $K_M$, $C_D$, $R_D$ values) are
\textit{not tightly determined} by these fits.

\subsection{Fitting $\boldsymbol{\alpha}$-Amylase and Allosteric Regulator Chlorine} \label{Appendix:FitAlphaAmylaseRegulator}

\fref[figAppendixInhibitorDataCollapse] shows three activity curves for
\textit{A. haloplanctis} $\alpha$-amylase titrating substrate at different
concentrations of the allosteric activator NaCl. This enzyme has one substrate
binding site and one allosteric site for binding chlorine ions. As discussed in
\sref{sec:ExperimentalData} of the main text, the $[S] / A$ curves are linear in $[S]$,
\begin{equation} \label{eqAppendixRegulatorForm}
\frac{[S]}{A} = \frac{e^{-\beta \epsilon_A}\left(1+\frac{[S]}{K_M^{A}}\right)\left(1+\frac{[R]}{R_D^{A}}\right)+e^{-\beta \epsilon_I}\left(1+\frac{[S]}{K_M^{I}}\right)\left(1+\frac{[R]}{R_D^{I}}\right)}{k_{cat}^A e^{-\beta \epsilon_A}\frac{1}{K_M^{A}}\left(1+\frac{[R]}{R_D^{A}} \right)+k_{cat}^I e^{-\beta \epsilon_I}\frac{1}{K_M^{I}}\left(1+\frac{[R]}{R_D^{I}} \right)}.
\end{equation}

Note that we are fitting the 7 parameters from this equation into a linear form
with 2 parameters (i.e. slope and intercept). Therefore, the individual
parameters are not themselves reliable; instead, these fits are intended to show
that the MWC model can account for the observed enzyme behavior. One possible
set of parameters that matches the data is given by $e^{-\beta \left(\epsilon
	_A-\epsilon _I\right)} = 7.8 \times 10^{-4}$, $K_M^A = 0.6 \,\text{mM}$, $K_M^I
= 0.2 \,\text{mM}$, $R_D^A = 0.03 \,\text{mM}$, $R_D^I = 7.9 \,\text{mM}$,
$k_{cat}^{A} = 14 \,\text{s}^{-1}$, and $k_{cat}^{I} = 0.01 \,\text{s}^{-1}$. To
find the value of an individual parameter, we would instead setup an experiment
where only that single parameter varies and fit the resulting data.

\begin{figure}[h!]
	\centering \includegraphics[scale=1]{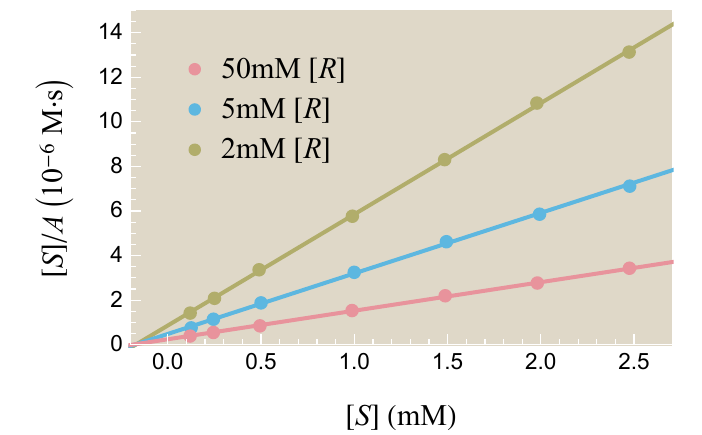} \caption{\textbf{Theoretically
		and experimentally probing the effects of an allosteric regulator on activity.}
		Data points show experimentally measured activity from Feller et al.~for the
		enzyme $\alpha$-amylase using substrate analog $[S]$ (EPS) and allosteric
		activator $[R]$ (NaCl), overlaid by theoretical curves of the form given in
		\eref[eqAppendixRegulatorForm] \cite{Bussy1996}. Reproduced from
		\fref[fig:amylaseRegulator] in the main
		text.}\label{figAppendixInhibitorDataCollapse}
\end{figure}

\subsection{Fitting $\boldsymbol{\alpha}$-Amylase and Competitive Inhibitor Isoacarbose} \label{Appendix:FitAlphaAmylaseInhibitor}

\fref[appendixFigInhibitorDataCollapse] shows three activity curves of human
pancreatic $\alpha$-amylase titrating competitive inhibitor at different
substrate concentrations. This enzyme has one active site which the substrate or
competitive inhibitor can bind to. As discussed in \sref{sec:ExperimentalData}
of the main text, the activity curves all take the form
\begin{equation} \label{eqAppendixInhibitorEqForm}
\left( \frac{d[P]}{dt} \right)^{-1} = \frac{1}{[E_{tot}]} \frac{e^{-\beta \epsilon_A}\left(1+\frac{[S]}{K_M^{A}}+\frac{[C]}{C_D^{A}}\right)+e^{-\beta \epsilon_I}\left(1+\frac{[S]}{K_M^{I}}+\frac{[C]}{C_D^{I}}\right)}{k_{cat}^Ae^{-\beta \epsilon_A}\frac{[S]}{K_M^{A}}+k_{cat}^Ie^{-\beta \epsilon_I}\frac{[S]}{K_M^{I}}}
\end{equation}
which is linear in $[C]$. 

As noted above, in fitting 6 parameters to a linear form, the best fit parameter
values are not reliable, but are only intended to show that the MWC model can
account for the observed enzyme behavior. One possible set of parameters that
matches the data is given by $e^{-\beta \left(\epsilon _A-\epsilon _I\right)} =
36$, $K_M^A = 0.9 \,\text{mM}$, $K_M^I = 2.6 \,\text{mM}$, $C_D^A = 12
\,\text{nM}$, $C_D^I = 260 \,\text{nM}$, and $\frac{k_{cat}^{A}}{k_{cat}^{I}} =
1.4$. Because units for activity were not included in original data, we instead
fit the dimensionless quantity $[E_{tot}] k_{cat}^A \left( \frac{d[P]}{dt}
\right)^{-1}$ which rescales the $y$-axis but does not change the form of the
activity curves \cite{Li2005}. 

\begin{figure}[h!]
	\centering \includegraphics[scale=1]{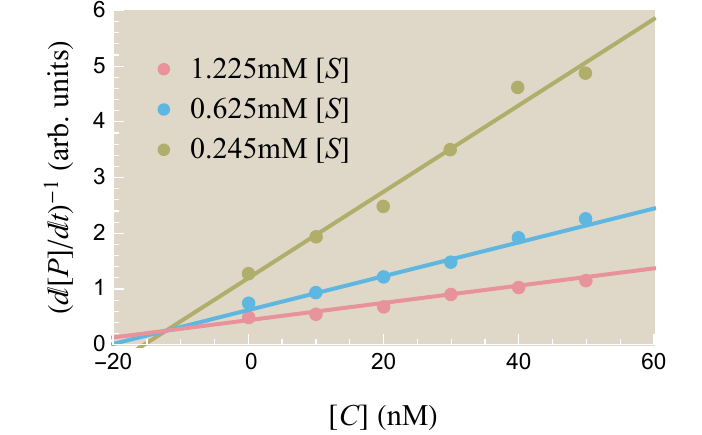} \caption{\textbf{Theoretically
		and experimentally probing the effects of a competitive inhibitor on activity.}
		Data points show experimentally measured activity in arbitrary units from Li et
		al.~for the enzyme $\alpha$-amylase using substrate analog $[S]$
		($\alpha$-maltotriosyl fluoride) and competitive inhibitor $[C]$ (isoacarbose),
		overlaid by theoretical curves of the form given by
		\eref[eqAppendixInhibitorEqForm] \cite{Li2005}. Best fit theoretical curves
		described by the inverse of \eref[eqDataActivity] are overlaid on the data.
		Reproduced from \fref[fig:dataCollapse]\letterParen{A} in the main text.}
	\label{appendixFigInhibitorDataCollapse}
\end{figure}

\subsection{Fitting Acetylcholinesterase Data} \label{Appendix:FittingDataAcetylcholinesterase}

\begin{figure}[h!]
	\centering \includegraphics[scale=1]{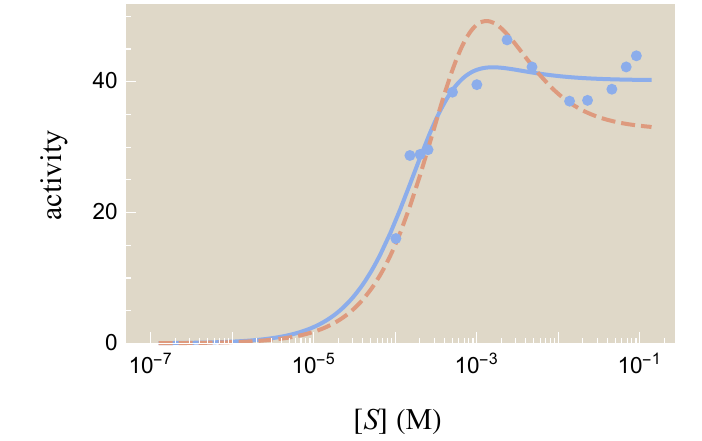} \caption{\textbf{The activity
		of acetylcholinesterase exhibits a peak. Activity for acetylcholinesterase is
		shown in units of $\boldsymbol{\text{(nanomoles product)}
		\cdot\text{min}^{-1}\cdot\text{(mL enzyme)}^{-1}}$ \cite{Changeux1966}.} The
		theoretical best-fit curve is shown (light blue) together with another theory
		curve which ignores the last three data points but better captures the height
		of the peak in the data (dashed, red).} \label{fig:AppendixSubstrateInhibitionFigure}
\end{figure}

The acetylcholinesterase data in \fref[fig:AppendixSubstrateInhibitionFigure]
was taken from \textit{Torpedo marmorata} \cite{Changeux1966}. Using our
framework from \sref{MoreComplexEnzymes}, activity is given by
\begin{equation} \label{eq:AppendixSubstrateInhibitionActivityCurveFitting}
A=N \frac{e^{-\beta \left(\epsilon _A-\epsilon _I\right)}k_{cat}^{A}\frac{[S]}{K_M^A}\left(1+\frac{[S]}{K_M^A}\right)^{N-1}+k_{cat}^{I}\frac{[S]}{K_M^I}\left(1+\frac{[S]}{K_M^I}\right)^{N-1}}{e^{-\beta \left(\epsilon _A-\epsilon _I\right)}\left(1+\frac{[S]}{K_M^A}\right)^N+\left(1+\frac{[S]}{K_M^I}\right)^N}
\end{equation}
where $N=2$ is the number of active sites. 

Activity is shown in units of $\text{(nanomoles product)}
\cdot\text{min}^{-1}\cdot\text{(mL enzyme)}^{-1}$. Using the density $3.6
\,\frac{\text{mg}}{\text{mL}}$ and molecular weight $2.3 \times 10^5
\,\frac{\text{g}}{\text{mol}}$ of the enzyme \cite{Changeux1966}, $1 \,\text{mL
	enzyme} = 1.6 \times 10^{-8} \,\text{mol}$. Therefore, 1 unit on the $y$-axis of
the figure corresponds to $10^{-3} \,\text{sec}^{-1}$.

The best fit parameters (light blue curve in
\fref[fig:AppendixSubstrateInhibitionFigure]) were $e^{-\beta \left(\epsilon
	_A-\epsilon _I\right)} = 0.5$, $K_M^A = 6.1 \times 10^{-3} \,\text{M}$, $K_M^I =
2.8 \times 10^{-4} \,\text{M}$, $k_{cat}^{A} = 3.1 \,\text{s}^{-1}$, and
$k_{cat}^{I} = 3.7 \times 10^{-2} \,\text{s}^{-1}$. The fitting is made
difficult by two factors. First, the data points are not evenly spaced, and the
three data points clumped together near $[S]=2 \times 10^{-4} \,\text{M}$ have
more weight on the fit than other points. Second, we suspect that the final
three data points in this figure have a significant amount of error and should
not curve back up - indeed, none of the other acetylcholinesterase substrate
inhibition curves from the same source exhibit this feature \cite{Changeux1966}.
To that end, we also show another theoretical curve (dashed, red) in order to
exemplify that the MWC model can capture the height of the peak in the data.
This latter curve has the parameters $e^{-\beta \left(\epsilon _A-\epsilon
	_I\right)} = 0.7$, $K_M^A = 7.4 \times 10^{-3} \,\text{M}$, $K_M^I = 5.9 \times
10^{-4} \,\text{M}$, $k_{cat}^{A} = 2.9 \,\text{s}^{-1}$, and $k_{cat}^{I} = 2.0
\times 10^{-2} \,\text{s}^{-1}$.

\subsection{Further Example Data} \label{Appendix:FittingDataATCase}

In this section, we present data on ATCase (not discussed in the main text)
which provides an excellent opportunity to combine all of the molecular players
and enzyme features we have analyzed - allosteric regulators, competitive
inhibitors, multiple substrate binding sites - in one complete model.



\begin{figure}[h!]
	\centering
	\includegraphics[scale=1]{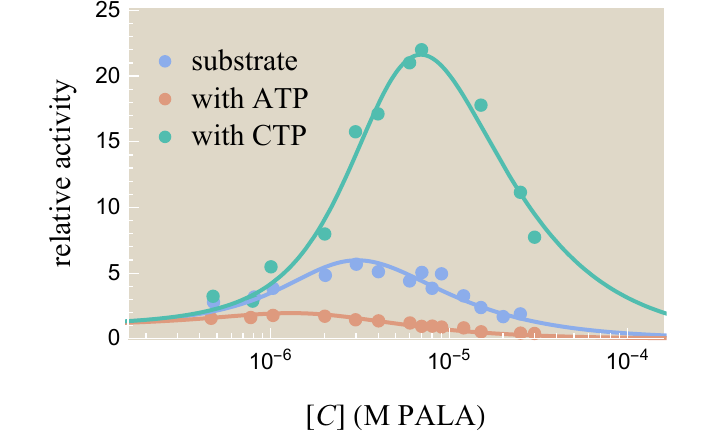}
	\caption{\textbf{Inhibitor
		activation in aspartate carbamoyltransferase (ATCase).} Activity curves from
		\textit{E. coli} ATCase are shown in the absence (blue circles) and the
		presence of allosteric effectors, either the activator ATP (yellow squares) or
		the inhibitor CTP (green diamonds) as a function of the competitive inhibitor
		\textit{N}-(phosphonacetyl)-{\small L}-aspartate (PALA). Data reproduced from
		Wales et al.~and fit to an MWC model \cite{Wales1999}.}
	\label{fig:AppendixLiteraturePlot}
\end{figure}

The ATCase data in \fref[fig:AppendixLiteraturePlot] was taken from
\textit{Escherichia coli} \cite{Wales1999}. ATCase is an allosteric enzyme with
6 active sites and 6 allosteric regulator sites. A competitive inhibitor PALA is
titrated, and the experiment is then repeated in the presence of the allosteric
activator ATP and the allosteric repressor CTP. Using our framework from
\sref{MoreComplexEnzymes}, the rate of product formation equals
\begin{equation} \label{eq:AppendixActivityCurveFitting}
\frac{d[P]}{dt}=N [E_{tot}] \frac{e^{-\beta \left(\epsilon _A-\epsilon _I\right)}k_{cat}^{A}\frac{[S]}{K_M^A}\left(1+\frac{[S]}{K_M^A}+\frac{[C]}{C_D^A}\right)^{N-1}+k_{cat}^{I}\frac{[S]}{K_M^I}\left(1+\frac{[S]}{K_M^I}+\frac{[C]}{C_D^I}\right)^{N-1}}{e^{-\beta \left(\epsilon _A-\epsilon _I\right)}\left(1+\frac{[S]}{K_M^A}+\frac{[C]}{C_D^A}\right)^N+\left(1+\frac{[S]}{K_M^I}+\frac{[C]}{C_D^I}\right)^N}
\end{equation}
where $N=6$ is the number of active sites. The plot in \fref[fig:AppendixLiteraturePlot] shows relative activity, which is defined as
\begin{equation}
\text{relative activity} = \frac{\frac{d[P]}{dt}}{\left(\frac{d[P]}{dt}\right)_{[C] \to 0}}.
\end{equation}

All three curves were carried out at a substrate concentration
$[S]=5\,\text{mM}$ of aspartate. In the absence of allosteric effectors (blue
curve), the best fit parameters were $e^{-\beta \left(\epsilon _A-\epsilon
	_I\right)} = 0.005$, $K_M^A = 1.1\,\text{mM}$, $K_M^I = 1.8\,\text{mM}$,
$k_{cat}^{A} = 400\,\text{s}^{-1}$, $k_{cat}^{I} = 0.02\,\text{s}^{-1}$, $C_D^A
= 0.3\,\mu\text{M}$, and $C_D^I = 1.8\,\mu\text{M}$. As per the theoretical
framework developed in \sref{AllostericEffectorSectionNew}, an allosteric
regulator such as ATP or CTP can be modeled by changing $e^{-\beta
	\left(\epsilon _A-\epsilon _I\right)} \to e^{-\beta \left(\epsilon _A-\epsilon
	_I\right)}\left(\frac{1+\frac{[R]}{R_D^A}}{1+\frac{[R]}{R_D^I}}\right)^N$ in
\eref[eq:AppendixActivityCurveFitting]. From \cite{Wales1999}, the
concentrations of ATP (gold curve) and CTP (green curve) were
$[R]=2\,\text{mM}$. Using the same MWC parameters as in the blue curve, the best
fit parameters for the allosteric activator ATP were $R_D^A = 0.07\,\text{mM}$
and $R_D^I = 0.10\,\text{mM}$; the best fit parameters for the allosteric
inhibitor CTP were $R_D^A = 0.14\,\text{mM}$ and $R_D^I = 0.10\,\text{mM}$.

\section{Data Collapse} \label{Appendix:DataCollapse}

In this section, we analyze the concept of data collapse, which allows us to map
the result of multiple activity curves onto a single curve using natural
parameters of the system. In section \ref{appendixD1}, we start by reviewing the
simplest case (presented in the main text) of an MWC enzyme with one active site
in the presence of a competitive inhibitor. We show that such an enzyme admits a
data collapse using a single parameter, so that all activity curves can be
collapsed onto a single curve. 
In section \ref{appendixD3}, we next consider the simplest MWC enzyme in the
presence of an allosteric regulator, with one active site and one allosteric
site. This case requires two parameters for a data collapse, and we show the
resulting collapse onto a sheet. We end with a general discussion of data
collapse theory in section \ref{appendixD4} which enables us to extend these
results to more complex enzymes (e.g. enzymes with more catalytic sites in the
presence of multiple species of allosteric regulators and competitive
inhibitors).

\subsection{Special Case: Enzyme with 1 Active Site and a Competitive Inhibitor} \label{appendixD1}

We start with a recap of the data collapse (discussed in
\sref{sec:ExperimentalData}) of an enzyme with a single active site in the
presence of a competitive inhibitor whose states and weights diagram is redrawn
in \fref[appendixRegulatorStatesWeights].
\begin{figure}[h!]
	\centering \includegraphics[scale=1]{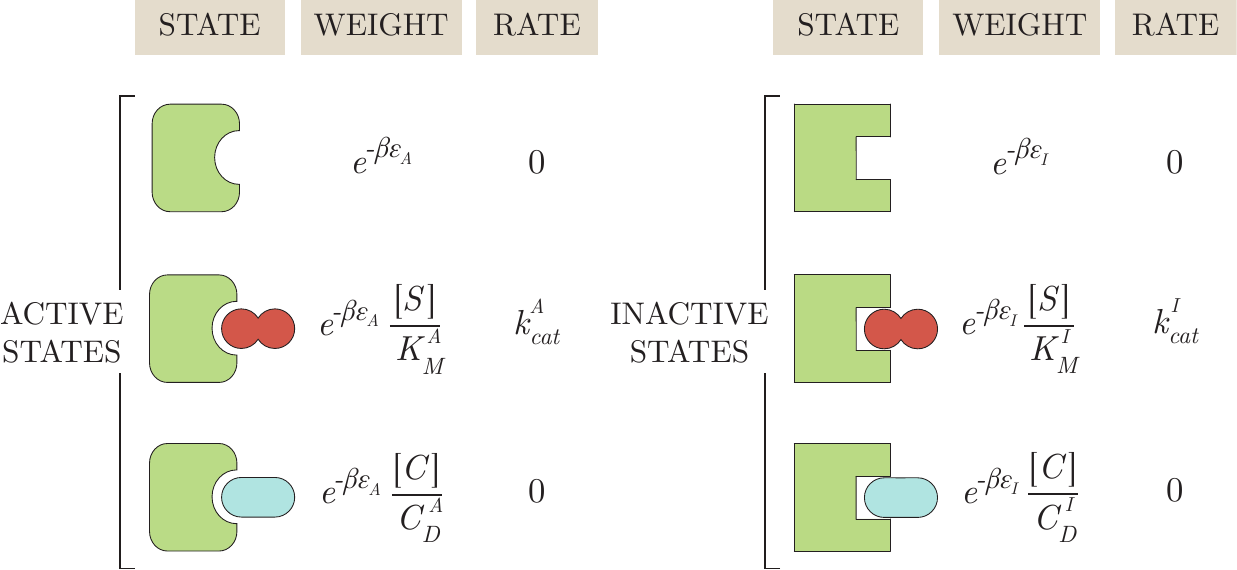} \caption{\textbf{States and
		weights for an MWC enzyme with an allosteric regulator.} Redrawn from
		\fref[fig:One_site_Enzyme_MWC_Reg] in the main text.}
	\label{appendixRegulatorStatesWeights}
\end{figure}
The activity $A = \frac{1}{[E_{tot}]}\frac{d[P]}{dt}$ for such an enzyme is
given by
\begin{equation}
A=\frac{k_{cat}^Ae^{-\beta  \Delta \epsilon }\frac{[S]}{K_M^A}+k_{cat}^I\frac{[S]}{K_M^I}}{e^{-\beta  \Delta \epsilon
	}\left(1+\frac{[S]}{K_M^A}+\frac{[C]}{C_D^A}\right)+\left(1+\frac{[S]}{K_M^I}+\frac{[C]}{C_D^I}\right)}
\end{equation}
where \(e^{-\beta  \Delta \epsilon }=e^{-\beta  \left(\epsilon _A-\epsilon
	_I\right)}\). Dividing the numerator and denominator by \(e^{-\beta  \Delta \epsilon}\left(1+\frac{[C]}{C_D^A}\right)+\left(1+\frac{[C]}{C_D^I}\right)\),
\begin{align}
A&=\frac{k_{cat}^A\left(\frac{e^{-\beta  \Delta \epsilon }\frac{[S]}{K_M^A}}{e^{-\beta  \Delta \epsilon }\left(1+\frac{[C]}{C_D^A}\right)+\left(1+\frac{[C]}{C_D^I}\right)}\right)+k_{cat}^I\left(\frac{\frac{[S]}{K_M^I}}{e^{-\beta
			\Delta \epsilon }\left(1+\frac{[C]}{C_D^A}\right)+\left(1+\frac{[C]}{C_D^I}\right)}\right)}{\frac{e^{-\beta  \Delta \epsilon }\frac{[S]}{K_M^A}}{e^{-\beta
			\Delta \epsilon }\left(1+\frac{[C]}{C_D^A}\right)+\left(1+\frac{[C]}{C_D^I}\right)}+\frac{\frac{[S]}{K_M^I}}{e^{-\beta  \Delta \epsilon }\left(1+\frac{[C]}{C_D^A}\right)+\left(1+\frac{[C]}{C_D^I}\right)}+1}
\nonumber \\
&=\frac{k_{cat}^Ae^{-\beta  \Delta F_{13}}+k_{cat}^Ie^{-\beta  \Delta F_{23}}}{e^{-\beta  \Delta F_{13}}+e^{-\beta  \Delta F_{23}}+1}
\end{align}
where we have defined the two \textit{Bohr parameters},
\begin{align}
\Delta F_{13}&=-\frac{1}{\beta }\text{Log}\left[\frac{e^{-\beta  \Delta \epsilon }\frac{[S]}{K_M^A}}{e^{-\beta  \Delta \epsilon }\left(1+\frac{[C]}{C_D^A}\right)+\left(1+\frac{[C]}{C_D^I}\right)}\right]\\
\Delta F_{23}&=-\frac{1}{\beta }\text{Log}\left[\frac{\frac{[S]}{K_M^I}}{e^{-\beta  \Delta \epsilon }\left(1+\frac{[C]}{C_D^A}\right)+\left(1+\frac{[C]}{C_D^I}\right)}\right]. \label{eq2}
\end{align}

Because both \(\Delta F_{13}\) and \(\Delta F_{23}\) have the exact same
dependence on \([S]\) and \([C]\), we can characterize the system by a
single natural variable. For example, since
\begin{equation}
e^{-\beta  \Delta F_{13}}=e^{-\beta  \Delta \epsilon }\frac{K_M^I}{K_M^A}e^{-\beta  \Delta F_{23}}
\end{equation}
we can rewrite \eref[eq2] using only \(\Delta F_{23}\),
\begin{equation}
A=\frac{k_{cat}^Ae^{-\beta  \Delta \epsilon }\frac{K_M^I}{K_M^A}e^{-\beta  \Delta F_{23}}+k_{cat}^Ie^{-\beta  \Delta
		F_{23}}}{e^{-\beta  \Delta \epsilon }\frac{K_M^I}{K_M^A}e^{-\beta  \Delta F_{23}}+e^{-\beta  \Delta F_{23}}+1}.
\end{equation}
For cleanliness, we can group the constants using
\begin{equation}
K\equiv e^{-\beta  \Delta \epsilon }\frac{K_M^I}{K_M^A},
\end{equation}
so that the activity becomes
\begin{equation} \label{eqAppendixDataCollapse}
A=\frac{\left(k_{cat}^AK +k_{cat}^I\right)e^{-\beta  \Delta F_{23}}}{(K+1)e^{-\beta  \Delta F_{23}}+1},
\end{equation}
matching \eref[eqActDataCollapse] from the text. As discussed in the text, this
form allows us to map any number of activity curves onto a single curve of
activity $A$ versus the natural variable of the system $\Delta F_{23}$. We
redraw such a plot from the main text in \fref[figAppendixDataCollapse].

\begin{figure}[h!]
	\centering \includegraphics[scale=1]{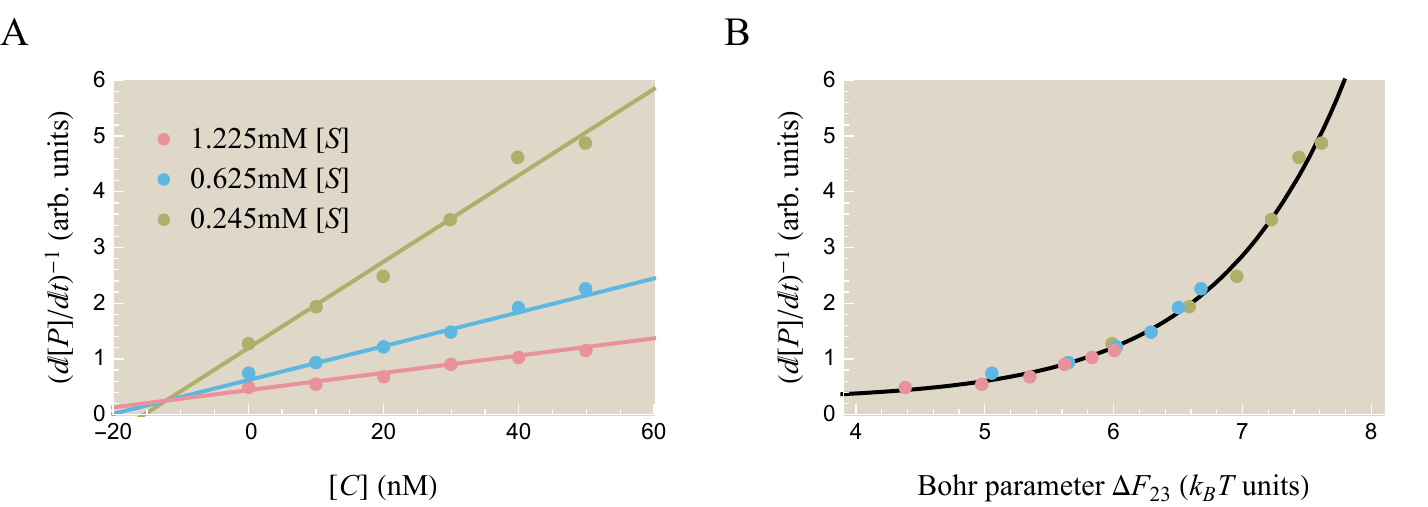} \caption{\textbf{Data from Li
		et al.~showing the effects of a competitive inhibitor $\boldsymbol{C}$ on the rate of
		product formation $\boldsymbol{\frac{d[P]}{dt}}$.} \letterParen{A} Individual activity curves
		are shown at various concentrations of the substrate $\alpha$-maltotriosyl
		fluoride ($\alpha$G3F) \cite{Li2005}. \letterParen{B} Curves are all data
		collapsed onto a single curve using the Bohr parameter $\Delta F_{23}$
		from \eref[eqAppendixDataCollapse].}\label{figAppendixDataCollapse}
\end{figure}

\subsection{Special Case: Enzyme with 1 Active Site and an Allosteric Regulator} \label{appendixD3}

Consider an enzyme with one active site and one allosteric site in the presence
of an allosteric regulator. The states and weights for such an enzyme are
redrawn in \fref[appendixInhibitorStatesWeights].
\begin{figure}[h!]
	\centering \includegraphics[scale=1]{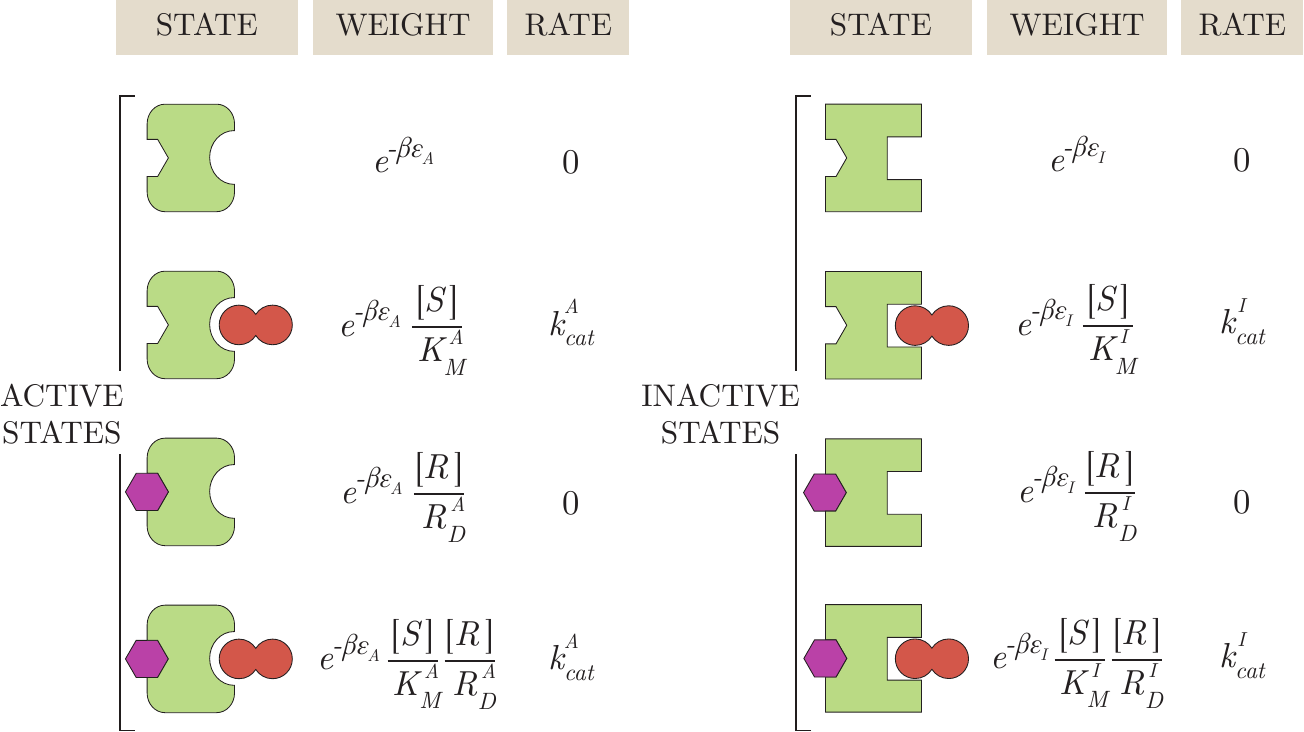} \caption{\textbf{States and
		weights for an MWC enzyme with a competitive inhibitor.} Redrawn from
		\fref[fig:CompetitorStatesWeights] in the main text.}
	\label{appendixInhibitorStatesWeights}
\end{figure}
The activity of such an enzyme is given by
\begin{equation}
A=\frac{k_{cat}^Ae^{-\beta  \Delta \epsilon }\frac{[S]}{K_M^A}\left(1+\frac{[R]}{R_D^A}\right)+k_{cat}^I\frac{[S]}{K_M^I}\left(1+\frac{[R]}{R_D^I}\right)}{e^{-\beta
		\Delta \epsilon }\left(1+\frac{[S]}{K_M^A}\right)\left(1+\frac{[R]}{R_D^A}\right)+\left(1+\frac{[S]}{K_M^I}\right)\left(1+\frac{[R]}{R_D^I}\right)}.
\end{equation}
where \(e^{-\beta  \Delta \epsilon }=e^{-\beta  \left(\epsilon _A-\epsilon
	_I\right)}\). We rewrite the numerator as
\begin{equation}
A=\frac{A_1e^{-\beta  \Delta \epsilon }\frac{[S]}{K_M^A}\left(1+\frac{[R]}{R_D^A}\right)+A_2\frac{[S]}{K_M^I}\left(1+\frac{[R]}{R_D^I}\right)}{e^{-\beta
		\Delta \epsilon }\left(1+\frac{[S]}{K_M^A}\right)\left(1+\frac{[R]}{R_D^A}\right)+\left(1+\frac{[S]}{K_M^I}\right)\left(1+\frac{[R]}{R_D^I}\right)}
\end{equation}
where 
\begin{align}
A_1&=k_{cat}^A\\
A_2&=k_{cat}^I.
\end{align}
Dividing the numerator and denominator by \(\left(1+\frac{[R]}{R_D^A}\right)+\left(1+\frac{[R]}{R_D^I}\right)\), we can rewrite the activity using
four natural variables,
\begin{equation}
A=\frac{A_1e^{-\beta  \Delta F_{13}}+A_2e^{-\beta  \Delta F_{23}}}{e^{-\beta  \Delta F_{13}}+e^{-\beta  \Delta F_{23}}+1}
\end{equation}
where
\begin{align}
\Delta F_{13}&=-\frac{1}{\beta }\text{Log}\left[\frac{e^{-\beta  \Delta \epsilon }\frac{[S]}{K_M^A}\left(1+\frac{[R]}{R_D^A}\right)}{e^{-\beta  \Delta \epsilon }\left(1+\frac{[R]}{R_D^A}\right)+\left(1+\frac{[R]}{R_D^I}\right)}\right] \label{eqAppendixDataCollapseRegulator1}\\
\Delta F_{23}&=-\frac{1}{\beta }\text{Log}\left[\frac{\frac{[S]}{K_M^I}\left(1+\frac{[R]}{R_D^I}\right)}{e^{-\beta  \Delta \epsilon }\left(1+\frac{[R]}{R_D^A}\right)+\left(1+\frac{[R]}{R_D^I}\right)}\right]. \label{eqAppendixDataCollapseRegulator2}
\end{align}
In this case, the two natural variables have a fundamentally different
dependence on \([R]\) and hence cannot be combined as in the case of a
competitive inhibitor. With two parameters, any number of activity curves can be
collapsed down upon a surface as shown in
\fref[figAppendixDataCollapseRegulator].

\begin{figure}[h!]
	\centering \includegraphics[scale=1]{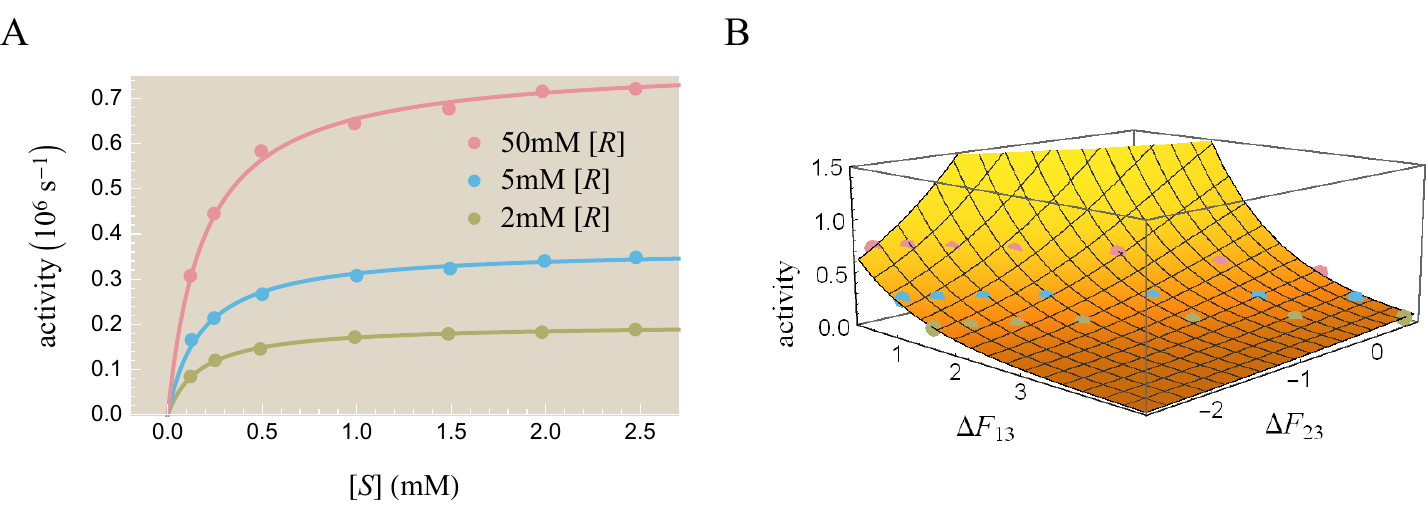} \caption{\textbf{Data from
		Feller et al.~demonstrating the rate of product formation $\boldsymbol{\frac{d[P]}{dt}}$ in
		the presence of an allosteric activator NaCl.} \letterParen{A} Individual
		activity curves of $\alpha$-amylase are shown at various concentrations of a
		substrate analog (EPS) \cite{Bussy1996}. Curves reproduced from
		\fref[fig:amylaseRegulator] in main text but with the $y$-axis showing
		$\frac{d[P]}{dt}$ rather than $\frac{[S]}{d[P]/dt}$. \letterParen{B} Curves are
		all data collapsed onto a surface using the Bohr parameters $\Delta F_{13}$ and
		$\Delta F_{23}$ from
		\eref[eqAppendixDataCollapseRegulator1][eqAppendixDataCollapseRegulator2].}\label{figAppendixDataCollapseRegulator}
\end{figure}

\subsection{General Theory} \label{appendixD4}

We now abstract the procedure used in the previous sections in order to
understand how to obtain a data collapse for any enzyme system. Suppose we
enumerate all of the states and weights of an enzyme, and that all of the states
pooled together only have three distinct catalytic rates $A_1$, $A_2$, and
$A_3$. (It is straightforward to generalize this argument to any number other
than three.)

Define \(S_1\), \(S_2\), and \(S_3\) to be the states that have catalytic rates
\(A_1\), \(A_2\), and \(A_3\). Then the activity of the enzyme is given by
\begin{equation}
A =\frac{A_1\sum _{j\in S_1}e^{-\beta  E_j}+A_2\sum _{j\in S_2}e^{-\beta  E_j}+A_3\sum _{j\in S_3}e^{-\beta  E_j}}{\sum _{j\in S_1}e^{-\beta
		E_j}+\sum _{j\in S_2}e^{-\beta  E_j}+\sum _{j\in S_3}e^{-\beta  E_j}}.
\end{equation}
Defining the free energies
\begin{align}
e^{-\beta  F_1} &\equiv \sum _{j\in S_1}e^{-\beta  E_j}\\
e^{-\beta  F_2} &\equiv \sum _{j\in S_2}e^{-\beta  E_j}\\
e^{-\beta  F_3} &\equiv \sum _{j\in S_3}e^{-\beta  E_j}
\end{align}
allows us to rewrite the activity as
\begin{align}
A &=\frac{A_1e^{-\beta  F_1}+A_2e^{-\beta  F_2}+A_3e^{-\beta  F_3}}{e^{-\beta  F_1}+e^{-\beta  F_2}+e^{-\beta  F_3}} \label{eq1} \nonumber\\
&=\frac{A_1e^{-\beta  \Delta F_{13}}+A_2e^{-\beta  \Delta F_{23}}+A_3}{e^{-\beta  \Delta F_{13}}+e^{-\beta  \Delta F_{23}}+1}
\end{align}
where \(\Delta F_{13} \equiv F_1-F_3\) and \(\Delta F_{23} \equiv F_2-F_3\) are
the two minimal parameters defining the system. Here, we see explicitly that
each Bohr parameter corresponds to a free energy difference between combinations
of states with the same activity (hence the notation $\Delta F$).

For example, in section \ref{appendixD1} above, the activity of an enzyme with one active site in the presence of a competitive inhibitor is given by 
\begin{equation}
A=\frac{k_{cat}^Ae^{-\beta  \Delta \epsilon
	}\frac{[S]}{K_M^A}+k_{cat}^I\frac{[S]}{K_M^I}}{e^{-\beta  \Delta \epsilon
}\left(1+\frac{[S]}{K_M^A}+\frac{[C]}{C_D^A}\right)+\left(1+\frac{[S]}{K_M^I}+\frac{[C]}{C_D^I}\right)}.
\end{equation}
To match the form of \eref[eq1], we rewrite this equation as 
\begin{equation}
A=\frac{k_{cat}^A\left(e^{-\beta  \Delta \epsilon }\frac{[S]}{K_M^A}\right)+k_{cat}^I\left(\frac{[S]}{K_M^I}\right)+0\left(e^{-\beta
		\Delta \epsilon }\left\{1+\frac{[C]}{C_D^A}\right\}+\left\{1+\frac{[C]}{C_D^I}\right\}\right)}{e^{-\beta  \Delta \epsilon }\left(1+\frac{[S]}{K_M^A}+\frac{[C]}{C_D^A}\right)+\left(1+\frac{[S]}{K_M^I}+\frac{[C]}{C_D^I}\right)},
\end{equation}
with \(A_1=k_{cat}^A\), \(A_1=k_{cat}^I\), and \(A_3=0\). Dividing the numerator and denominator by \(e^{-\beta  \Delta \epsilon}\left(1+\frac{[C]}{C_D^A}\right)+\left(1+\frac{[C]}{C_D^I}\right)\) yields the data collapse equation
\begin{align}
A&=\frac{k_{cat}^A\left(\frac{e^{-\beta  \Delta \epsilon }\frac{[S]}{K_M^A}}{e^{-\beta  \Delta \epsilon }\left(1+\frac{[C]}{C_D^A}\right)+\left(1+\frac{[C]}{C_D^I}\right)}\right)+k_{cat}^I\left(\frac{\frac{[S]}{K_M^I}}{e^{-\beta
			\Delta \epsilon }\left(1+\frac{[C]}{C_D^A}\right)+\left(1+\frac{[C]}{C_D^I}\right)}\right)}{\frac{e^{-\beta  \Delta \epsilon }\frac{[S]}{K_M^A}}{e^{-\beta
			\Delta \epsilon }\left(1+\frac{[C]}{C_D^A}\right)+\left(1+\frac{[C]}{C_D^I}\right)}+\frac{\frac{[S]}{K_M^I}}{e^{-\beta  \Delta \epsilon }\left(1+\frac{[C]}{C_D^A}\right)+\left(1+\frac{[C]}{C_D^I}\right)}+1}
\nonumber \\
&=\frac{k_{cat}^Ae^{-\beta  \Delta F_{13}}+k_{cat}^Ie^{-\beta  \Delta F_{23}}}{e^{-\beta  \Delta F_{13}}+e^{-\beta  \Delta F_{23}}+1}
\end{align}
with the two Bohr parameters
\begin{align}
\Delta F_{13}&=-\frac{1}{\beta }\text{Log}\left[\frac{e^{-\beta  \Delta \epsilon }\frac{[S]}{K_M^A}}{e^{-\beta  \Delta \epsilon }\left(1+\frac{[C]}{C_D^A}\right)+\left(1+\frac{[C]}{C_D^I}\right)}\right]\\
\Delta F_{23}&=-\frac{1}{\beta }\text{Log}\left[\frac{\frac{[S]}{K_M^I}}{e^{-\beta  \Delta \epsilon }\left(1+\frac{[C]}{C_D^A}\right)+\left(1+\frac{[C]}{C_D^I}\right)}\right].
\end{align}

\section{Inhibitor Acceleration: ATCase} \label{Appendix:FittingData}

This section will examine the phenomenon of inhibitor acceleration. The analysis
will closely follow \sref{secSubstrateInhibition} in the text. We first
demonstrate that inhibitor acceleration (having a peak in activity as a function
of competitive inhibitor concentration) cannot occur for any enzyme with one
active site and then show that it can occur for an MWC enzyme with two (or more)
active sites.

\subsection{Inhibitor Acceleration Does Not Occur for an Enzyme with One Active Site}

Consider an enzyme with a single active site in the presence of a competitive
inhibitor, as in \fref[fig:CompetitorStatesWeights]. We start by rewriting the
activity for such an enzyme from \eref[eq:MWCActCompdPdt],
\begin{equation} \label{eq:MWCActCompdPdtAppendix}
A = \frac{1}{[E_{tot}]}\frac{d[P]}{dt} = \frac{k_{cat}^Ae^{-\beta \epsilon_A}\frac{[S]}{K_M^{A}}+k_{cat}^Ie^{-\beta \epsilon_I}\frac{[S]}{K_M^{I}}}{e^{-\beta \epsilon_A}\left(1+\frac{[S]}{K_M^{A}}+\frac{[C]}{C_D^{A}}\right)+e^{-\beta \epsilon_I}\left(1+\frac{[S]}{K_M^{I}}+\frac{[C]}{C_D^{I}}\right)}.
\end{equation}
The derivative of activity with respect to inhibitor concentration $[C]$ is given by
\begin{equation} \label{1SubMWCActivityDerivativeC}
\frac{dA}{d[C]}=-\frac{\left(e^{-\beta \epsilon_A}\frac{1}{C_D^A}+e^{-\beta \epsilon_I}\frac{1}{C_D^I}\right) \left( e^{-\beta \epsilon_A} \frac{k_{cat}^A}{K_M^A} + e^{-\beta \epsilon_I} \frac{k_{cat}^I}{K_M^I} \right) [S]}{\left( e^{-\beta \epsilon_A}\left(1+\frac{[S]}{K_M^{A}}+\frac{[C]}{C_D^{A}}\right)+e^{-\beta \epsilon_I}\left(1+\frac{[S]}{K_M^{I}}+\frac{[C]}{C_D^{I}}\right) \right)^2}.
\end{equation}
Since the numerator cannot equal zero for any value of $[C]$, a peak cannot
occur when the competitive inhibitor is added. Instead, $\frac{dA}{d[C]}$ is
negative, indicating that adding more competitive inhibitor will decrease the
activity, as is typically expected from an inhibitor.

\subsection{Inhibitor Acceleration for an Enzyme with Two Active Sites} \label{AcceleratedActivityInhibitionSection}

\begin{figure}
	\centering \includegraphics[scale=1]{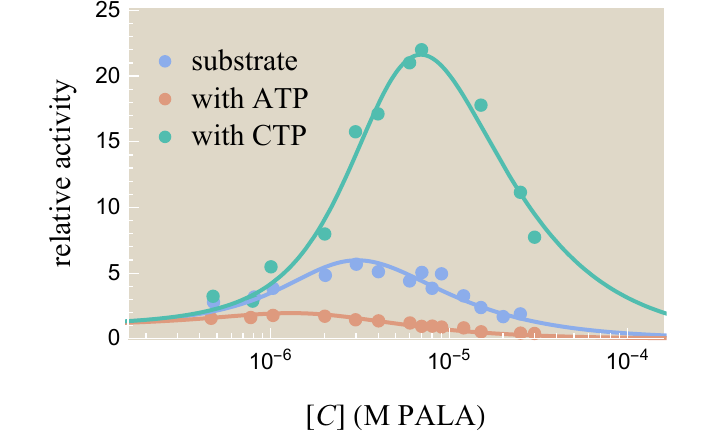}
	\caption{\textbf{The activity of aspartate carbamoyltransferase (ATCase) exhibits a peak.}
		Reproduced from \fref[fig:AppendixLiteraturePlot].}
	\label{fig:literaturePlot}
\end{figure}

Some allosteric enzymes exhibit an increase in activity when a small amount of
competitive inhibitor $C$ is introduced, as shown in \fref[fig:literaturePlot].
The simplest enzyme model which allows such a peak has two substrate binding
sites and includes allostery. For simplicity, we work in the limit
$k_{cat}^I=0$. Combining the results from sections
\ref{competitiveInihbitorSectionNew} and
\ref{sectionMultipleSubstrateBindingSites}, the activity for such an enzyme is
given by
\begin{equation} \label{eq:twoSubstrateSiteCompetitorActivityEquation}
A = k_{cat}^A e^{-\beta \epsilon_A} \frac{2 \frac{[S]}{K_M^A} \left( 1 + \frac{[C]}{C_D^A}+\frac{[S]}{K_M^A} \right)}{e^{-\beta \epsilon _A} \left(1+\frac{[S]}{K_M^A}+\frac{[C]}{C_D^A}\right)^2+e^{-\beta \epsilon _I} \left(1+\frac{[S]}{K_M^I}+\frac{[C]}{C_D^I}\right)^2}.
\end{equation}
A peak will occur provided that $\frac{dA}{d[C]}=0$ for a positive value of
$[C]$. For now, we skip the details of solving such a root (discussed in
Appendix \ref{appendixCompetitiveInhibitor2SitesSection}) and move straight to the results.
\eref[eq:twoSubstrateSiteCompetitorActivityEquation] will have a positive root
for $[C]$ provided the following relation holds,
\begin{equation} \label{eq:peakConditionCompetitorActivityEquation}
e^{-\beta \left(\epsilon_A - \epsilon_I \right)} < \left( \frac{1+\frac{[S]}{K_M^I}}{1+\frac{[S]}{K_M^A}}\right)^2 - 2 \frac{C_D^A}{C_D^I} \frac{1+\frac{[S]}{K_M^I}}{1+\frac{[S]}{K_M^A}} \;\;\;\;\;\;\;\; (k_{cat}^I=0).
\end{equation}

Acceleration by an inhibitor has historically been explained by a competitive
inhibitor binding to one active site of an enzyme, forcing it into the active
state \cite{Howlett1977}. This is indeed part of the story. Consider an enzyme
that natively favors the inactive state when no inhibitor is present, as shown
in the $[C] \ll C_D^A$ region of \fref[fig:pictoralCompetitorAcceleration]. As
$[C]$ increases, many enzymes will bind inhibitor in one active site, leaving
the remaining active site free to bind substrate. If the inhibitor favors
binding to the active-state enzyme, the ratio of active to inactive
enzymes will increase which will generate a peak in activity. When $[C] \gg
C_D^A$, the inhibitor will fill nearly all active sites and quash product
formation. This story suggests that having a smaller $\frac{C_D^A}{C_D^I}$ value
(i.e. having an inhibitor which strongly prefers binding to an active-state
enzyme) will increase the likelihood of generating a peak. This is confirmed by
the peak condition \eref[eq:peakConditionCompetitorActivityEquation] where
decreasing $\frac{C_D^A}{C_D^I}$ increases the right-hand side of the
inequality.

\begin{figure}[h!]
	\centering \includegraphics[scale=1]{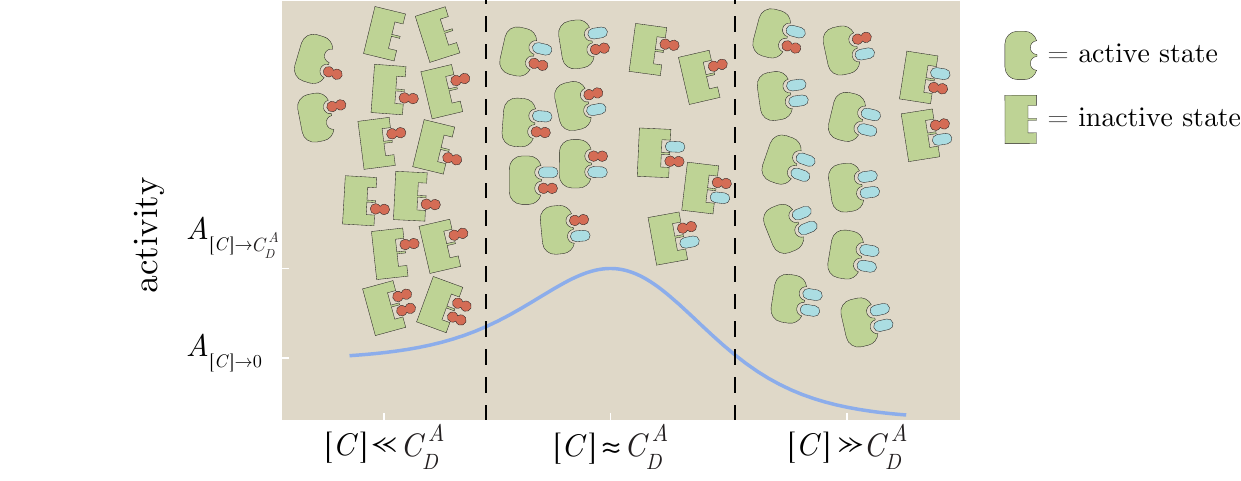} \caption{\textbf{Mechanism
		underlying peak in activation by a competitive inhibitor $\boldsymbol{C}$.} At low inhibitor
		concentrations, $[C] \ll C_D^A$, most enzymes are in the inactive form (sharp,
		green). As the amount of inhibitor increases, it will begin to compete with the
		substrate for active sites. At medium concentrations, $[C] \gg C_D^A$, some
		enzymes will have one site filled with a competitive inhibitor which prefers to
		bind in an active-state (rounded, green) enzyme complex. This increased
		probability of having active-state enzyme-substrate complexes (albeit with one
		enzyme site filled with an inhibitor) yields a larger activity compared to the
		low inhibitor concentrations. At large inhibitor concentrations, $[C] \gg
		C_D^A$, the inhibitor outcompetes the substrate for active sites and enzyme
		activity is suppressed.} \label{fig:pictoralCompetitorAcceleration}
\end{figure}

However, the complete story behind activation by inhibitor is more nuanced. To
gain some intuition, we first consider the limit $\frac{C_D^A}{C_D^I}\approx 0$
where the inhibitor binds exclusively to the active rather than the inactive
state. This limit maximizes the right-hand side of
\eref[eq:peakConditionCompetitorActivityEquation] which we can rewrite as
\begin{equation} \label{eq:peakConditionCompetitorActivityEquationNoCd}
e^{-\beta \epsilon_A} \left(1+\frac{[S]}{K_M^A} \right)^2 < e^{-\beta \epsilon_I} \left(1+\frac{[S]}{K_M^I}\right)^2 \;\;\;\;\;\;\;\; (k_{cat}^I=0, \, \frac{C_D^A}{C_D^I} = 0).
\end{equation}
\begin{figure}[h!]
	\centering
	\includegraphics[scale=1]{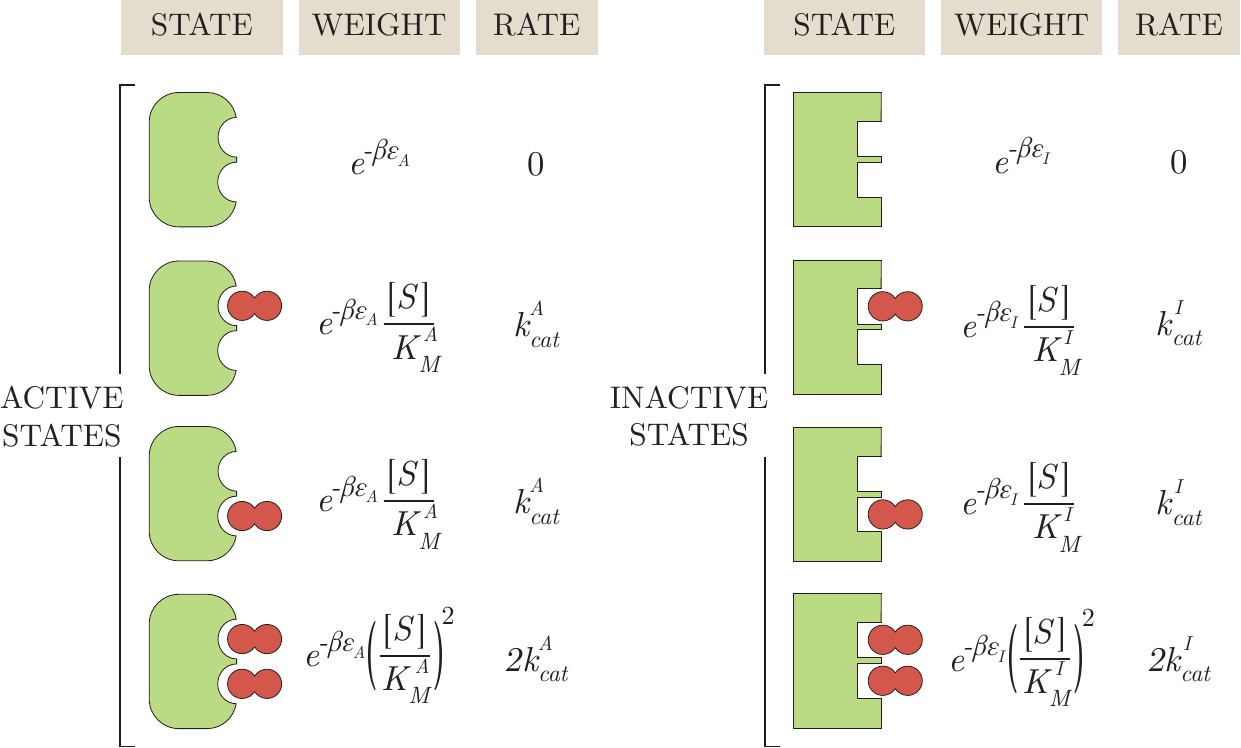}
	\caption{\textbf{States and weights for an MWC enzyme with two substrate binding sites.}
		Reproduced from \fref[fig:Two-site_MWC].} \label{figTwoSiteMWCAppendix}
\end{figure}
This inequality tells us about the nature of the enzyme. Let us return
momentarily to the states and weights of an allosteric enzyme with two substrate
binding sites in the absence of competitive inhibitor which we reproduce here in
\fref[figTwoSiteMWCAppendix]. The total weights of the enzyme being in any
active state is given by the sum of the weights in the left column,
\begin{equation}
w_{A} = e^{-\beta \epsilon_A} + e^{-\beta \epsilon_A} \frac{[S]}{K_M^A} + e^{-\beta \epsilon_A} \frac{[S]}{K_M^A} + e^{-\beta \epsilon_A} \left( \frac{[S]}{K_M^A} \right)^2 = e^{-\beta \epsilon_A} \left(1+\frac{[S]}{K_M^A} \right)^2.
\end{equation}
Similarly, the total weight of the enzyme being in any inactive state is given by 
\begin{equation}
w_{I} = e^{-\beta \epsilon_I} + e^{-\beta \epsilon_I} \frac{[S]}{K_M^I} + e^{-\beta \epsilon_I} \frac{[S]}{K_M^I} + e^{-\beta \epsilon_I} \left( \frac{[S]}{K_M^I} \right)^2 = e^{-\beta \epsilon_I} \left(1+\frac{[S]}{K_M^I} \right)^2.
\end{equation}
Therefore, the relation \eref[eq:peakConditionCompetitorActivityEquationNoCd]
states that the total weight of the active states is smaller than the total
weight of the inactive states, $w_A < w_I$, or equivalently that the enzyme (in
the absence of a competitive inhibitor) is more likely to be in an inactive
state.

We now return to the more general case when $\frac{C_D^A}{C_D^I} > 0$. Recall
that as $\frac{C_D^A}{C_D^I}$ increases, so does the relative affinity of the
competitive inhibitor to the inactive states over the active states. We can
rewrite the peak condition when $\frac{C_D^A}{C_D^I} > 0$ from
\eref[eq:peakConditionCompetitorActivityEquation] as
\begin{equation} \label{eq:peakConditionCompetitorActivityEquationModified}
e^{-\beta \epsilon_A} \left( 1+\frac{[S]}{K_M^A} \right)^2 < e^{-\beta \epsilon_I} \left( 1+\frac{[S]}{K_M^I}\right)^2 - \bigg\{ 2 e^{-\beta \epsilon_I} \frac{C_D^A}{C_D^I} \left( 1+\frac{[S]}{K_M^I} \right) \left( 1+\frac{[S]}{K_M^A} \right) \bigg\} \;\;\;\;\;\; (k_{cat}^I=0).
\end{equation}
The term in curly braces $\{ \cdot \cdot \cdot \}$ on the right is positive and
increases with $\frac{C_D^A}{C_D^I}$. Compared to the special case
$\frac{C_D^A}{C_D^I} = 0$ in
\eref[eq:peakConditionCompetitorActivityEquationNoCd], an enzyme satisfying
\eref[eq:peakConditionCompetitorActivityEquationModified] must favor the
inactive states over the active states to a greater extent. More formally, the
maximal ratio $\frac{w_A}{w_I}$ of the active state weights to inactive state
weights that permits a peak decreases as $\frac{C_D^A}{C_D^I}$ increases.

Second, consider the limit $C_D^A = C_D^I$ where the competitive inhibitor
equally favors the active and inactive states. According to
\eref[eq:peakConditionCompetitorActivityEquation], a peak can still occur
provided that
\begin{equation} \label{eq:peakConditionCompetitorActivityEquationLimitCDZero}
1+e^{-\beta \left(\epsilon _A-\epsilon _I\right)}<\left( \frac{\frac{[S]}{K_M^A}}{1+\frac{[S]}{K_M^A}} \right)^2 \left(\frac{K_M^A}{K_M^I}-1\right)^2 \;\;\;\;\;\;\;\; (k_{cat}^I=0, \, C_D^A=C_D^I).
\end{equation}
It may seem surprising that an inhibitor which binds equally well to the active
and inactive enzyme states can increase the amount of active state enzymes as
per \fref[fig:pictoralCompetitorAcceleration]. However,
\eref[eq:peakConditionCompetitorActivityEquationNoCd] shows that any enzyme that
exhibits inhibitor acceleration must favor the inactive states more in the
absence of inhibitor. Relative to this pool of enzyme which are mostly in the
inactive states, the presence of an inhibitor with $C_D^A=C_D^I$ will increase
the fraction of enzymes in the active states. 


Finally, we consider the case where introducing a competitor keeps the same
fraction of enzymes in the active and inactive states, and we expect that this case
cannot generate a peak in activity. Drawing on the states and weights in
\fref[fig:CompetitorStatesWeights] (but recalling that our enzyme has two active
sites), the dissociation constants $C_D^A$ and $C_D^I$ of such a competitive
inhibitor must satisfy
\begin{equation} \label{eq:competitorEqualProbabilityScenario}
\frac{e^{-\beta \epsilon_A} \left(1+\frac{[S]}{K_M^A}+\frac{[C]}{C_D^A} \right)^2}{e^{-\beta \epsilon_I} \left(1+\frac{[S]}{K_M^I}+\frac{[C]}{C_D^I}\right)^2} = \frac{e^{-\beta \epsilon_A} \left(1+\frac{[S]}{K_M^A} \right)^2}{e^{-\beta \epsilon_I} \left(1+\frac{[S]}{K_M^I}\right)^2}.
\end{equation}
The only solution to this equation occurs when 
\begin{equation} \label{eq:competitorKeepsSameFraction}
\frac{C_D^A}{C_D^I} = \frac{1+\frac{[S]}{K_M^I}}{1+\frac{[S]}{K_M^A}},
\end{equation}
which upon substitution into \eref[eq:peakConditionCompetitorActivityEquation] yields
the expected result that a peak cannot occur when the competitive inhibitor does
not change the balance between the active and inactive states. One might expect
that for all values of $\frac{C_D^A}{C_D^I}$ smaller than this (where the
inhibitor does push more enzymes into the active state), a peak could occur.
However, \eref[eq:peakConditionCompetitorActivityEquation] indicates that a can only
occur provided that a stronger constraint holds, namely
\begin{equation} \label{eq:competitorNecessaryPeakCriteria}
2\frac{C_D^A}{C_D^I} < \frac{1+\frac{[S]}{K_M^I}}{1+\frac{[S]}{K_M^A}}.
\end{equation}

\begin{figure}[h!]
	\centering \includegraphics[scale=1]{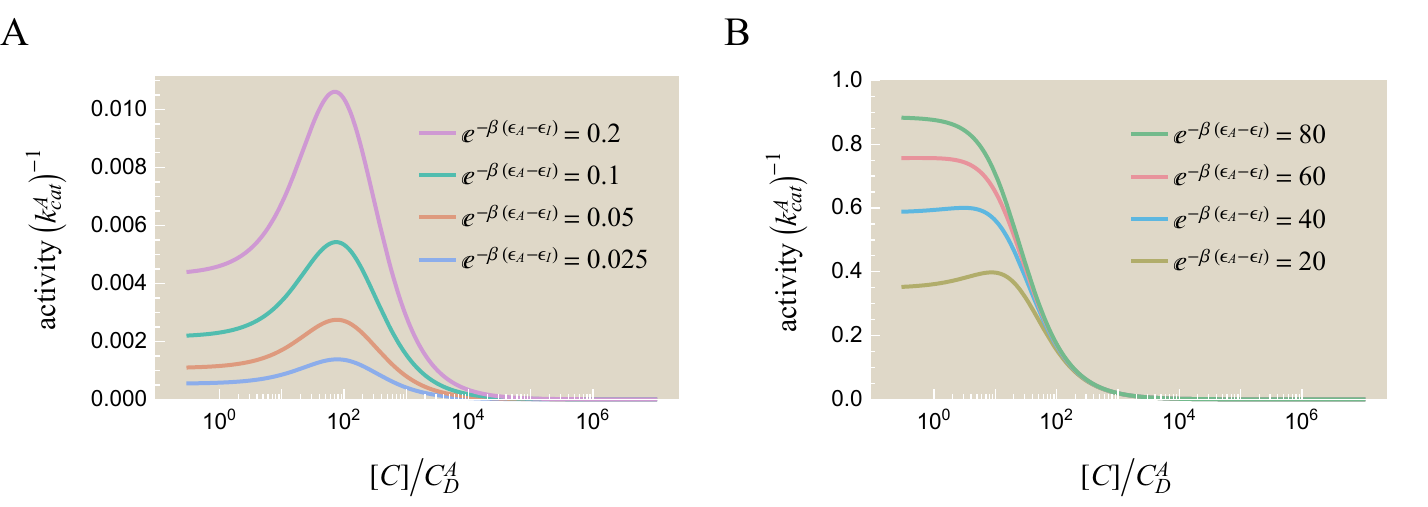} \caption{\textbf{Peak in enzyme
		activity $\boldsymbol{A = \frac{1}{E_{tot}}\frac{d[P]}{dt}}$ as a function of
		\textit{competitive inhibitor} concentration $\boldsymbol{[C]}$.} As shown in
		\fref[fig:peakIntroductoryCartoons]\letter{B}, with Michaelis-Menten kinetics
		adding a competitive inhibitor can only slow down activity, but an MWC enzyme
		can be activated by an inhibitor which results in a peak. Peak are shown for
		\letterParen{A} small and \letterParen{B} large ratios of the enzyme's energy
		in the active versus inactive state, $e^{-\beta \left(\epsilon _A-\epsilon
			_I\right)}$. As in the case of substrate inhibition, the height of the peak
		increases with $e^{-\beta \left(\epsilon _A-\epsilon _I\right)}$. The activity
		is computed from \eref[eq:twoSubstrateSiteCompetitorActivityEquation] using the
		parameters $\frac{[S]}{K_M^A}=10$, $\frac{C_D^A}{C_D^I}=10^{-2}$, the
		parameters from \fref[fig:TwoSiteMWCPeakPlot], and the different values of
		$e^{-\beta \left(\epsilon _A-\epsilon _I\right)}$ shown. 
		As predicted by \eref[eq:peakConditionCompetitorActivityEquation], for the
		parameters chosen every value in the range $e^{-\beta \left(\epsilon
			_A-\epsilon _I\right)}<65$ will yield a peak in activity.}
	\label{fig:TwoSiteMWCPeakPlotCompetitor}
\end{figure}

Having analyzed these specific cases, we now turn to some general
characteristics of this peak. Having calculated the concentration $[C]_0$ in
Appendix \ref{appendixCompetitiveInhibitor2SitesSection} where the peak occurs,
it is straightforward to compute the maximum height of the activity curve,
\begin{equation} \label{eq:peakHeightInhibitorAcceleration}
A_{peak} = k_{cat}^A \frac{[S]}{K_M^A} \frac{\left( \sqrt{\left(\frac{C_D^A}{C_D^I}\right)^2+e^{-\beta \left(\epsilon _A-\epsilon _I\right)}} - \frac{C_D^A}{C_D^I} \right)}{\left(1+\frac{[S]}{K_M^I}\right) - \frac{C_D^A}{C_D^I} \left(1+\frac{[S]}{K_M^A}\right)}.
\end{equation}
Substituting in the peak condition
\eref[eq:peakConditionCompetitorActivityEquation] we obtain
\begin{equation}
A_{peak} < k_{cat}^A \frac{\frac{[S]}{K_M^A}}{1+\frac{[S]}{K_M^A}}.
\end{equation}
The enzyme can approach the maximum possible activity $k_{cat}^A$ in the limit
$1 \ll \frac{[S]}{K_M^A}$ when the active state enzyme dominates, analogous to
the result for substrate inhibition
\eref[eq:peakHeightSubstrateAccelerationRelation]. We can also compare the peak
height to the activity when no inhibitor is present,
\begin{equation} \label{eq:zeroInhibitorConcentration}
A_{[C] \to 0} = 2k_{cat}^A \frac{e^{-\beta \left(\epsilon _A-\epsilon _I\right)} \left(\frac{[S]}{K_M^A} + \frac{[S]}{K_M^A}^2\right)}{e^{-\beta \left(\epsilon _A-\epsilon _I\right)} \left(1+\frac{[S]}{K_M^A}\right)^2 + \left(1+\frac{[S]}{K_M^I}\right)^2}.
\end{equation}
Examples of such peaks are shown in \fref[fig:TwoSiteMWCPeakPlotCompetitor]. As
in the case of substrate inhibition, the peak height $A_{peak}$
monotonically increases and the relative peak height $\frac{A_{peak}}{A_{[C] \to
		0}}$ monotonically decreases with the energy difference between the active and
inactive state, $e^{-\beta \left(\epsilon _A-\epsilon _I\right)}$.

The enzyme ATCase offers an example of inhibitor acceleration. ATCase is an
allosteric enzyme with 6 active sites and 6 regulatory sites
\cite{Cockrell2013}. In the absence of ligand, ATCase exists in an equilibrium
between the unbound active and unbound inactive states, the latter being more
energetically favorable \cite{Fetler2007}. When the inhibitor PALA binds to
ATCase, it strongly induces a transition from inactive to active state
\cite{Mendes2010}, in line with our theoretical prediction. It has been shown
that by adding allosteric regulators, the peak in ATCase activity can be
increased or prevented altogether \cite{Wales1999}. It would be interesting to
undertake the converse experiment and induce inhibitor activation in an enzyme
that typically does not show a peak in activity.

\section{Derivations} \label{derivingPeakCondition}

\subsection{Substrate Inhibition} \label{derivingPeakConditionTwoBindingSites}

We now derive the general peak condition for substrate inhibition without the
extra assumption $k_{cat}^I = 0$ used in the text. Recall that we define the
active state of an enzyme as the state with the greater catalytic rate so that
$k_{cat}^A > k_{cat}^I$. We start by rewriting the full form of the activity
equation \bareEq{eq:twoSubstrateSiteActivityEquation} from
\sref{twoSiteMWCEnzymeSection},
\begin{equation} \label{eq:AppendixtwoSubstrateSiteActivityEquation}
A = \frac{2k_{cat}^A e^{-\beta \epsilon_A}\frac{[S]}{K_M^{A}}\left(1+\frac{[S]}{K_M^{A}} \right)+2k_{cat}^I e^{-\beta \epsilon_I}\frac{[S]}{K_M^{I}}\left(1+\frac{[S]}{K_M^{I}} \right)}{e^{-\beta \epsilon_A}\left(1+\frac{[S]}{K_M^{A}}\right)^2+e^{-\beta \epsilon_I}\left(1+\frac{[S]}{K_M^{I}}\right)^2},
\end{equation}
we derive the peak condition \eref[eq:TwoSiteMWCPeak1a]. We define the numerator
and denominator of the activity as
\begin{equation}
A \equiv \frac{Z_S}{Z_{tot}}
\end{equation}
where, from states and weights in \fref[fig:Two-site_MWC],
\begin{equation}
Z_S=2k_{cat}^Ae^{-\beta \epsilon _A}\frac{[S]}{K_M^A}\left(1+\frac{[S]}{K_M^A}\right)+2k_{cat}^Ie^{-\beta \epsilon _I}\frac{[S]}{K_M^I}\left(1+\frac{[S]}{K_M^I}\right)
\end{equation}
is the sum of all weights multiplied by their rate of product formation and
\begin{equation}Z_{tot}=e^{-\beta \epsilon _A}\left(1+\frac{[S]}{K_M^A}\right)^2+e^{-\beta \epsilon _I}\left(1+\frac{[S]}{K_M^I}\right)^2
\end{equation}
is the sum of all weights. By varying the substrate concentration \([S]\), we
find a peak in the activity \(A\) provided that
\begin{equation} \label{eqSubstrateInhibitionFormulaPeak}
\frac{dA}{d[S]}=\frac{\frac{dZ_S}{d[S]}Z_{tot}-Z_S\frac{dZ_{tot}}{d[S]}}{Z_{tot}^2}=0.
\end{equation}
Thus, a peak occurs if the numerator
$\frac{dZ_S}{d[S]}Z_{tot}-Z_S\frac{dZ_{tot}}{d[S]}$ equals zero. Because $Z_S$
and $Z_{tot}$ are quadratic in $[S]$, the terms $\frac{dZ_S}{d[S]}Z_{tot}$ and
$Z_S\frac{dZ_{tot}}{d[S]}$ in the numerator are cubic in $[S]$. However, the
cubic terms exactly cancel each other, so that
\eref[eqSubstrateInhibitionFormulaPeak] becomes a quadratic equation, 
\begin{equation}
0=\frac{dZ_S}{d[S]}Z_{tot}-Z_S\frac{dZ_{tot}}{d[S]}\equiv 2\left(K_M^AK_M^I\right)^4\left(a[S]^2+b[S]+c\right),
\end{equation}
where we have pulled out the prefactor $2\left(K_M^AK_M^I\right)^4$ for
convenience and
\begin{align}
a=\left(e^{-\beta \epsilon _A}+e^{-\beta \epsilon _I}\right)&\left(\frac{e^{-\beta \epsilon _A}k_{cat}^A}{\left(K_M^A\right)^3}+\frac{e^{-\beta \epsilon _I}k_{cat}^I}{\left(K_M^I\right)^3}\right)\\
-e^{-\beta \epsilon _A}e^{-\beta \epsilon _I}&\left(\frac{1}{K_M^A}-\frac{1}{K_M^I}\right)^2\left(\frac{k_{cat}^A}{K_M^A}+\frac{k_{cat}^I}{K_M^I}\right)\\
b=2 \left(e^{-\beta \epsilon _A}+e^{-\beta \epsilon _I}\right)& \left(\frac{e^{-\beta \epsilon _A} k_{cat}^A}{ \left(K_M^A\right)^2} + \frac{e^{-\beta \epsilon _I} k_{cat}^I}{ \left(K_M^I\right)^2}\right)\\
c=\left(e^{-\beta \epsilon _A}+e^{-\beta \epsilon _I}\right)& \left(\frac{e^{-\beta \epsilon _A} k_{cat}^A}{K_M^A} + \frac{e^{-\beta \epsilon _I} k_{cat}^I}{K_M^I}\right).
\end{align}
The roots of this equation are given by
\begin{equation} \label{eq:quadraticRoots}
[S]_0=\frac{-b\pm \sqrt{b^2-4a c}}{2a}.
\end{equation}
Since \(b,c>0\), there will only be a positive real root \([S]_0>0\) if 
\begin{equation}
a<0.
\end{equation}
Writing this inequality out as
\begin{equation}
\left(e^{-\beta \epsilon _A}+e^{-\beta \epsilon _I}\right)\left(\frac{e^{-\beta \epsilon _A}k_{cat}^A}{\left(K_M^A\right)^3}+\frac{e^{-\beta \epsilon _I}k_{cat}^I}{\left(K_M^I\right)^3}\right) < e^{-\beta \epsilon _A}e^{-\beta \epsilon _I}\left(\frac{1}{K_M^I}-\frac{1}{K_M^A}\right)^2\left(\frac{k_{cat}^A}{K_M^A}+\frac{k_{cat}^I}{K_M^I}\right),
\end{equation}
we multiply by $\frac{\left( K_M^A\right)^3}{e^{-\beta \epsilon _A}e^{-\beta
		\epsilon _I}k_{cat}^I}$ to obtain
\begin{equation}
\left(1+\frac{e^{-\beta \epsilon _A}}{e^{-\beta \epsilon _I}}\right)\left(\frac{k_{cat}^A}{k_{cat}^I}+\frac{e^{-\beta \epsilon _I}}{e^{-\beta \epsilon _A}} \left( \frac{K_M^A}{K_M^I}\right)^3\right) < \left(\frac{K_M^A}{K_M^I}-1\right)^2\left(\frac{k_{cat}^A}{k_{cat}^I}+\frac{K_M^A}{K_M^I}\right)
\end{equation}
and move the \(\frac{k_{cat}^A}{k_{cat}^I}\) terms to one side, 
\begin{equation} \label{eq:aaabbb1}
\left(\frac{K_M^A}{K_M^I}\right)^3\left(\left(1+\frac{e^{-\beta \epsilon _I}}{e^{-\beta \epsilon _A}}\right)-\left(\frac{K_M^I}{K_M^A}-1\right)^2\right)<\frac{k_{cat}^A}{k_{cat}^I}\left(\left(\frac{K_M^A}{K_M^I}-1\right)^2-\left(1+\frac{e^{-\beta \epsilon _A}}{e^{-\beta \epsilon _I}}\right)\right).
\end{equation}
There are now two cases to consider. If the term on the right hand side is positive, 
\begin{equation}
1+\frac{e^{-\beta \epsilon _A}}{e^{-\beta \epsilon _I}}<\left(\frac{K_M^A}{K_M^I}-1\right)^2,
\end{equation}
then we can divide by this term on both sides to obtain the peak condition
\begin{equation}
-\frac{\left(1+\frac{e^{-\beta \epsilon _I}}{e^{-\beta \epsilon _A}}\right)-\left(\frac{K_M^I}{K_M^A}-1\right)^2}{\left(1+\frac{e^{-\beta \epsilon
			_A}}{e^{-\beta \epsilon _I}}\right)-\left(\frac{K_M^A}{K_M^I}-1\right)^2}\left(\frac{K_M^A}{K_M^I}\right)^3<\frac{k_{cat}^A}{k_{cat}^I}.
\end{equation}
On the other hand, if the term on the right-hand side of \eref[eq:aaabbb1] is
negative, then the term on the left-hand side must also be negative,
\begin{align}
1+\frac{e^{-\beta \epsilon _A}}{e^{-\beta \epsilon _I}}&>\left(\frac{K_M^A}{K_M^I}-1\right)^2\label{eq:aaabbb2a}\\
1+\frac{e^{-\beta \epsilon _I}}{e^{-\beta \epsilon _A}}&<\left(\frac{K_M^I}{K_M^A}-1\right)^2\label{eq:aaabbb2b},
\end{align}
and because $e^{-\beta \epsilon _A},e^{-\beta \epsilon _I},K_M^A,K_M^I>0$ this implies 
\begin{equation} \label{eq:aaabbb3}
0<\frac{K_M^A}{K_M^I}<\frac{1}{2}.
\end{equation}
Solving \eref[eq:aaabbb1] for $\frac{k_{cat}^A}{k_{cat}^I}$ (and flipping the
sign of the inequality because of \eref[eq:aaabbb2a]) yields the relation
\begin{equation} \label{eq:aaabbb4}
-\frac{\left(1+\frac{e^{-\beta \epsilon _I}}{e^{-\beta \epsilon _A}}\right)-\left(\frac{K_M^I}{K_M^A}-1\right)^2}{\left(1+\frac{e^{-\beta \epsilon_A}}{e^{-\beta \epsilon _I}}\right)-\left(\frac{K_M^A}{K_M^I}-1\right)^2}\left(\frac{K_M^A}{K_M^I}\right)^3>\frac{k_{cat}^A}{k_{cat}^I}.
\end{equation}
Assuming \eref[eq:aaabbb3], the term on the left-hand side can be at most
$\frac{1}{2}$, so that for an enzyme that satisfies $k_{cat}^A>k_{cat}^I$
\eref[eq:aaabbb4] can never be satisfied. Hence, for a two substrate binding
site enzyme assuming \(k_{cat}^I<k_{cat}^A\), a peak in activity as a function
of substrate concentration \([S]\) will occur if and only if
\begin{align}
\left(1+\frac{e^{-\beta \epsilon _A}}{e^{-\beta \epsilon _I}}\right)<&\left(\frac{K_M^A}{K_M^I}-1\right)^2\label{eq:appendixRef1}\\
-\frac{\left(1+\frac{e^{-\beta \epsilon _I}}{e^{-\beta \epsilon _A}}\right)-\left(\frac{K_M^I}{K_M^A}-1\right)^2}{\left(1+\frac{e^{-\beta \epsilon_A}}{e^{-\beta \epsilon _I}}\right)-\left(\frac{K_M^A}{K_M^I}-1\right)^2}&\left(\frac{K_M^A}{K_M^I}\right)^3<\frac{k_{cat}^A}{k_{cat}^I}.\label{eq:appendixRef2}
\end{align}
In the text, we assumed $k_{cat}^I=0$ so that the second condition
\eref[eq:appendixRef2] is automatically satisfied and \eref[eq:appendixRef1] became
the only necessary condition for a peak. In the general case when
$k_{cat}^I$ is not negligible, the second constraint \eref[eq:appendixRef2]
ensures that the contribution of product formation from the inactive state does
not destroy the peak which would be formed by the active states alone.


Activity curves that exhibit a peak with a non-zero $k_{cat}^I$ value are shown
in \fref[fig:AppendixTwoSiteMWCPeakPlot]. Although these curves look very similar
to those shown in \fref[fig:TwoSiteMWCPeakPlot] for the case $k_{cat}^I=0$, one
important difference is that given $K_M^{A,I}$ and $k_{cat}^{A,I}$ values, there
is now a \textit{lower} bound for $e^{-\beta \left(\epsilon _A-\epsilon
	_I\right)}$ given by the second peak condition \eref[eq:appendixRef2].

\begin{figure}[h!]
	\centering \includegraphics[scale=1]{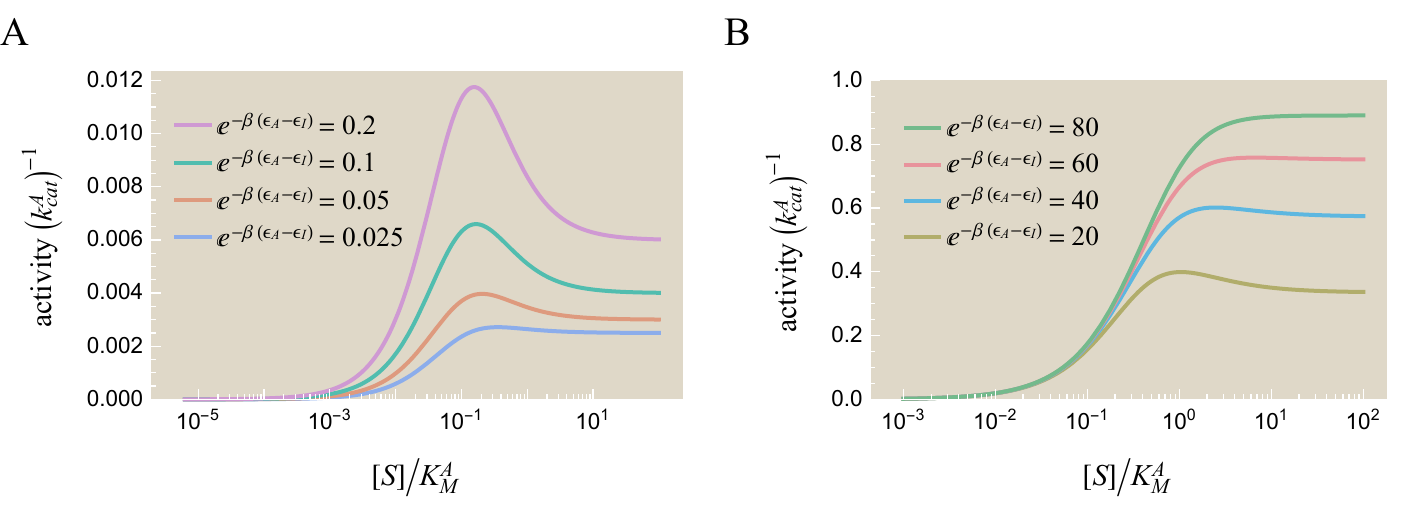}
	\caption{\textbf{Peak in enzyme activity $\boldsymbol{A = \frac{1}{E_{tot}}\frac{d[P]}{dt}}$ as a
		function of \textit{substrate} concentration $\boldsymbol{[S]}$.} As shown in
		\fref[fig:peakIntroductoryCartoons]\letter{A}, with Michaelis-Menten kinetics
		adding substrate can only increase enzyme activity, but an MWC enzyme can
		exhibit a peak due to the interactions between the active and inactive state.
		Peaks are shown for \letterParen{A} small and \letterParen{B} large ratios of
		the enzyme's energy in the active versus inactive state, $e^{-\beta
			\left(\epsilon _A-\epsilon _I\right)}$. The activity is computed from
		\eref[eq:AppendixtwoSubstrateSiteActivityEquation] using the same parameter
		values from \fref[fig:TwoSiteMWCPeakPlot] except that
		$\frac{k_{cat}^A}{k_{cat}^I}=10^3$. The curves with small $e^{-\beta
			\left(\epsilon _A-\epsilon _I\right)}$ values shown in \letterParen{A} vary
		appreciably from those in \fref[fig:TwoSiteMWCPeakPlot] (where $k_{cat}^I = 0$)
		because the inactive state catalyzes substrate. This changes both the shape and
		the height of the activity curves.} \label{fig:AppendixTwoSiteMWCPeakPlot}
\end{figure}

It is straightforward to substitute the positive root for substrate
concentration \eref[eq:quadraticRoots] into the activity
\eref[eq:AppendixtwoSubstrateSiteActivityEquation] to find the height of the
peak, resulting in
\begin{equation}
A_{peak} = \frac{k_{cat}^I K_M^A-k_{cat}^A K_M^I + \sqrt{\left(\frac{1}{e^{-\beta \epsilon_I}} + \frac{1}{e^{-\beta \epsilon_A}} \right) \left( e^{-\beta \epsilon_I} \left(k_{cat}^I K_M^A\right)^2 + e^{-\beta \epsilon_A} \left(k_{cat}^A K_M^I\right)^2 \right)}}{K_M^A-K_M^I}.
\end{equation}
In the limit $k_{cat}^I = 0$ discussed in the text, this simplifies to
\begin{equation}
A_{peak} = k_{cat}^A \frac{K_M^I}{K_M^A-K_M^I} \left( \sqrt{1+\frac{e^{-\beta \epsilon_A}}{e^{-\beta \epsilon_I}}} - 1 \right).
\end{equation}

Lastly, we note that adding a fixed amount of competitive inhibitor \([C]\) to a
system may induce a peak in activity as a function of substrate concentration
\([S]\), as shown in \fref[fig:explainingPeak4]. In the language of the MWC
model
(\eref[eq:competitorEnzymeEquation112a][eq:competitorEnzymeEquation112b][eq:competitorEnzymeEquation112c][eq:competitorEnzymeEquation112d] in the text), adding the inhibitor tunes the MWC parameters so that the peak conditions \eref[eq:appendixRef1][eq:appendixRef2] apply.

\begin{figure}[h!]
	\centering \includegraphics[scale=1]{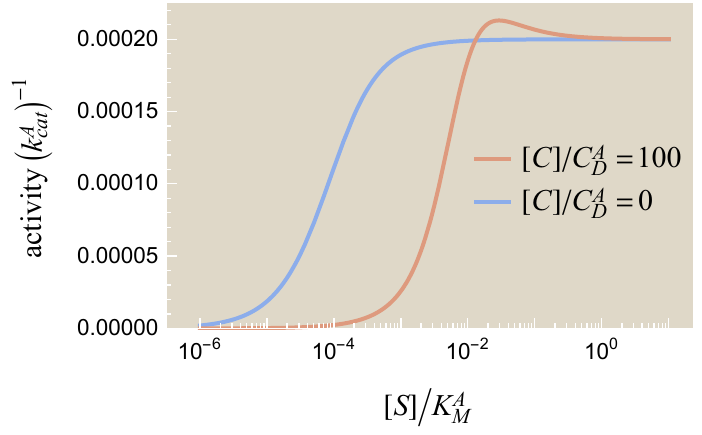} \caption{\textbf{Peaks in
		activity can be induced by a competitive inhibitor.} Adding a competitive
		inhibitor can induce a peak in activity $\frac{d[P]}{dt}$ versus substrate
		concentration $[S]$. Curves are shown for an enzyme with two active sites using the
		parameters $\frac{k_{cat}^A}{k_{cat}^I}=10^4$, $\frac{K_M^A}{K_M^I}=10^{4}$,
		$\frac{C_D^A}{C_D^I}=10^{-1}$, and $e^{-\beta \left(\epsilon _A - \epsilon _I
			\right)}=\frac{1}{2}$.} \label{fig:explainingPeak4}
\end{figure}

\subsection{Inhibitor Acceleration} \label{appendixCompetitiveInhibitor2SitesSection}

We now derive the peak condition for Inhibitor Acceleration discussed in
Appendix \ref{AcceleratedActivityInhibitionSection}. For an enzyme with two substrate binding
sites and a competitive inhibitor $C$, enzyme activity is given by 
\begin{align}
A=k_{cat}^A&\left(2p_{E_AS}\right)+k_{cat}^A\left(2p_{E_ASC}\right)
+2k_{cat}^A\left(p_{E_AS^2}\right) \nonumber \\
&+k_{cat}^I\left(2p_{E_IS}\right)+k_{cat}^I\left(2p_{E_ISC}\right)
+2k_{cat}^I\left(p_{E_IS^2}\right).
\end{align}
Assuming $k_{cat}^I=0$ for simplicity, this equation takes the form 
\begin{align} \label{eq:AppendixD1}
A &= 2 k_{cat}^A e^{-\beta \epsilon_A} \frac{\frac{[S]}{K_M^A}+\frac{[S]}{K_M^A} \frac{[C]}{C_D^A}+\left( \frac{[S]}{K_M^A} \right)^2}{e^{-\beta \epsilon _A} \left(1+\frac{[S]}{K_M^A}+\frac{[C]}{C_D^A}\right)^2+e^{-\beta \epsilon _I} \left(1+\frac{[S]}{K_M^I}+\frac{[C]}{C_D^I}\right)^2}\nonumber\\
&\equiv 2 k_{cat}^A e^{-\beta \epsilon_A} \frac{Z_C}{Z_{tot}}
\end{align}
where 
\begin{align}
Z_C&=\frac{[S]}{K_M^A}+\frac{[S]}{K_M^A} \frac{[C]}{C_D^A}+\left( \frac{[S]}{K_M^A} \right)^2\\
Z_{tot}&=e^{-\beta \epsilon _A} \left(1+\frac{[S]}{K_M^A}+\frac{[C]}{C_D^A}\right)^2+e^{-\beta \epsilon _I} \left(1+\frac{[S]}{K_M^I}+\frac{[C]}{C_D^I}\right)^2.
\end{align}
A peak in activity will occur provided that
\begin{equation}
\frac{dA}{d[C]}=2 k_{cat}^A e^{-\beta \epsilon_A} \frac{\frac{dZ_C}{d[C]}Z_{tot}-Z_C\frac{dZ_{tot}}{d[C]}}{Z_{tot}^2}=0,
\end{equation}
or equivalently that the numerator
$\frac{dZ_C}{d[C]}Z_{tot}-Z_C\frac{dZ_{tot}}{d[C]}$ equals zero. We can rewrite
the numerator 
as
\begin{equation}
0=\frac{dZ_C}{d[C]}Z_{tot}-Z_C\frac{dZ_{tot}}{d[C]}\equiv \frac{[S]}{K_M^A}\left(a[C]^2+b[C]+c\right)
\end{equation}
where
\begin{align}
a&=-\frac{1}{C_D^A} \left( \frac{e^{-\beta \epsilon _A}}{\left(C_D^A\right)^2} + \frac{e^{-\beta \epsilon _I}}{\left(C_D^I\right)^2} \right)\\
b&=-2 \left( \frac{e^{-\beta \epsilon _A}}{\left(C_D^A\right)^2} + \frac{e^{-\beta \epsilon _I}}{\left(C_D^I\right)^2} \right) \left(1+ \frac{[S]}{K_M^A} \right) [C]\\
c&= \frac{e^{-\beta \epsilon _I}}{C_D^A} \left(1+ \frac{[S]}{K_M^I} \right)^2 - \frac{e^{-\beta \epsilon _A}}{C_D^A} \left(1+ \frac{[S]}{K_M^A} \right)^2 - 2 \frac{e^{-\beta \epsilon _I}}{C_D^I} \left(1+ \frac{[S]}{K_M^A} \right) \left(1+ \frac{[S]}{K_M^I} \right).
\end{align}
The roots of this equation are given by
\begin{equation}
[C]_0=\frac{-b\pm \sqrt{b^2-4a c}}{2a}.
\end{equation}
Since \(a,b<0\), there will only be a positive real root \([C]_0>0\) if 
\begin{equation}
c>0.
\end{equation}
Therefore, the peak condition can be written as 
\begin{equation}
2 \, \frac{C_D^A}{C_D^I} \left(1+ \frac{[S]}{K_M^A} \right) \left(1+ \frac{[S]}{K_M^I} \right) < \left(1+ \frac{[S]}{K_M^I} \right)^2 - \frac{e^{-\beta \epsilon _A}}{e^{-\beta \epsilon _I}} \left(1+ \frac{[S]}{K_M^A} \right)^2
\end{equation}
or equivalently,
\begin{equation}
\frac{e^{-\beta \epsilon _A}}{e^{-\beta \epsilon _I}} < \left(\frac{1+ \frac{[S]}{K_M^I}}{1+ \frac{[S]}{K_M^A}} \right)^2 - 2 \, \frac{C_D^A}{C_D^I} \frac{1+ \frac{[S]}{K_M^I}}{1+ \frac{[S]}{K_M^A}}
\end{equation}
which matches \eref[eq:peakConditionCompetitorActivityEquation], as desired.

\subsection{Michaelis-Menten Enzymes Do Not Exhibit Peaks} \label{derivingNoPeakNonAllostericEnzymes} 

In this section, we show that a Michaelis-Menten enzyme with an arbitrary number
of substrate binding sites cannot exhibit substrate inhibition nor
inhibitor acceleration. This implies that the interplay between the active
and inactive MWC states were necessary to produce the peaks in activity
discussed in \sref{twoSiteMWCEnzymeSection} and Appendix
\ref{AcceleratedActivityInhibitionSection}.

Consider a Michaelis-Menten enzyme with $N$ binding sites where either a
substrate $S$ or a competitive inhibitor $C$ can bind. Using the general
formulation from \sref{MoreComplexEnzymes}, we will assume that the enzyme only
has an active state and drop the $A$ superscripts. Each binding site can be
either be empty, occupied by substrate, or occupied by competitor, which would
contribute a factor of $1$, $\frac{[S]}{K_M}$, or $\frac{[C]}{C_D}$,
respectively, to its weight. A state with $j$ bound substrates forms product at
a rate of $j k_{cat}$. Therefore, the activity
$A=\frac{1}{E_{tot}}\frac{d[P]}{dt}$ equals
\begin{align}
A=&\frac{\sum _{j=0}^N\sum _{k=0}^{N-j}\left(j k_{cat}\right)\frac{N!}{j!k!(N-j-k)!}\left(\frac{[S]}{K_M}\right)^j\left(\frac{[C]}{C_D}\right)^k}{\left(1+\frac{[S]}{K_M}+\frac{[C]}{C_D}\right)^N} \nonumber \\
=& N k_{cat} \frac{\frac{[S]}{K_M}\left(1+\frac{[C]}{C_D}+\frac{[S]}{K_M}\right)^{N-1}}{\left(1+\frac{[S]}{K_M}+\frac{[C]}{C_D}\right)^N} \nonumber \\
=& N k_{cat} \frac{\frac{[S]}{K_M}}{1+\frac{[S]}{K_M}+\frac{[C]}{C_D}}. \label{NoAllosteryNoSoup}
\end{align}
Taking the derivative of the activity with respect to the substrate concentration $[S]$ and the inhibitor concentration $[C]$,
\begin{equation}
\frac{dA}{d[S]} = \frac{N k_{cat}}{K_M} \frac{1+\frac{[C]}{C_D}}{\left(1+\frac{[S]}{K_M}+\frac{[C]}{C_D}\right)^2}
\end{equation}
and
\begin{equation}
\frac{dA}{d[C]} = -\frac{N k_{cat}}{C_D} \frac{\frac{[S]}{K_M}}{\left(1+\frac{[S]}{K_M}+\frac{[C]}{C_D}\right)^2},
\end{equation}
we find that neither derivative can be zero. Therefore, inhibitor acceleration
cannot occur for a non-MWC enzyme.

\end{singlespace}

\begin{singlespace}
\providecommand{\latin}[1]{#1}
\providecommand*\mcitethebibliography{\thebibliography}
\csname @ifundefined\endcsname{endmcitethebibliography}
{\let\endmcitethebibliography\endthebibliography}{}

\end{singlespace} 				
\setcounter{page}{32}

\begin{singlespace}
\providecommand{\latin}[1]{#1}
\providecommand*\mcitethebibliography{\thebibliography}
\csname @ifundefined\endcsname{endmcitethebibliography}
{\let\endmcitethebibliography\endthebibliography}{}

\end{singlespace} 			

\end{document}